\documentclass[aps,prl,groupedaddress,fleqn,twocolumn]{revtex4}
\pdfoutput=1
\usepackage{graphicx}
\usepackage{amsmath}
\usepackage{gensymb}

\begin{document}
\title{Collective modes and thermodynamics of the liquid state}
\author{K. Trachenko$^{1}$}
\author{V. V. Brazhkin$^{2}$}
\address{$^1$ School of Physics and Astronomy, Queen Mary University of London, Mile End Road, London, E1 4NS, UK}
\address{$^2$ Institute for High Pressure Physics, RAS, 142190, Moscow, Russia}

\begin{abstract}
Strongly interacting, dynamically disordered and with no small parameter, liquids took a theoretical status between gases and solids, with the historical tradition of hydrodynamic description as the starting point. We review different approaches to liquids as well as recent experimental and theoretical work, and propose that liquids do not need classifying in terms of their proximity to gases and solids or any categorizing for that matter. Instead, they are a unique system in their own class with a notably mixed dynamical state in contrast to pure dynamical states of solids and gases. We start with explaining how the first-principles approach to liquids is an intractable, exponentially complex problem of coupled non-linear oscillators with bifurcations. This is followed by a reduction of the problem based on liquid relaxation time $\tau$ representing non-perturbative treatment of strong interactions. On the basis of $\tau$, solid-like high-frequency modes are predicted and we review related recent experiments. We demonstrate how the propagation of these modes can be derived by generalizing either hydrodynamic or elasticity equations. We comment on the historical trend to approach liquids using hydrodynamics and compare it to an alternative solid-like approach. We subsequently discuss how collective modes evolve with temperature and how this evolution affects liquid energy and heat capacity as well as other properties such as fast sound. Here, our emphasis is on understanding experimental data in real, rather than model, liquids. Highlighting the dominant role of solid-like high-frequency modes for liquid energy and heat capacity, we review a wide range of liquids: subcritical low-viscous liquids, supercritical state with two different dynamical and thermodynamic regimes separated by the Frenkel line, highly-viscous liquids in the glass transformation range and liquid-glass transition. We subsequently discuss the fairly recent area of liquid-liquid phase transitions, the area where the solid-like properties of liquids have become further apparent. We then discuss gas-like and solid-like approaches to quantum liquids and theoretical issues that are similar to the classical case. Finally, we summarize the emergent view of liquids as a unique system in a mixed dynamical state, and list several areas where interesting insights may appear and continue the extraordinary liquid story.
\end{abstract}

\maketitle
\tableofcontents

\section{Introduction}

Condensed matter physics as a term originated from adding liquids to the then-existing field of solid state physics. Proposals to do so precede what is often thought, and date back to the 1930s when J. Frenkel proposed to develop liquid theory as a generalization of solid state theory and unify the two states under the term ``condensed bodies'' \cite{frenkel}. At the same time, the seeming similarity of liquids and gases in terms of their ability to flow has led to the unified term ``fluids''. Such a dual classification of liquids is more than just semantics: it has given rise to two fundamentally different ways of describing liquids theoretically in hydrodynamic and solid-like approaches. The phase diagram of matter in Figure 1 highlights the intermediate location of liquids between solids and gases and hints at the duality of their physical properties that will come out in our detailed analysis.

\begin{figure}
\begin{center}
{\scalebox{0.5}{\includegraphics{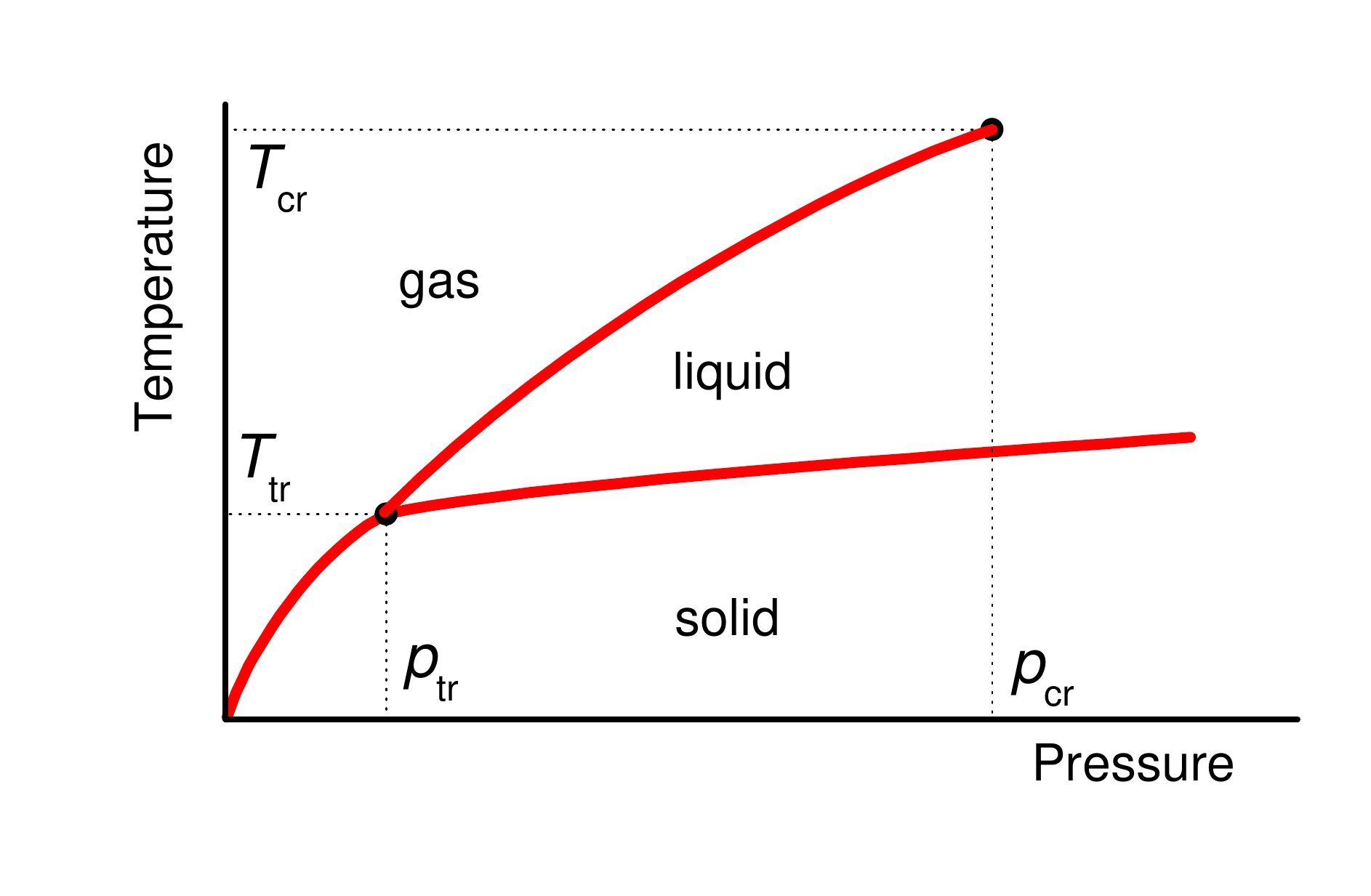}}}
\end{center}
\caption{Colour online. Phase diagram of matter with triple and critical points shown. Schematic illustration.}
\label{double}
\end{figure}

It is the intermediate state of liquids which has ultimately resulted in great difficulties when developing liquid theory because well-developed theoretical tools for the two limiting states of gases and solids failed. It is also the intermediate state of liquids and the combination of solid-like and gas-like properties which continues to be remarkably intriguing for theorists. According to Figure 1, one can start in the gas state above the critical point, move to the liquid state and end up in the solid glass state (if crystallization is avoided) in a seemingly continuous way and without any qualitative changes of physical properties. This is a surprising observation from a theoretical point of view and signifies the intermediate state of liquids and the duality of their physical properties.

At the end of this review, we will see that liquids need not be thought of in terms of their proximity to solids or gases and do not require any other categorization: they are self-contained systems with interesting, unique and rich dynamical and thermodynamic properties. In fact, understanding this richness helps better understand the properties of gases and solids by delineating them as two limiting states of matter in terms of dynamics and thermodynamics.

The long and extraordinary history of liquid research includes several notable discouraging assertions. One of the most important properties crucial to properly understanding liquids is that they are strongly-interacting systems. Particles in liquids are close enough to be within the reach of interatomic forces as in solids, resulting in the condensed liquid state. The energy of a system with $N$ particles and pair-wise interaction energy $U(r)$ can be written as

\begin{equation}
E=\frac{3}{2}Nk_{\rm B}T+\frac{N\rho}{2}\int U(r)g(r)dV
\label{01}
\end{equation}

\noindent where $\rho$ is number density and $g(r)$ is pair distribution function.

$U(r)$ is strong and system-dependent; consequently, $E$ or other thermodynamic properties of the liquid are strongly system-dependent. For this reason, Landau and Lifshitz assert \cite{landau} (twice, in paragraphs 66 and 74) that it is impossible to derive any general equation describing liquid properties or their temperature dependence. Whatever approximation scheme or method used, any approach aimed at deriving a generally applicable result using Eq. (\ref{01}), or evaluating the configurational part of the partition function, is destined to fail.

The above problem does not originate in strongly-interacting solids because the smallness of atomic vibrations around the fixed reference lattice, crystalline or amorphous, enables expansion of the potential energy in Taylor series. The harmonic term in this expansion, combined with the kinetic term, gives the phonon energy of the solid consistent with experimental heat capacities. These can be corrected by the next-order terms in the Taylor series for potential energy. Traditionally, this approach is deemed inapplicable to liquids due to the absence of fixed reference points around which an expansion can be made. The problem also does not originate in weakly-interacting gases: they have no fixed reference points but interactions are small so that the perturbation theory is warranted.

Liquids have neither the small displacements of solids nor the small interactions of gases. Summarized aptly by Landau, liquids have no small parameter.

For this reason, we are seemingly compelled to treat liquids as general strongly-interacting disordered systems, where disorder is both static and dynamic, with no simplifying assumptions. In this spirit, large amount of work was aimed at elucidating the structure and dynamics of liquids. In comparison, the discussion of liquid thermodynamic properties such as heat capacity is nearly non-existent. Indeed, physics textbooks have very little, if anything, to say about liquid specific heat, including textbooks dedicated to liquids \cite{landau,hydro,ziman,boonyip,march,march1,baluca,zwanzig,faber,hansen1,hansen2}. In an amusing story about his teaching experience in the University of Illinois (UIUC), Granato recalls living in fear about a potential student question about liquid heat capacity \cite{granato}. Observing that the question was never asked by a total of 10000 students, Granato proposes that ``...an important deficiency in our standard teaching method is a failure to mention sufficiently the unsolved problems in physics. Indeed, there is nothing said about liquids [heat capacity] in the standard introductory textbooks, and little or nothing in advanced texts as well. In fact, there is little general awareness even of what the basic experimental facts to be explained are.'' It is probably fair to say that the question of liquid heat capacity would be out of the comfort zone not only for general condensed matter practitioners but also for many working in the area related to the liquid state such as soft matter.

Historically, thermodynamic properties of liquids have been approached from the gas state, a seemingly appropriate approach in view of liquid fluidity. For example, common approaches start with the kinetic energy of the gas and aim to calculate the potential energy using the perturbation theory. The dynamical properties of liquids are discussed on the basis of hydrodynamic theory where the elements of solid-like behaviour are introduced as a subsequent correction \cite{ziman,boonyip,march,march1,baluca,zwanzig,faber,hansen1,hansen2,enskog}. This is in interesting contrast to experiments informing us that liquids not far from melting points are close to solids in terms of density, bulk moduli, heat capacity and other main properties, but are very different from gases.

\begin{figure}
\begin{center}
{\scalebox{0.4}{\includegraphics{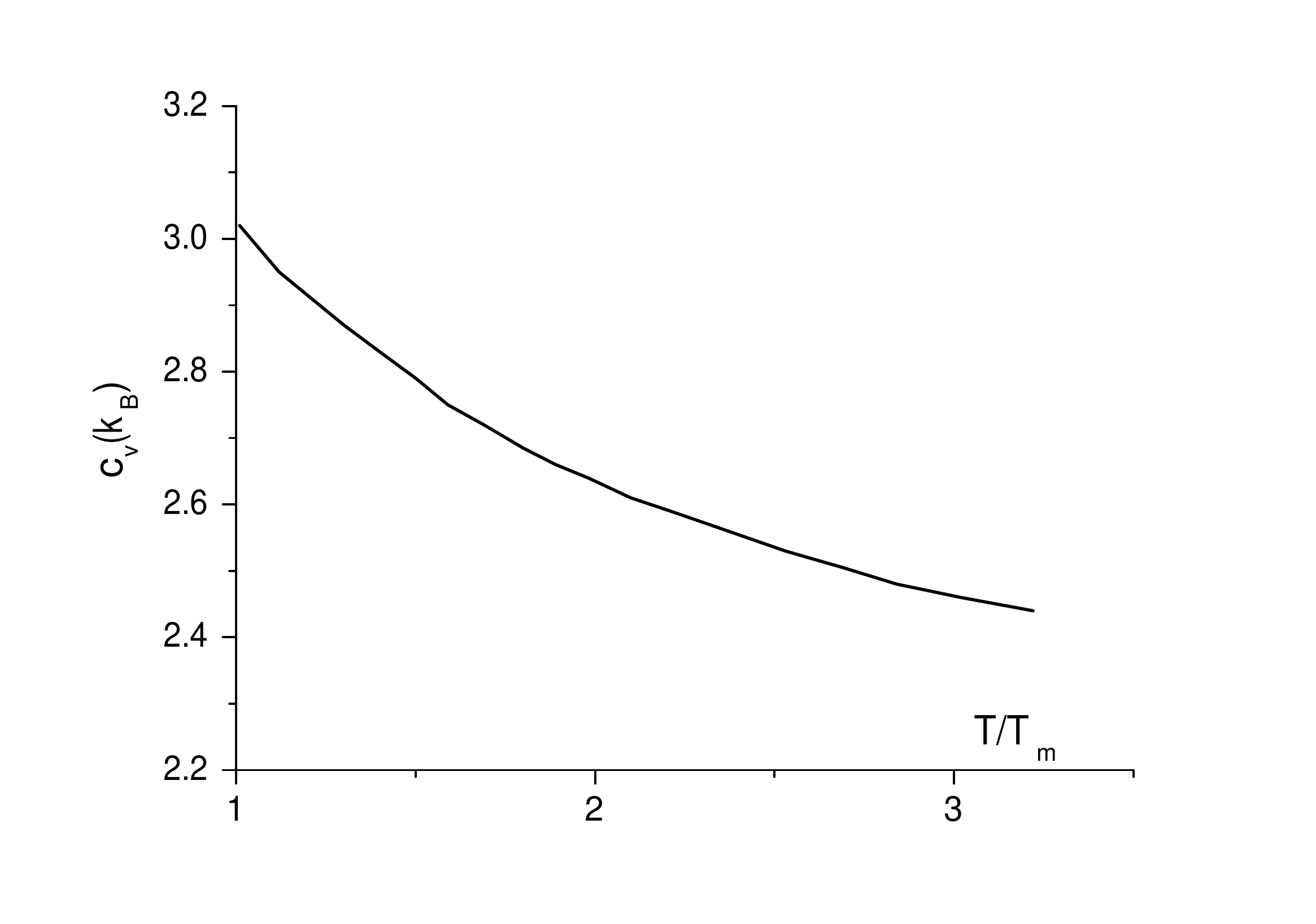}}}
\end{center}
\caption{Experimental specific heat of liquid mercury in $k_{\rm B}$ units \cite{grimvall,wallace}. The $x$-axis is in the relative temperature units where $T_m$ is the melting temperature.}
\label{mercury}
\end{figure}

The focus of this review is on understanding liquid thermodynamic properties such as heat capacity and their relationship to collective modes. To be more specific and set the stage early, we show the experimental specific heat of liquid mercury in Figure \ref{mercury}. We observe that $c_v$ starts from around $3k_{\rm B}$ just above the melting point and decreases to about $2k_{\rm B}$ at high temperature. As discussed below, this effect is very common and operates in over 20 different liquids we analyzed, including metallic, noble, molecular and network liquids, and is present in complex liquids. The decrease of $c_v$ interestingly contrasts the temperature dependence of $c_v$ in solids which is either constant in the classical harmonic case or increases due to anharmonicity or due to phonon excitations at low temperature. We also observe that liquid $c_v$ is significantly larger than the gas value of $\frac{3}{2}k_{\rm B}$ in a wide temperature range in Figure \ref{mercury}.

Notably, the commonly discussed Van Der Waals model of liquids gives $c_v=\frac{3}{2}k_{\rm B}$ \cite{landau}, the ideal gas value. The same result holds for another commonly discussed model of liquids, hard spheres, as well as for several other more elaborate models. Clearly, real liquids have an important mechanism at operation that significantly affects their $c_v$ and that is missed by several common liquid models.

Notwithstanding the theoretical difficulties involved in treating liquids, we rely on the known result that low-energy states of a strongly-interacting system are collective excitations or modes (throughout this review, we use terms ``phonons'', ``modes'' and ``collective excitations'' interchangeably depending on context and common usage). In solids, collective modes, phonons, play a central role in the theory, including the theory of thermodynamic properties. Can collective modes in liquids play the same role? It is from this perspective that we review collective modes in liquids. In our review, we emphasize the main different approaches to collective modes in liquids and list starting equations in each approach. We do not discuss details of how the field has branched out over time; that formidable task is outside the scope of this paper. To a large extent, this was done in earlier textbooks and reviews
\cite{ziman,boonyip,march,march1,baluca,zwanzig,faber,hansen1,hansen2,enskog}.

We focus on real rather than model liquids, measurable effects and take a pragmatic approach to understand the main experimental properties of liquids such as heat capacity and provide relationships between different physical properties. Throughout this review, we seek to make connections between different areas of physics that help understand the problem. We are not trying to be completely comprehensive, focusing instead on providing a pedagogical introduction, interpreting previous basic results and fundamental equations and explaining recent advances. Our discussion includes original work not reported previously as well as results from our published work.

As already mentioned, the long and extraordinary history of liquid research is related to problems of theoretical description. The fundamental problem of the first-principles description of liquids is not generally discussed, so we start with explaining that this problem is due to the intractability of the exponential complexity of finding bifurcations and stationary points in the system of coupled non-linear oscillators. We then discuss how the problem can be reduced using Frenkel's idea of liquid relaxation time. On this basis, several important assertions can be made regarding the continuity of liquid and solid states and the propagation of solid-like collective modes in liquids. We subsequently review how collective modes can be studied by either incorporating elastic effects in hydrodynamic equations or viscous effects in elasticity equations. We find the same results in both approaches, supporting the view that the historical hydrodynamic description of liquids is not unique and that a solid-like description is equally justified. This assertion becomes more specific when we review and comment on generalized hydrodynamics. As far as liquid thermodynamics is concerned, it turns out that the solid-like elastic regime is the relevant one because high-frequency solid-like collective modes contribute most to the energy.

We then proceed to review recent experimental evidence for high-frequency solid-like collective modes in liquids and discuss their similarity to those in solids.

We subsequently discuss how the evolution of collective modes in liquids can be related to liquid energy and heat capacity in widely different liquid regimes: low-viscous subcritical liquids; high-temperature supercritical gas-like fluids; highly-viscous liquids in the glass transformation range; and systems at the liquid-glass transition. In all cases, high-frequency modes govern the main thermodynamic properties of liquids such as energy and heat capacity and affect other interesting effects such as fast sound.

The solid-like properties of liquids have additionally become apparent in the recently accumulated and reviewed data on liquid-liquid phase transitions. We finally discuss the gas-like and solid-like approach in quantum liquids and interesting issues regarding the operation of Bose-Einstein condensates in real liquids.

At the end of this review we will see that most important properties of liquids and supercritical fluids can be consistently understood in the picture in which these systems are in notably {\it mixed} dynamical state. Therefore, the emergent picture of liquids is that they do not need classifying on the basis of their proximity to fluid gases or solids, or any other compartmentalizing for that matter. Instead, they should be considered as distinct systems in the mixed state of particle dynamics, the state that should serve as a starting point for liquid description. Moreover, we will see that appreciating the mixed state of particle dynamics in liquids helps understand gases and solids better as two limiting and dynamically {\it pure} states. This point is particularly useful for understanding the supercritical matter.

We conclude with possible future work which may bring new understanding and advance the remarkable liquid story.

Before we start, we comment on several terms used in this review. Traditionally, the term ``liquids'' is used for subcritical conditions on the phase diagram. The systems above the critical point are often referred to as ``supercritical fluids''. We continue to use these terms in our review where we also propose that the supercritical system in fact consists of two states in terms of particle dynamics and physical properties: a ``rigid liquid-like'' state below the Frenkel line and a ``non-rigid gas-like fluid'' state above the line. We use the term ``glass'' to commonly denote a very viscous liquid which stops flowing at the typical experimental time scale. The term ``viscous liquid'' commonly refers to liquids in the glass transformation range, implying viscosity considerably higher than that in, for example, water at ambient conditions. The term is quantitatively defined at the beginning of the section ``Viscous liquids''.

\section{First-principles approach and its failure}

The absence of a small parameter in liquids pointed out in the Introduction, is one perceived reason that makes the theoretical description of liquids difficult. It tells us why perturbation-based approaches that are successful in solids and gases do not work in liquids. Yet it is interesting to explore the actual reason for the difficulty of constructing a first-principles theory of liquids using the same microscopic approach as in the solid theory. As far as we know, this point is not discussed in textbooks \cite{frenkel,landau,hydro,ziman,boonyip,march,march1,baluca,zwanzig,faber,hansen1,hansen2}.

Below we show that the challenge for the first-principles description of liquids can be well formulated in the language of non-linear theory where it acquires a specific meaning. In this language, the challenge is related to the intractability of the exponentially complex problem involved in solving a large number of coupled non-linear equations.

First-principles treatment of collective modes in a solid is based on solving coupled Newton equations of motion for $N$ atoms. We assume that the atoms oscillate around fixed lattice points $q_{i0}$, and introduce atomic coordinates $q_i$ and displacements $x_i=q_i-q_{i0}$. The potential energy is expanded in series as far as quadratic terms:

\begin{equation}
U=U_0+\frac{1}{2}\sum_{ij}k_{ij}x_j x_k
\label{harmonic}
\end{equation}

Writing the equations of motion as

\begin{equation}
\sum_i m_i \ddot{x}_i+k_{ji}x_i=0
\label{haeq}
\end{equation}

\noindent and seeking the solutions as $x_k=b_k\exp(i\omega t)$ gives the characteristic equation for the eigenfrequencies

\begin{equation}
\left|k_{ij}-\omega^2m_i\right|=0
\label{eigen}
\end{equation}

Eq. (\ref{eigen}) gives most detailed information about collective modes in the system, and returns $3N$ eigenfrequencies, ranging from the lowest frequency set by the system size to the largest frequency in the system, often referred to as Debye frequency (note that Debye frequency is the result of quadratic approximation to the energy spectrum, and is somewhat lower than the maximal frequency of the real spectrum). Each atomic coordinate can be expressed as a superposition of normal coordinates as

\begin{equation}
x_k=\sum\limits_\alpha\Delta_{k\alpha}\Theta_\alpha
\end{equation}

\noindent where $\Theta_\alpha={\rm Re}\left(C_\alpha\exp(i\omega_{\alpha}t)\right)$ are normal coordinates, $C_\alpha$ are arbitrary complex constants and $\Delta_{k\alpha}$ are minors of the determinant (\ref{eigen}) \cite{landmech}. This result is central for the development of many areas in the solid state theory.

Note that the above treatment does not assume a crystalline lattice. Crystallinity, if present, is the next step in the treatment enabling to write the solution as a set of plane waves with $x\propto\exp(ikan)$, where $k$ is the wavenumber and $a$ is the shortest interatomic separation, and derive dispersion curves for model systems.

To continue to use the first-principles description of liquids, we need to account for particle rearrangements in liquids. As discussed in the next section, particle dynamics in the liquid consists of small solid-like oscillations around quasi-equilibrium positions and diffusive jumps to new neighbouring locations. This corresponds to potential energy of the double-well form shown in Figure \ref{double} which endows particles with both oscillatory motion and thermally-induced jumps between different minima. Note that in an equilibrium liquid, each diffusing particle visits many minima, hence the potential energy is multi-well, however the minima and energy profiles can be assumed to be close to their averages in a homogeneous system so that the double-well potential in Figure \ref{double} suffices.

To model the double-well energy, the harmonic expansion (\ref{harmonic}) needs to be extended to include higher terms, at which point the equations of motion become non-linear. The simpler form often considered includes the third and fourth powers of $x_i$, ``$-x^3+x^4$'':

\begin{figure}
\begin{center}
{\scalebox{0.35}{\includegraphics{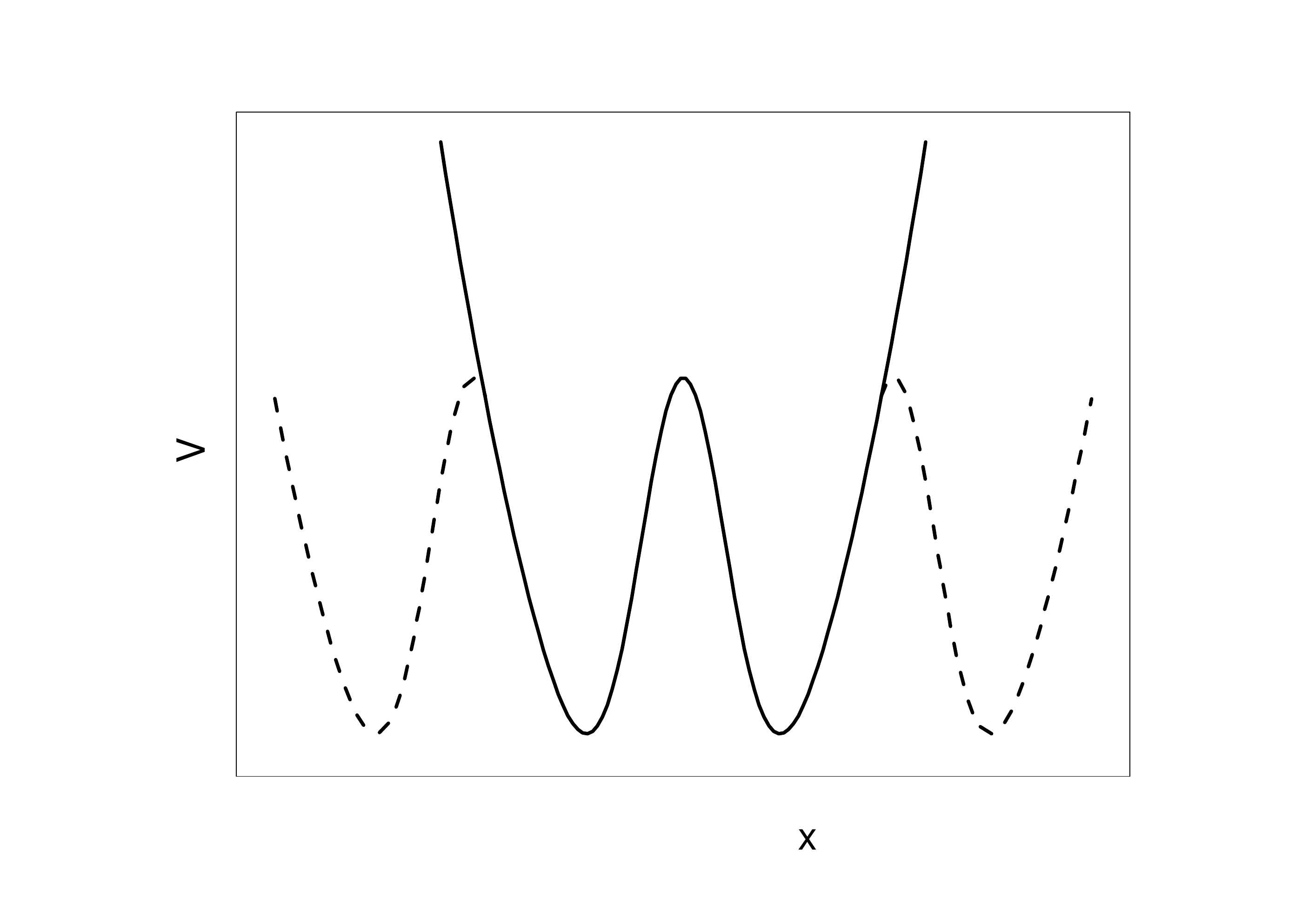}}}
\end{center}
\caption{Double-well potential describing the particle motion in liquids and involving jumps between different quasi-equilibrium positions.}
\label{double}
\end{figure}

\begin{equation}
\begin{aligned}
&U=U_0+\frac{1}{2}\sum_{ij}k_{ij}x_j x_k+\\
&\sum_{ijl}k_{ijl}x_ix_jx_l+\sum_{ijlm}k_{ijl}x_ix_jx_lx_m+...
\end{aligned}
\label{anh1}
\end{equation}

\noindent or, if a symmetric form of $U$ is preferred, the higher-order potential can be written in ``$-x^4+x^6$'' or similarly symmetric form as in Figure \ref{double}.

At small enough energy or temperature of particles motion, Eq. (\ref{anh1}) is used to describe the effects of anharmonicity of atomic motion in solids using the perturbation theory. The main results include the correction to the Dulong-Petit result of solids, thermal expansion and modification of the phonon spectrum, phonon scattering and so on. Unfortunately, the quantitative evaluation of anharmonicity effects has remained a challenge, with the frequent result that the accuracy of leading-order anharmonic perturbation theory is unknown and the magnitude of anharmonic terms is challenging to justify \cite{cowley,marad,grimvall0,wallace1,fultz}. Experimental data such as phonon lifetimes and frequency shifts can provide quantitative estimates for anharmonicity effects and expansion coefficients in particular, although this involves complications and limits the predictive power of the theory \cite{cowley}.

The real problem is at higher energy where the anharmonicity in Eq. (\ref{anh1}) is not small and jumps between different minima in Fig. \ref{double} become operative, as they do in liquids. Here, the perturbation approach does not apply, and we enter the realm of non-linear physics \cite{nonlinear,nonlinear1}. The illustrative example is the simplest system of two coupled Duffing oscillators with the energy (see, e.g. \cite{nonlinear}):

\begin{equation}
E=\sum_{i=1}^2\left(\frac{1}{2}\dot{x}_i^2+\frac{\alpha}{2}x_i^2-\frac{\beta}{4}x_i^4\right)+\frac{\epsilon^2}{2}(x_1-x_2)^2
\end{equation}

and the equations of motion

\begin{equation}
\begin{aligned}
&\ddot{x_1}+\alpha x_1+\epsilon^2(x_1-x_2)-\beta x_1^3=0\\
&\ddot{x_2}+\alpha x_2+\epsilon^2(x_2-x_1)-\beta x_2^3=0
\end{aligned}
\label{nonl1}
\end{equation}

\noindent where $\epsilon$ is the coupling strength.

This model is not integrable, and can not be solved analytically but using approximations only. However, a simpler model can written in terms of variables $\psi_n=\frac{1}{\sqrt{2\omega_0}}\left(\omega_0x_n+i\frac{dx_n}{dt}\right)$, where $\omega_0$ is the frequency of the uncoupled oscillator:

\begin{equation}
\begin{aligned}
&i\frac{d\psi_1}{dt}=\omega_0\psi_1+\frac{\Omega}{2}(\psi_1-\psi_2)-\alpha\left|\psi_1\right|^2\psi_1\\
&i\frac{d\psi_2}{dt}=\omega_0\psi_2+\frac{\Omega}{2}(\psi_2-\psi_1)-\alpha\left|\psi_2\right|^2\psi_2
\end{aligned}
\label{nonl2}
\end{equation}

\noindent where the last terms represent the non-linearity and $\Omega=\frac{\epsilon^2}{\omega_0}$.

\begin{figure}
\begin{center}
{\scalebox{0.55}{\includegraphics{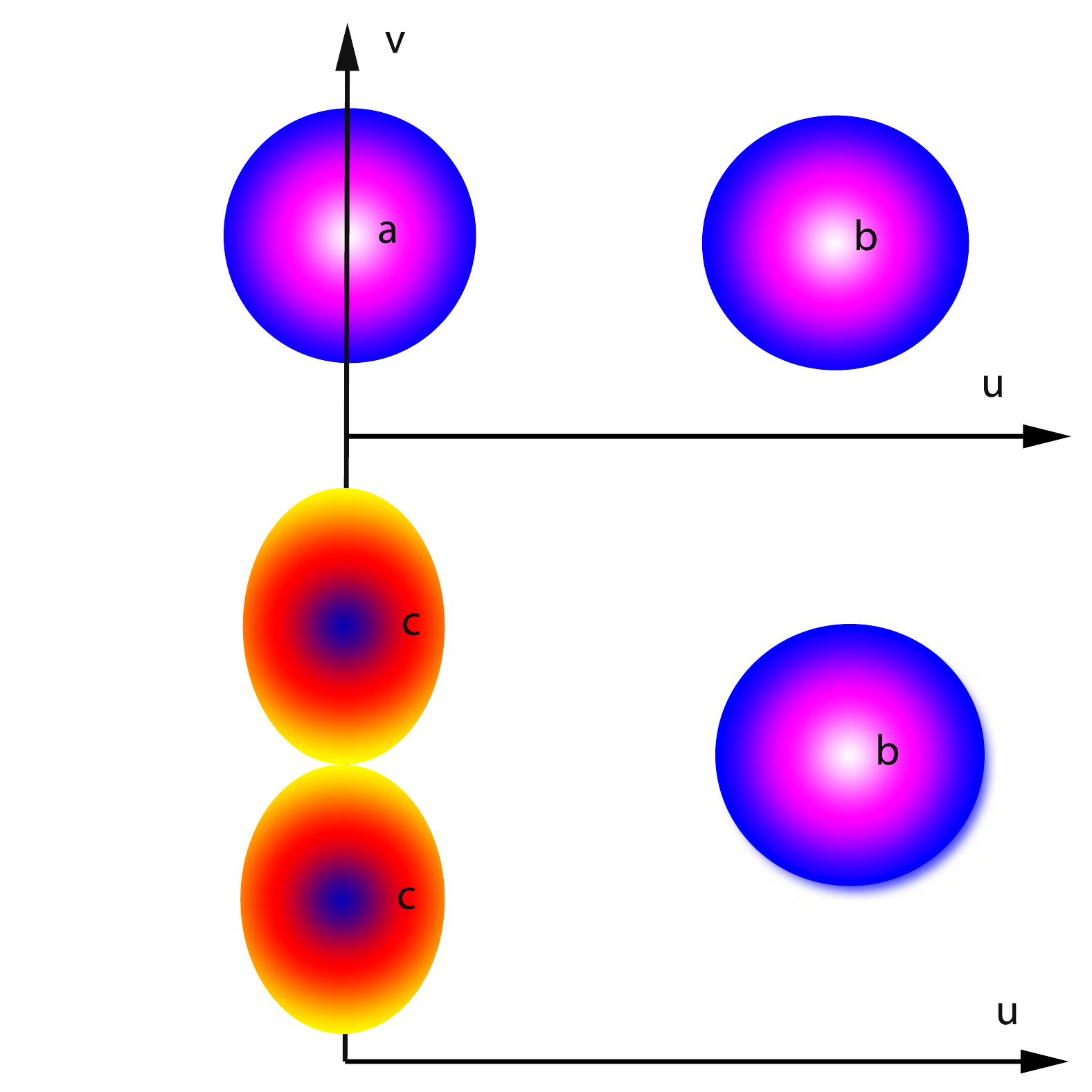}}}
\end{center}
\caption{Colour online. The phase portrait of the discrete trapping model (\ref{nonl2}) at two different energies. Independent variables $u$ and $v$ are functions of $\psi_1$ and $\psi_2$ in Eq. (\ref{nonl2}), and describe the trajectories of two coupled non-linear oscillators. At low energy (top), two stationary points ``a'' and ``b'' remain unchanged as in the case of two coupled linear oscillators, corresponding to weak non-linearity. At high energy (bottom), the bifurcation takes place: point ``a'' becomes an unstable saddle point and two new stationary points ``c'' appear. Schematic illustration, adapted from Ref. \cite{nonlinear}.}
\label{bifur}
\end{figure}

Eqs. (\ref{nonl2}) is known as the discrete self-trapping (DST) model, and is one of the rare examples in non-linear physics that are exactly solvable analytically. The important results can be summarized as follows. At low energy, the stationary points on the map ($x,\dot{x}$) (or on the map of two other independent dynamical variables) do not change, and the motion remains oscillatory and similar to the linear case. The character of oscillations qualitatively changes at a certain energy that depends on $\frac{\Omega}{\alpha}$: the old stationary point becomes an unstable saddle point, and a new {\it pair} of stable stationary points emerge, separated by the energy barrier \cite{nonlinear}. This corresponds to the bifurcation point, the emergence of new solutions as a result of changes of parameters in the dynamical system. This is illustrated in Figure \ref{bifur}.

An accompanying interesting insight is that contrary to the linear harmonic case, the energy is not equally partitioned between the oscillating points but can localize at one point, reflecting the more general insight that the superposition principle no longer works in non-linear systems in general.

The importance of the above result is that it demonstrates that the first-principles treatment of the non-linear equations of motion gives rise, via the bifurcation at high energy, a new qualitatively different solution: instead of oscillating around a fixed position at low energy as in a solid, a particle starts to move between two stable stationary points at high energy, corresponding to the liquid-like motion of particles between two minima in Figure \ref{double}. It proves that in the most simple non-linear system, the liquid-like motion emerges as a {\it bifurcation} of the solid-like solution.

The DST model (\ref{nonl2}) is not identical to the original simple system of coupled Duffing oscillators (\ref{nonl1}). The difference with the DST model is that, due to the non-integrability of (\ref{nonl1}), islands of chaotic dynamics appear on the phase map and grow with the system energy. The excitations in the original model (\ref{nonl1}) can only be found using approximate techniques. However, the DST model is close to (\ref{nonl1}) for small oscillation amplitudes and small couplings $\epsilon$. This proximity between the two models is used to assert the same qualitative result, the emergence of the bifurcation of solutions.

We note the result from this discussion to which we return below: the bifurcation in the original model (\ref{nonl1}) emerges at energies $E\propto\epsilon^2$ or amplitudes $x\propto\epsilon$, the result which is not unexpected: the energy of coupling needs to be surmounted in order to break away from the low-energy solid-like solution.

The real problem appears when the number of non-linear oscillators, $N$, increases. The analysis of $N=3$ non-linear coupled oscillators is complicated from the outset by the fact that the corresponding DST model is non-integrable to begin with. The approximations involved in the increasingly complicated analysis of stationary states, new bifurcations emerging from these states and corresponding collective modes become harder to control. The results from computer modeling indicate the emergence of many unanticipated modes and chaotic behaviour at higher energy. The problem significantly increases for $N=4$, including finding new stationary points and related collective modes, analyzing non-trivial branching of next-generation bifurcations and so on. For larger $N$, only approximate qualitative observations can be made regarding the energy spectrum, energy localization and emerging collective modes. This is done on the basis of approximations and insights from $N=2-4$ \cite{nonlinear}.

Importantly, the number of stationary states and bifurcations exponentially increases with $N$. The problem of finding stationary states, bifurcations, collective modes and their evolution with the system's energy is exponentially complex and intractable for arbitrary $N$ \cite{nonlinear}.

Therefore, the failure of the first-principles treatment of liquids at the same level as Eq. (\ref{haeq}) for solids has its origin in the intractability of the exponentially complex problem of calculating bifurcations, stationary points and collective modes in a large system of coupled non-linear equations.

\section{Relaxation time and phonon states in liquids: Frenkel's reduction}

It is fitting to discuss terms such as ``collective modes'', ``phonons'' and other quasi-particles in relation to J. Frenkel's work because he was involved in coining and disseminating these terms. For example, the term ``phonon'', attributed to Tamm, first appeared in print in Frenkel's 1932 publication \cite{kozhevn}.

Frenkel's ideas occupy a significant part of our discussion. This might appear unusual to the reader, in view that this is not the case in other liquid textbooks \cite{landau,hydro,ziman,boonyip,march,march1,baluca,zwanzig,faber,hansen1,hansen2}. Frenkel's work is not unknown but why would we want to delve into it in detail now? We find that many discussions of liquids either do not mention Frenkel's work (see, e.g., Refs. \cite{eyring,zwanzig1,zwanzig2,wall1,wall2}) or mention it in an irrelevant context, yet they develop many ideas which, when stripped of details, are essentially due to Frenkel to a large extent. This will become apparent in this review. More importantly, we find that, combined with recent experimental evidence, Frenkel's work related to collective modes in liquids gives a constructive tool to develop a predictive thermodynamic theory of liquids.

Frenkel proposed a number of new ideas of how to understand liquids emphasizing their ``gas-like'' and ``solid-like'' properties \cite{frenkel}. Some of the ideas such as the ``hole theory'' of liquids were not followed or developed, perhaps for the reason that the picture was qualitative and without links to experimental data. It should be noted that the experimental data on liquids at the time was only very basic so Frenkel's theoretical work was truly pioneering. However, other ideas discussed in Frenkel book and his earlier papers on liquids transformed the field in a way which is not fully appreciated even today.

This transformation proceeded slowly and sporadically over the last 80--90 years since Frenkel's work, during which alternative approaches to liquids were developing and Frenkel's ideas forgotten and surfaced anew more than once (see, e.g., Refs. \cite{eyring,zwanzig2,wall2}). In our view, Frenkel was too ahead of his time. A transformative idea, proposed and experimentally confirmed within a generation of scientists has a larger chance of succeeding as compared to the Frenkel's case where the new idea was proposed long before its confirmation. For example, his proposal that liquids are able to support solid-like longitudinal and transverse modes with frequencies extending to the highest Debye frequency implies that liquids are just like solids (solid glasses) in terms of their ability to sustain collective modes. Therefore, main liquid properties such as energy and heat capacity can be described using the same first-principles approach based on collective modes as solids - an assertion that is considered very unusual. The evidence for this has come only recently because liquids turned out to be too hard to probe experimentally. The evidence has started to mount only after powerful X-ray synchrotrons were deployed, and more than 80 years after Frenkel's first published paper on the subject.

Frenkel's work on liquids is interestingly described by Sir N. F. Mott \cite{mott}:

``Frenkel was a theoretical physicist. By this I am stressing that he was primarily and most of all interested in what is happening in real systems, and the mathematics he used served his physics and not otherwise as is sometimes the case for the modern generation of scientists... He asks: `what is really happening and how can this be explained?' ''

\subsection{Liquid relaxation time and phonon states}

Throughout this paper, we are using terms such as collective modes and phonons inter-changeably. Their meaning will be clarified in the later sections where we will also comment on the issue of dissipation of harmonic waves in disordered systems including glasses and liquids.

Dating to 1926 \cite{frenkel26} and developed in his later book \cite{frenkel}, main ideas of Frenkel on liquids preceded the advance of the non-linear theory discussed earlier. Frenkel's discussion includes many important ideas, of which we review only those relevant to understanding collective modes and different regimes of wave propagation in liquids.

\begin{figure}
\begin{center}
{\scalebox{0.6}{\includegraphics{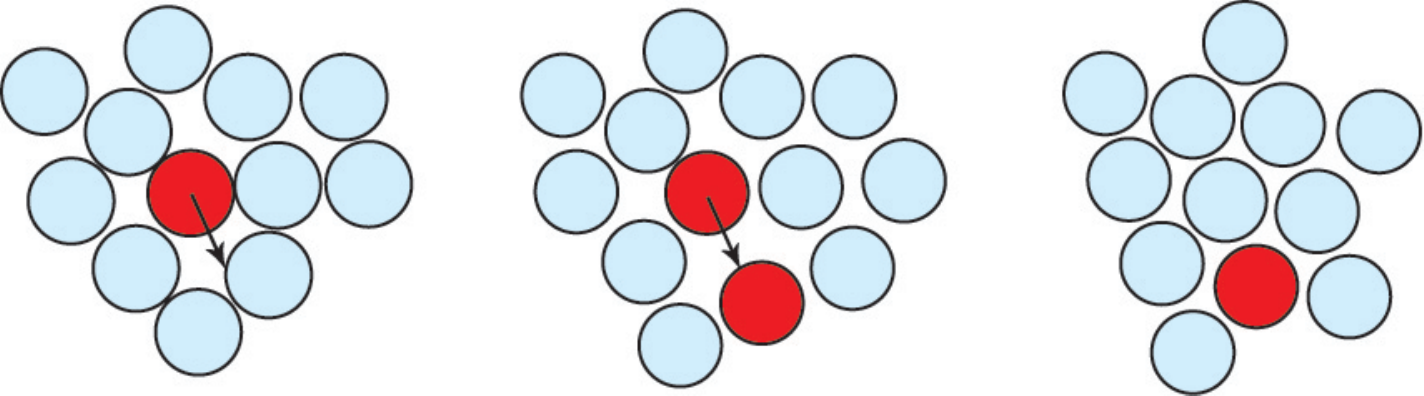}}}
\end{center}
\caption{Colour online. Illustration of a particle jump between two quasi-equilibrium positions in a liquid. These jumps take place with a period of $\tau$ on average.}
\label{jump}
\end{figure}

Frenkel was naturally led to liquid dynamics by his work on defect migration in solids, and viewed the two processes as sharing important qualitative properties. The migration rate of defects in a solid (crystalline or amorphous) is governed the potential energy barrier $U$ set by the surrounding atoms. At fixed volume of the ``cage'' formed by the nearest neighbours, $U$ is very large for the diffusion event to occur in any reasonable time. However, the cage thermally oscillates and periodically opens up fast local diffusion pathways. If $\Delta r$ is the increase of the cage radius required for the atom to jump from its case (see Figure \ref{jump}), $U$ is \cite{frenkel}:

\begin{equation}
U=8\pi G r\Delta r^2
\label{acten}
\end{equation}

\noindent where $r$ is the cage radius and $G$ is shear modulus. Note that when a sphere expands in a static elastic medium, no compression takes place at any point. Instead, the system expands by the amount equal to the increase of the sphere volume \cite{frenkel}, resulting in a pure shear deformation. The strain components $u$ from an expanding sphere (noting that $u\rightarrow$0 as $r\rightarrow\infty$) are $u_{rr}=-2b/r^3$, $u_{\theta\theta}=u_{\phi\phi}=b/r^3$ \cite{lanstat}, giving pure shear $u_{ii}=0$. As a result, the energy to statically expand the sphere depends on shear modulus $G$ only.

Frenkel considered the above picture applicable to liquids as well as solids, and introduced liquid relaxation time $\tau$ as the average time between particle jumps at one point in space in a liquid.

The range of $\tau$ is bound by two important values. If crystallization is avoided, $\tau$ increases at low temperature until it reaches the value at which the liquid stops flowing at the experimental time scale, corresponding to $\tau=10^2-10^{3}$ s and the liquid-glass transition \cite{dyre,angell}. At high temperature, $\tau$ approaches its minimal value given by Debye vibration period, $\tau_{\rm D}\approx 0.1$ ps, when the time between the jumps becomes comparable to the shortest vibrational period. Frenkel's picture has been confirmed in numerous molecular dynamics simulations of liquids which, since early days of computer modeling \cite{stil}, observed and studied particle jumps and transitions between different minima. The operation of particle jumps in liquids is often referred to as ``relaxation process''.

With a remarkable physical insight, Frenkel proposed the following simple picture of vibrational states in the liquid. At times significantly shorter than $\tau$, no particle rearrangements take place. Hence, the system is a solid glass describable by Eqs. (\ref{haeq},\ref{eigen}) and supports one longitudinal mode and two transverse modes. At times longer than $\tau$, the system is a flowing liquid, and hence does not support shear stress or shear modes but one longitudinal mode only as any elastic medium (in a dense liquid, the wavelength of this mode extends to the shortest wavelength comparable to interatomic separations as discussed below). This is equivalent to asserting that the only difference between a liquid and a solid glass is that the liquid does not support all transverse modes as the solid, but only those above the Frenkel frequency $\omega_{\rm F}$:

\begin{equation}
\omega>\omega_{\rm F}=\frac{1}{\tau}
\label{tau}
\end{equation}

\noindent where we omit the factor of $2\pi$ in $\omega=\frac{2\pi}{\tau}$ for brevity and for the reason that in liquids the range of $\tau$ spans 16 orders of magnitude, making a small constant factor unimportant.

Eq. (\ref{tau}) implies that liquids have solid-like ability to support shear stress, with the only difference that this ability exists not at zero frequency as in solids but at frequency larger than $\omega_{\rm F}$ (below we often use the term ``solid-like'' to denote the property in (\ref{tau})). This was an unexpected insight at the time, and took many decades to prove experimentally as discussed below. It also posed a fundamental question about the difference between solids and liquids: liquids are different from solids by the value of $\omega_{\rm F}$ only which is a quantitative difference rather than a qualitative one. In Frenkel's view, this reflected the continuity of liquid and solid states, the question that is still debated in the context of the problem of liquid-glass transition. We will discuss this in the next sections.

The longitudinal mode remains propagating in the Frenkel's picture based on $\tau$: density fluctuations exist in any interacting system. We will see below that in real dense liquids, experiments have ascertained that the longitudinal vibrations extend to the largest Debye frequency as in solids. However, the presence of relaxation process and $\tau$ differently affects the propagation of the longitudinal collective modes in different regimes $\omega>\frac{1}{\tau}$ and $\omega<\frac{1}{\tau}$, as discussed in the next sections.

We note that the separation of particle motion in the liquid into oscillatory and diffusive jump motion works well for liquids with large $\tau$ (or viscosity, see next section). For smaller $\tau$ at high temperature, jumps can become less pronounced and oscillations increasingly anharmonic. The disappearance of oscillatory component of particle motion can be related to the Frenkel line discussed in the later section.

We also note that the concept of $\tau$ implies average relaxation time. In real liquids, there is a distribution of relaxation times as is widely established in experiments such as dielectric spectroscopy (see, e.g., Ref. \cite{logar2}).

\subsection{Relationship to Maxwell relaxation theory}

Here, we discuss the important relationship between the analysis of Frenkel and Maxwell. Maxwell proposed that a body is generally capable of both elastic and viscous deformation and, under external perturbation such as shear stress, the total strain is the sum of viscous and elastic strains \cite{maxwell}. The $y-$gradient of horizontal velocity $v_x$ due to viscous deformation is $\frac{\partial{v_x}}{\partial{y}}=\frac{P_{xy}}{\eta}$, where $P_{xy}$ is shear stress and $\eta$ is viscosity. The gradient of velocity due to elastic deformation is $\frac{\partial{v_x}}{\partial{y}}=\frac{1}{G}\frac{dP_{xy}}{dt}$ where $G$ is shear modulus. When both viscous and elastic deformations are present, the velocity of a layer $v_x$ is the sum of the two velocities, giving:

\begin{equation}
\frac{\partial{v_x}}{\partial{y}}=\frac{1}{G}\frac{dP_{xy}}{dt}+\frac{P_{xy}}{\eta}
\label{interp}
\end{equation}

The presence of both viscous and elastic response has been subsequently called ``viscoelastic'' response, and is commonly used at present.

When external perturbation stops and $v_x=0$, Eq. (\ref{interp}) gives

\begin{equation}
\begin{aligned}
&P_{xy}=P_0\exp\left(-\frac{t}{\tau_{\rm M}}\right)\\
&\tau_{\rm M}=\frac{\eta}{G}
\end{aligned}
\label{decay}
\end{equation}

\noindent where $\tau_{\rm M}$ is Maxwell relaxation time.

Frenkel has proposed that the time constant in Eq. (\ref{decay}), $\tau_{\rm M}$, is related to liquid relaxation time $\tau$ he introduced (the time between particle rearrangements), and concluded that $\tau_{\rm M}\approx\tau$. Then, relaxation of shear stress in a viscoelastic liquid is exponential with Frenkel's liquid relaxation time $\tau$:

\begin{equation}
P_{xy}=P_0\exp\left(-\frac{t}{\tau}\right)
\label{decay1}
\end{equation}

The second equation in (\ref{decay}) where $\tau$ is used instead of $\tau_{\rm M}$ is often called the Maxwell relationship:

\begin{equation}
\eta=G_\infty\tau
\label{maxrel}
\end{equation}
\noindent

Here, $G_\infty$ is the ``instantaneous'' shear modulus. $G_\infty$ is understood to be the shear modulus at high frequency at which the liquid supports shear stress. In practice, this frequency can be taken as the maximal frequency of shear waves present in the liquid, comparable to Debye frequency \cite{puosi}.

The activation energy for particle jumps in the liquid can be calculated using Eq. (\ref{acten}), but with the proviso that $G$ is the shear modulus at high-frequency.

Experimentally, shear stress and various correlations in viscous liquids (liquids where $\tau\gg\tau_{\rm D}$; see Section ``Viscous liquids'' for more detailed discussion below) and glasses decay according to the stretched-exponential relaxation (SER) law rather than pure exponential as in Eq. (\ref{decay1}):

\begin{equation}
f\propto\exp\left(-\left(\frac{t}{\tau}\right)^\beta\right)
\label{ser}
\end{equation}

\noindent where $f$ is a decaying function such as $P_{xy}$ in (\ref{decay1}) and $\beta$ is the stretching parameter conforming to $0<\beta<1$.

First observed by Kohlrausch around the time of development of Maxwell relaxation theory \cite{kohl}, the physical origin of SER has been widely discussed \cite{dyre,ngai,phillips}. It is believed that SER is as a result of cooperativity of molecular relaxation emerging in the viscous regime. Here, ``cooperativity'' is not well-defined but can be identified with the elastic interaction between particle rearrangement events via high-frequency waves they induce \cite{ourser1,ourser2}. Regardless of whether the relaxation is exponential or stretched-exponential, the decay of shear stress and other correlation function takes place with a characteristic time $\tau$ in both (\ref{decay1}) and (\ref{ser}).

\subsection{Frenkel reduction}

It is interesting to discuss the meaning of Frenkel's theory from the point of view of a first-principles description of liquids. This theory is not a first-principles description at level (\ref{haeq}) and (\ref{eigen}) but, as discussed in the earlier section, the first-principles treatment of liquid collective modes is exponentially complex and not tractable. Instead, this approach singles out the main physical property of liquids ($\tau$, or viscosity, see Eq. (\ref{maxrel})) which governs the relative contributions of oscillatory and diffusive motion and which ultimately controls the phonon states in the liquid. This reduces the exponentially large problem to one physically relevant parameter. We call it the ``Frenkel reduction'' \cite{annals}.

Implicit in this reduction is a physically reasonable assumption that quasi-equilibrium states and the local particle surroundings of jumping atoms in a homogeneous liquid are equivalent, and that fluctuations in a statistically large system can be ignored \cite{landau}. In the language of non-linear theory, the reduction lies in assuming that emerging new bifurcations and stationary states at all generations produce physically equivalent states on average. This implies that as temperature (or energy) increases, the conditions governing particle jumps can be considered approximately the same everywhere in the system. Therefore, particle dynamics is governed by the temperature-activated jumps as the dynamics of point defects in solids:

\begin{equation}
\tau=\tau_{\rm D}\exp\left(\frac{U}{T}\right)
\label{activ}
\end{equation}

\noindent where $U$ is given by (\ref{acten}).

It is generally agreed that $\tau$ and viscosity in liquids are indeed governed by the temperature-activated process, with a caveat that $U$ can include an additional temperature-dependent term due to cooperativity of molecular relaxation, in which case $\tau$ grows faster-than Arrhenius (``super-Arrhenius'') as discussed below. This cooperative process is of the same nature as the one governing the non-exponentiality of relaxation in (\ref{ser}).

We recall the result from the non-linear theory that a bifurcation emerges when the energy of the particle becomes comparable to the coupling energy between two non-linear oscillators. Noting that this result is derived approximately, we can relate the coupling energy to the activation energy given in (\ref{acten}). Indeed, the coupling energy in the system of non-linear equations is the energy that a particle needs to escape a bound state with another particle. This energy is of the same nature and order of magnitude as that needed to break the atomic cage shown in Figure \ref{jump}. Therefore, the approximation in the Frenkel theory is of the same nature as the one in the non-linear theory.

In our discussion of generalized hydrodynamics below, we will see that the introduction of relaxation process and solid-like features in the hydrodynamic equations is done at the same level as in Eq. (\ref{decay1}) in the Frenkel theory, by assuming the exponential decay of different correlation functions with the decay time $\tau$.

\begin{figure}
\begin{center}
{\scalebox{0.35}{\includegraphics{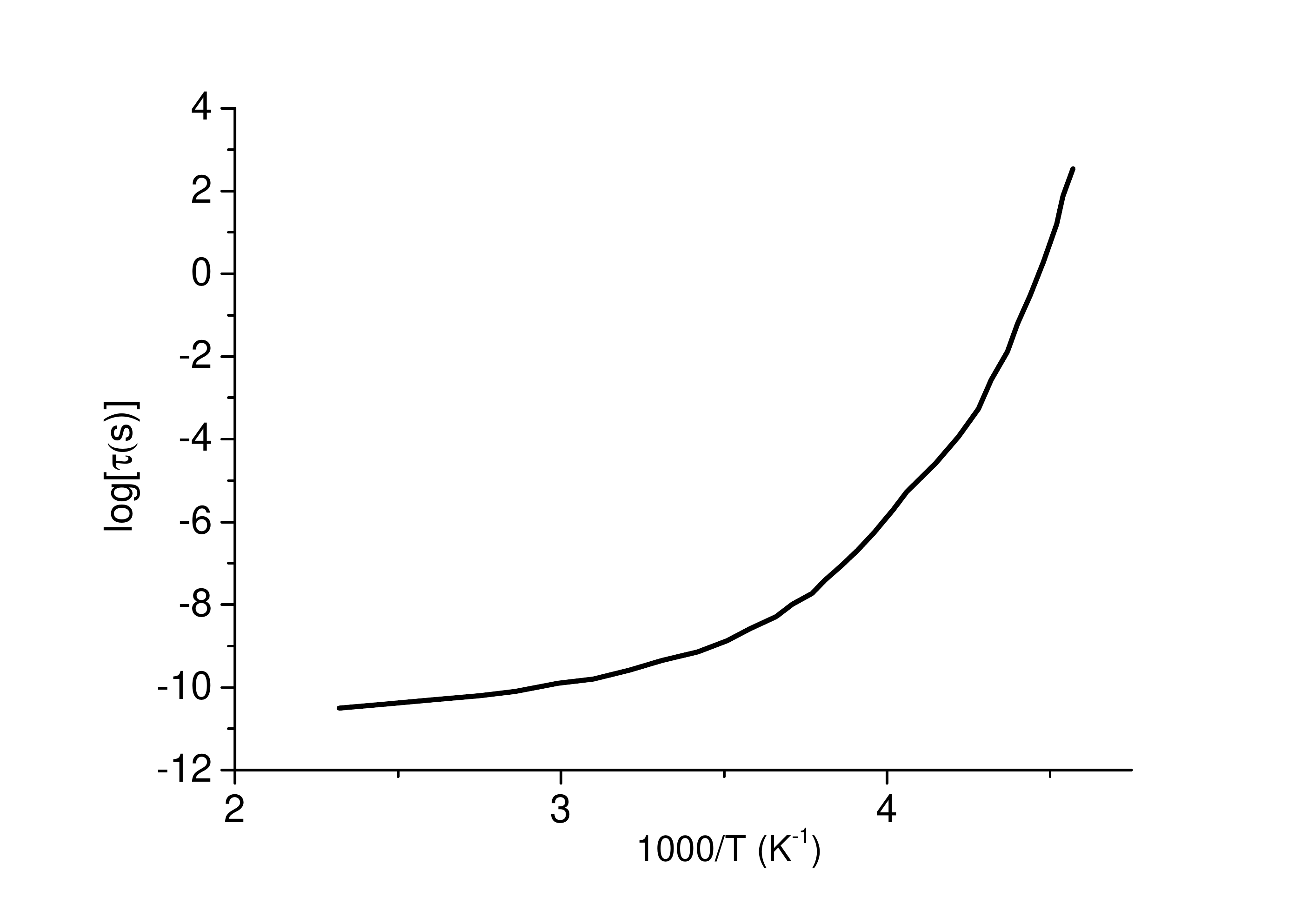}}}
\end{center}
\caption{Relaxation time of salol measured in dielectric relaxation experiments \cite{casa}.}
\label{tausal}
\end{figure}

We emphasize that $\tau$ is readily measured using several well-established experiments including dielectric relaxation experiments, NMR, positron annihilation spectroscopy and so on. $\tau$ can also be derived from viscosity measurements using Eq. (\ref{maxrel}) using widely available techniques including the classic Stokes experiments applicable to many types of liquids including at high pressure and temperature \cite{braball}. $\tau$ can also be calculated in molecular dynamics simulations as, for example, time decay of various correlation functions. In Figure \ref{tausal} we show $\tau$ measured in salol over many orders of magnitude as an example, and comment on it in the next section.

In essence, Frenkel reduction introduces a cutoff frequency $\omega_{\rm F}$ (see \ref{tau}) above which the liquid can be described by the same first-principles equations of motion as the solid in Eqs. (\ref{haeq}) and (\ref{eigen}). Therefore, liquid collective modes include both longitudinal and transverse modes with frequency above $\omega_{\rm F}$ in the solid-like elastic regime and one longitudinal hydrodynamic mode with frequency below $\omega_{\rm F}$ (shear mode is non-propagating below frequency $\omega_{\rm F}$ as discussed below).

Recall Landau's assertion that a thermodynamic theory of liquids can not be developed because liquids have no small parameter. How is this fundamental problem addressed here? According to Frenkel reduction, liquids behave like solids with small oscillating particle displacements serving as a small parameter. Large-amplitude diffusive particle jumps continue to play an important role, but do not destroy the existence of the small parameter. Instead, the jumps serve to modify the phonon spectrum: their frequency, $\omega_{\rm F}$, sets the minimal frequency above which the small-parameter description applies and solid-like modes propagate.

This approach is therefore a method of {\it non-perturbative} treatment of strong interactions, the central problem in field theories and other areas of physics \cite{annals}. It is markedly different from any other method of treating strong interactions contemplated in areas outside of liquids.

\section{Continuity of solid and liquid states and liquid-glass transition}

The picture of liquid based on relaxation time $\tau$ has a notable consequence for liquid-solid transitions. In 1935, Frenkel published an article in Nature entitled, ``Continuity of the solid and the liquid states'', \cite{debate1} where he proposed and later developed \cite{frenkel} an argument that liquids and solids are qualitatively the same. This follows from the concept of $\tau$: as $\tau$ increases on lowering the temperature beyond the experimental time frame, the liquid becomes frozen glass, and supports shear modes at all frequencies including at zero frequency. Hence, liquids and solids are different in terms of $\tau$ only, i.e. quantitatively, but not qualitatively. Frenkel subsequently stated that ``classification of condensed bodied into solids and liquids [has] a relative meaning convenient for practical purposes but devoid of scientific value'' \cite{frenkel}, an assertion that many would find unusual today let alone then.

This idea was quickly criticized by Landau \cite{debate2,debate3} on the basis that the liquid-crystal transition involves symmetry changes and therefore can not be continuous according to the phase transitions theory. This debate unfortunately reflected a misunderstanding because Frenkel was emphasizing supercooled liquids that becomes glasses on cooling, rather than crystals \cite{frenkel}.

Remarkably, essentially the same debate is still continuing in the area of liquid-glass transition where one of the main discussed questions is whether a phase transition is involved \cite{dyre,angell,gl4,gl5,gl6,gl7,gl8}? According to the large set of experimental data, liquids and glasses are structurally identical, and liquid-glass transition does not involve structural changes. Yet at the glass transition temperature $T_g$ the heat capacity changes with a jump, seemingly providing support to the thermodynamic signature of the glass transition. Here, $T_g$ is defined as temperature at which $\tau$ exceeds the experimental time frame, $\tau=10^2-10^3$ s, corresponding to liquid becoming frozen in terms of particle rearrangements during the observation period.

We will return to the question of heat capacity jump at $T_g$ when we discuss thermodynamic properties of viscous liquids. Here we note that although few consider the jump of heat capacity at $T_g$ as a phase transition, versatile proposals were related to a possible phase transition at lower temperature $T_0<T_g$ \cite{dyre,angell,gl4,gl5,gl6,gl7,gl8}. The possibility of this was suggested by the Vogel-Fulcher-Tammann (VFT) temperature dependence of $\tau$:

\begin{equation}
\tau\propto\exp\left(\frac{A}{T-T_0}\right)
\label{vft}
\end{equation}

\noindent where $A$ and $T_0$ are constants.

$\tau$ in the VFT law diverges at $T_0$, and the same applies to viscosity $\eta$ according to Eq. (\ref{maxrel}). This led to proposals that the ``ideal'' glass transition takes place at $T_0$ to the ideal glass state. The transition and the state are ostensibly not seen because its observation is suppressed by very slow relaxation process below $T_g$, and remain to be of unknown nature \cite{dyre,angell,gl4,gl5,gl6,gl7,gl8}.

An example of the super-Arrhenius behaviour is shown in Figure \ref{tausal} for a commonly measured glass-forming system, salol \cite{casa}. Here and in other cases, $\tau$ is described by the VFT dependence fairly well, although a more careful experimental analysis revealed that on lowering the temperature, $\tau$ crosses over from the VFT to Arrhenius (or nearly Arrhenius) behavior \cite{cro1,cro2,cro3,cro31,cro4,cro5}. This takes place at about midway of the glass transformation range where $\tau\approx 10^{-6}$ s, i.e. above $T_g$ and hence well above $T_0$. Known more widely in the experimental community as compared to theorists, the crossover removes the basis for considering divergences and a possible thermodynamic phase transition at $T_0$.

An interesting question is what causes the crossover from the VFT law at high temperature to nearly Arrhenius at low. A useful insight comes from the observation that a sudden local jump event such as the one shown in Figure \ref{jump} induces an elastic wave with a wavelength comparable to interatomic separation and cage size. This wave propagates in the system and affects relaxation of other events, setting the cooperativity of molecular relaxation. As discussed in the next section, being a high-frequency wave, it propagates in the solid-like elastic regime with the propagation length given by Eq. (\ref{dis1}):

\begin{equation}
d=\lambda\omega\tau\approx c\tau
\end{equation}

\noindent where $c$ is the speed of sound, $\omega$ is frequency and $\lambda$ is wavelength. As discussed in the next section, $d$ increases with $\tau$ in this regime, in contrast to the propagation length of the commonly considered hydrodynamic waves \cite{hydro}.

At high temperature when $\tau\approx\tau_{\rm D}$, $d=c\tau_{\rm D}\approx a$, where $a$ is interatomic separation. This means that the wave does not propagate beyond the nearest neighbors and that the relaxation is non-cooperative (independent) and is Arrhenius and exponential as a result. Importantly, $d$ increases on lowering the temperature because $\tau$ increases. This increases the cooperativity of molecular relaxation \cite{ngai} but only until $d$ reaches system size $L$. Therefore, the crossover from the VFT law to Arrhenius takes place at $\tau=\frac{L}{c}$, in quantitative agreement with experiments \cite{ourser2}.

%An interesting insight into the possibility of a complete dynamical arrest of the viscous liquid due to the infinite increase of the activation barrier at a finite temperature (such as $T_0$ in Eq. (\ref{vft})) comes from the non-linear theory. Recall that the energy is not equally partitioned between coupled non-linear oscillators and can localize at certain points in the system. This means that a bifurcation of the solid-like solutions, the particle jump into the new position, can emerge at any low temperature. This contradicts the picture where a phase transition takes place at a finite temperature into the state with no particle jumps.

\section{Hydrodynamic and solid-like elastic regimes of wave propagation}

As discussed above, liquids behave differently depending on observation time or frequency. Frequencies $\omega>\omega_{\rm F}$ and $\omega<\omega_{\rm F}$ correspond to solid-like elastic regime ($\omega\tau>1$) and hydrodynamic regime ($\omega\tau<1$), respectively. The two regimes are described by different equations, those of elasticity \cite{lanstat} and hydrodynamics \cite{hydro}. The transition between the two regimes can be most easily seen by considering the response of the right-hand side of Eq. (\ref{interp}) to a periodic force $P=A\exp(i\omega t)$, giving

\begin{equation}
\left(\frac{1}{G}\frac{dP_{xy}}{dt}+\frac{P_{xy}}{\eta}\right)\exp(i\omega t)=\frac{1}{\eta}(1+i\omega\tau)P
\label{inter1}
\end{equation}

\noindent where we used $\eta=G\tau$.

For $\omega\tau>1$, (\ref{inter1}) gives $\frac{1}{\eta}(i\omega\tau)P=\frac{P}{G}i\omega=\frac{1}{G}\frac{dP}{dt}$, or purely elastic response. For $\omega\tau<1$, (\ref{inter1}) returns purely viscous response, $\frac{P}{\eta}$.

To discuss liquid's ability to operate in both regimes depending on $\omega_{\rm F}$, we can either start with hydrodynamic equations and introduce the solid-like elastic response or start with elasticity equations and introduce the hydrodynamic response. The first method has received most attention in the history of liquid research, and generally forms the basis for a variety of approaches collectively known as ``Generalized Hydrodynamics'' discussed in the later section. The second method is not commonly discussed and its implications are less understood.

Below we consider important examples of the difference in which collective modes operate in the hydrodynamic and solid-like elastic regimes, and start with the second method.

\subsection{Modifying elasticity: including hydrodynamics in elasticity equations}

Condition (\ref{tau}), $\omega\tau>1$ (sometimes written as $\omega\tau\gg1$) corresponds to wave propagation in the liquid with frozen structure (as in a solid), where the microscopic equations are Newton equations for all particles (\ref{haeq},\ref{eigen}). This is the solid-like elastic regime of wave propagation. Modifying elasticity equations by including hydrodynamics enables us to address our first case study, the difference of wave propagation in regimes $\omega\tau>1$ and $\omega\tau<1$. We will see that dissipation length, the length over which an induced wave is dissipated due to viscous effects, behaves qualitatively differently in the two regimes.

We consider both elastic and viscous response in the form equivalent to Eq. (\ref{interp})

\begin{equation}
\frac{ds}{dt}=\frac{P}{2\eta}+\frac{1}{2G}\frac{dP}{dt}
\label{a1}
\end{equation}
\noindent where $s$ is shear strain and introduce the operator

\begin{equation}
A=1+\tau\frac{d}{dt}
\label{a2}
\end{equation}
\noindent where $\tau=\frac{\eta}{G}$ from (\ref{maxrel}). Then, Eq. (\ref{a1}) can be written as

\begin{equation}
\frac{ds}{dt}=\frac{1}{2\eta}AP
\label{a3}
\end{equation}

If $A^{-1}$ is the reciprocal operator to $A$, $P=2\eta A^{-1}\frac{ds}{dt}$. Because $\frac{d}{dt}=\frac{A-1}{\tau}$ from Eq. (\ref{a2}), $P=2G(1-A^{-1})s$. Comparing this with $P=2Gs$, we find that the presence of relaxation process is equivalent to the substitution of $G$ by the operator $M=G(1-A^{-1})$.

The above constitutes the modification of the constituent elasticity equations by introducing the relaxation process in the liquid and $\tau$, i.e. approach to liquids from the solid elastic state:

\begin{equation}
P=2Gs\rightarrow P=2G(1-A^{-1})s
\label{modif1}
\end{equation}

Let us now consider the propagation of the wave of $P$ and $s$ with time dependence $\exp(i\omega t)$. Differentiation gives multiplication by $i\omega$. Then, $A=1+i\omega\tau$, and $M$ is:

\begin{equation}
M=\frac{G}{1+\frac{1}{i\omega\tau}}
\label{a4}
\end{equation}

If $M=R\exp(i\phi)$, the inverse complex velocity is $\frac{1}{v}=\sqrt\frac{\rho}{M}=\sqrt\frac{\rho}{R}(\cos\frac{\phi}{2}-i\sin\frac{\phi}{2})$, where $\rho$ is density. $P$ and $s$ depend on time and position $x$ as $f=\exp(i\omega(t-x/v))$. Using the above expression for $\frac{1}{v}$, $f=\exp(i\omega t)\exp(-ikx)\exp(-\beta x)$, where $k=\omega\sqrt\frac{\rho}{R}\cos\frac{\phi}{2}$ and absorbtion coefficient $\beta=\omega\sqrt\frac{\rho}{R}\sin\frac{\phi}{2}$. Combining the last two expressions for $k$ and $\beta$, we write $\beta=\frac{2\pi\tan\frac{\phi}{2}}{\lambda}$, where $\lambda=\frac{2\pi}{k}$ is the wavelength.

From Eq. (\ref{a4}), $\tan\phi=\frac{1}{\omega\tau}$. For high-frequency waves $\omega\tau\gg 1$, $\tan{\phi}\approx\phi=\frac{1}{\omega\tau}$, giving $\beta=\frac{\pi}{\lambda\omega\tau}$. Lets introduce the propagation length $d=1/\beta$ so that $f\propto\exp(-x/d)$. Then, $d=\frac{\lambda\omega\tau}{\pi}$. Therefore, this theory gives propagating shear waves in the solid-like elastic regime $\omega\tau\gg 1$, with the propagation length

\begin{equation}
\begin{aligned}
&d\approx\lambda\cdot\omega\tau\\
&(\omega\tau\gg 1)
\label{dis1}
\end{aligned}
\end{equation}

We note that this result is derived for plane waves, and it approximately holds in disordered systems for wavelengths that are large enough. At smaller wavelengths comparable to structural inhomogeneities, $d$ is reduced due to the dissipation of plane waves in the disordered medium. The dissipation is related to how well the eigenstates of the disordered system can be approximated by plane waves (for more detailed discussion, see Ref. (\cite{jphyschem}).

In the hydrodynamic regime $\omega\tau\ll 1$, we find $\phi=\frac{\pi}{2}$ and $d=\frac{\lambda}{2\pi}$. Different from the high-frequency case, this means that low-frequency shear waves are not propagating (because they are dissipated over the distance comparable to the wavelength), a result that is also known from hydrodynamics \cite{hydro}.

The consideration of the propagation velocity of longitudinal waves involves the bulk modulus which can be written in the form $L=K_1+\frac{K_2}{1+\frac{1}{i\omega\tau}}$ containing the non-zero static part as well as the frequency-dependent part as in (\ref{a4}). Repeating the same steps as above, the propagation length in the solid-like elastic regime $\omega\tau\gg 1$ is the same as in Eq. (\ref{dis1}). In the hydrodynamic regime $\omega\tau\ll 1$, the propagation length becomes

\begin{equation}
\begin{aligned}
&d\approx\frac{\lambda}{\omega\tau}\\
&(\omega\tau\ll 1)
\label{dis2}
\end{aligned}
\end{equation}

Comparing Eqs. (\ref{dis1}) and (\ref{dis2}), we see that the two different regimes give qualitatively different character of waves dissipation: the propagation length increases with $\tau$ and viscosity in the former, but decreases with $\tau$ and viscosity in the latter.

The decrease of the propagation length with liquid viscosity in the commonly discussed hydrodynamic regime is a familiar result from fluid mechanics \cite{hydro}. On the other hand, the increase of propagation length in the solid-like elastic regime is less known.

An important insight from this discussion is that the two regimes of waves propagation are different from the physical point of view and yield qualitatively different results, including directly opposite results for the propagation length. This implies that essential physics in the hydrodynamic regime and its underlying equations can not be {\it extrapolated} to the solid-like elastic regime (and vice versa). By extrapolating here we mean extending the hydrodynamic regime to large $k$ and $\omega$ while keeping the underlying physics and associated equations qualitatively the same. We will return to this point below when we discuss the approach to liquids based on generalized hydrodynamics.

Our second case study is related to the crossover between two regimes of propagation. In the solid-like elastic regime, the propagation velocity in the isotropic medium is $v=\sqrt{\frac{B+\frac{4}{3}G}{\rho}}$ \cite{lanstat}, where $B$ and $G$ are bulk and shear moduli, respectively. This is the case for solids as well as liquids in the solid-like elastic regime where shear waves above $\omega_{\rm F}$ are propagating. In the hydrodynamic regime where no shear waves propagate as discussed above, the propagation speed is $v=\sqrt{\frac{B}{\rho}}$, corresponding to $G=0$. Therefore, Frenkel argued, the transition between the two regimes results in the noticeable increase of the propagation speed by a factor $\sqrt{1+\frac{4}{3}\frac{G}{B}}$. The transition can be achieved by either changing $\tau$ at a given frequency by altering temperature or pressure, or by changing frequency at fixed temperature and pressure.

In the later section, ``Fast sound'', we will revisit this effect on the basis of recent experimental results.

\subsection{Modifying hydrodynamics: including elasticity in hydrodynamic equations}

Eqs. (\ref{a2})-(\ref{modif1}) modify (generalize) elasticity equations by including relaxation and viscous effects in the liquid in the form of viscous flow at times longer than $\tau$. Equally, Frenkel argued \cite{frenkel}, one can generalize hydrodynamic equations by endowing the system with the solid-like property to sustain shear stress at times shorter than $\tau$. This idea is generally similar in its spirit to the approach of Generalized Hydrodynamics that appeared later (see ``Generalized Hydrodynamics'' section below), although Frenkel implemented the idea differently. Apart from the general interest, this implementation deserves attention because it is not discussed in traditional generalized hydrodynamics approaches \cite{boonyip,march1,baluca}.

Lets write the Navier-Stokes equation as

\begin{equation}
\nabla^2{\bf v}=\frac{1}{\eta}\left(\rho\frac{d{\bf v}}{dt}+\nabla p\right)
\label{navier2}
\end{equation}

\noindent where ${\bf v}$ is velocity, $p$ is pressure, $\eta$ is shear viscosity, $\rho$ is density and the full derivative is $\frac{d}{dt}=\frac{\partial}{\partial t}+{\bf v\nabla}$.

Eqs. (\ref{a2})-(\ref{modif1}) account for both long-time viscosity and short-time elasticity. From (\ref{a1})-(\ref{a3}), we see that accounting for both effects is equivalent to making the substitution $\frac{1}{\eta}\rightarrow\frac{1}{\eta}+\frac{1}{G}\frac{d}{dt}$. Using $\eta=G\tau$ from Eq. (\ref{maxrel}), the substitution becomes:

\begin{equation}
\frac{1}{\eta}\rightarrow\frac{1}{\eta}\left(1+\tau\frac{d}{dt}\right)
\label{sub2}
\end{equation}

Using (\ref{sub2}) in Eq. (\ref{navier2}) gives

\begin{equation}
\eta\nabla^2{\bf v}=\left(1+\tau\frac{d}{dt}\right)\left(\rho\frac{d{\bf v}}{dt}+\nabla p\right)
\label{gener}
\end{equation}

Having proposed Eq. (\ref{gener}), Frenkel did not analyze it or its solutions. We do it below.

We consider the absence of external forces, $p=0$ and the slowly-flowing fluid so that $\frac{d}{dt}=\frac{\partial}{\partial t}$. Then, Eq. (\ref{gener}) reads

\begin{equation}
\eta\frac{\partial^2v}{\partial x^2}=\rho\tau\frac{\partial^2v}{\partial t^2}+\rho\frac{\partial v}{\partial t}
\label{gener2}
\end{equation}

\noindent where $v$ can be $y$ or $z$ velocity components perpendicular to $x$.

In contrast to the Navier-Stokes equation, the generalized hydrodynamic equation Eq. (\ref{gener2}) contains the second time derivative of $v$ and hence allows for propagating waves. Indeed, Eq. (\ref{gener2}) without the last term reduces to the wave equation for propagating shear waves with velocity $c_s=\sqrt{\frac{\eta}{\tau\rho}}=\sqrt{\frac{G}{\rho}}$. The last term represents dissipation. Using $\eta=G\tau=\rho c_s^2\tau$, we re-write Eq. (\ref{gener2}) as

\begin{equation}
c_s^2\frac{\partial^2v}{\partial x^2}=\frac{\partial^2v}{\partial t^2}+\frac{1}{\tau}\frac{\partial v}{\partial t}
\label{gener3}
\end{equation}

Seeking the solution of (\ref{gener3}) as $v=v_0\exp\left(i(kx-\Omega t)\right)$ gives the quadratic equation for $\Omega$:

\begin{equation}
\Omega^2+\Omega\frac{i}{\tau}-c_s^2k^2=0
\label{quadr}
\end{equation}

Equation (\ref{quadr}) has purely imaginary roots if $c_sk<\frac{1}{2\tau}$, approximately corresponding to the hydrodynamic regime $\omega\tau<1$. Therefore, we find that shear waves are not propagating in the hydrodynamic regime $\omega\tau<1$, which is the same result as the one derived in the previous section where elasticity equations were modified to include viscous effects.

If $c_sk>\frac{1}{2\tau}$ (corresponding to the solid-like elastic regime $\omega\tau>1$), Eq. (\ref{quadr}) gives $\Omega=-\frac{i}{2\tau}\pm\sqrt{c_s^2k^2-\frac{1}{4\tau^2}}$, and we find

\begin{equation}
\begin{aligned}
&v\propto\exp\left(-\frac{t}{2\tau}\right)\exp(i\omega t)\\
&\omega=\sqrt{c_s^2k^2-\frac{1}{4\tau^2}}
\end{aligned}
\label{hydro-gen}
\end{equation}

Eq. (\ref{hydro-gen}) describes propagating shear waves, contrary to the original Navier-Stokes equation. We therefore find that shear waves are propagating in the solid-like elastic regime $\omega\tau>1$, the same result we derived in the previous section where elasticity equations were modified to incorporate fluidity.

According to Eq. (\ref{hydro-gen}), the increase of $\tau$ or viscosity gives smaller wave dissipation (larger lifetime) in the solid-like elastic regime $\omega\tau>1$, contrary to the hydrodynamic regime \cite{hydro}. This is the same effect that we have discussed in the previous section where we found the increase of the propagation length of shear waves with $\tau$ and viscosity (see Eq. (\ref{dis1})).

We note that for large $\tau$ or viscosity, $\omega$ in Eq. (\ref{hydro-gen}) becomes $\omega=c_sk$ as in the case of shear waves with no dissipation at all. These are solid-like elastic waves with wavelengths extending to the shortest interatomic separations and frequencies up to the highest Debye frequency as predicted in the solid-like elastic approach by Eq. (\ref{tau}).

We also note that $\omega$ of shear waves in Eq. (\ref{hydro-gen}) does not increase from 0 to its linear branch $\omega=c_sk$ in a jump-like manner as follows from (\ref{tau}). Instead, starting from about $\omega=\omega_{\rm F}=\frac{1}{\tau}$, $\omega$ gradually increases from the square-root dependence to the linear dependence $\omega=c_sk$ at large $\tau$. This is consistent with the experimental result showing a gradual increase of the speed of sound and shear rigidity with the wave frequency \cite{jeong}. We will revisit this point when we discuss the phonon approach to liquid thermodynamics.

To derive the propagation of longitudinal waves, we need to include the longitudinal viscosity in the Navier-Stokes equations and modify it similarly to (\ref{sub2}), remembering that bulk viscosity is related to the bulk modulus which, in addition to frequency-depending term, always has non-zero static term \cite{frenkel}. This will give propagating longitudinal waves in both solid-like elastic regime and in the hydrodynamic regime, in agreement with the results in the previous section. We will not pursue this derivation here.

Therefore, we find that as far as wave propagation is concerned, equations of hydrodynamics modified (generalized) to include solid-like elastic effects give the same results as equations of elasticity modified to include viscous effects.

Interestingly, it is the approach of ``Generalized hydrodynamics'' which historically received wide attention and development and has become a distinct area of research \cite{boonyip,march1,baluca}. We will discuss this approach in the later section. This reflects the historical trend we alluded to in the introduction: the community largely viewed liquids as systems conforming to the hydrodynamic equation at the fundamental level, with possible solid-like elastic effects to be introduced, if needed, on top. To some extent, this view was consistent with existing experiments at the time that mostly probed low-energy properties of liquids. As discussed in the next Section, high-energy experiments uncovering solid-like properties of liquids have emerged relatively recently.

It can be argued that the approach to liquids starting with the solid-like elastic description contains more information about structure and dynamics and, therefore, is more suited to discuss high-frequency dynamics of liquids. This becomes particularly important for constructing the phonon theory of liquid thermodynamics where high-frequency modes govern system's energy and heat capacity as discussed in the later section.

\section{Experimental evidence for high-frequency collective modes in liquids}

Low-frequency collective modes, including familiar sound waves, are well understood in liquids. Yet these modes make a negligible contribution to liquid energy and heat capacity. Indeed, the liquid energy is almost entirely governed by high-frequency modes due to the approximately quadratic density of phonon states. However, the prediction of high-frequency solid-like modes in liquids in the regime $\omega\tau>1$ was non-trivial, and was outside the commonly discussed hydrodynamic approach where $\omega\tau<1$ \cite{landau,hydro,ziman,boonyip,march,march1,baluca,zwanzig,faber,hansen1,hansen2}.

The experimental evidence supporting the propagation of high-frequency modes in liquids includes inelastic X-ray, neutron and Brillouin scattering experiments. Most of the evidence is recent and follows the deployment of powerful synchrotron sources of X-rays.

Early experiments detected the presence of high-frequency propagating modes and mapped dispersion curves which were in striking resemblance to those in solids \cite{copley}. This and similar results were generated at temperature around melting. The measurements were later extended to high temperatures considerably above the melting point, confirming the same result. It is now well established that liquids sustain propagating modes extending to wavelengths comparable to interatomic separations \cite{pilgrim,burkel,pilgrim2,ruocco,water,rec-review,hoso,hoso3,mon-na,mon-ga,sn,disu1,disu2}. More recently, the same result has been asserted for supercritical fluids \cite{water,disu1,disu2}.

Importantly, the propagating modes in liquids include transverse modes. Initially detected in highly viscous liquids (see, e.g., Refs. \cite{grim,scarponi}), transverse modes have been later studied in low-viscous liquids on the basis of positive dispersion \cite{pilgrim,burkel,pilgrim2,rec-review} (recall our previous discussion that the presence of high-frequency transverse modes increases sound velocity from the hydrodynamic to the solid-like value). These studies included water \cite{water-fast}, where it was found that the onset of transverse excitations coincides with the inverse of liquid relaxation time \cite{water-tran}, as predicted by (\ref{tau}). More recently, transverse modes in liquids were directly measured in the form of distinct dispersion branches and verified on the basis of computer modeling \cite{hoso,mon-na,mon-ga,sn,hoso3}.

\begin{figure}
\begin{center}
{\scalebox{0.5}{\includegraphics{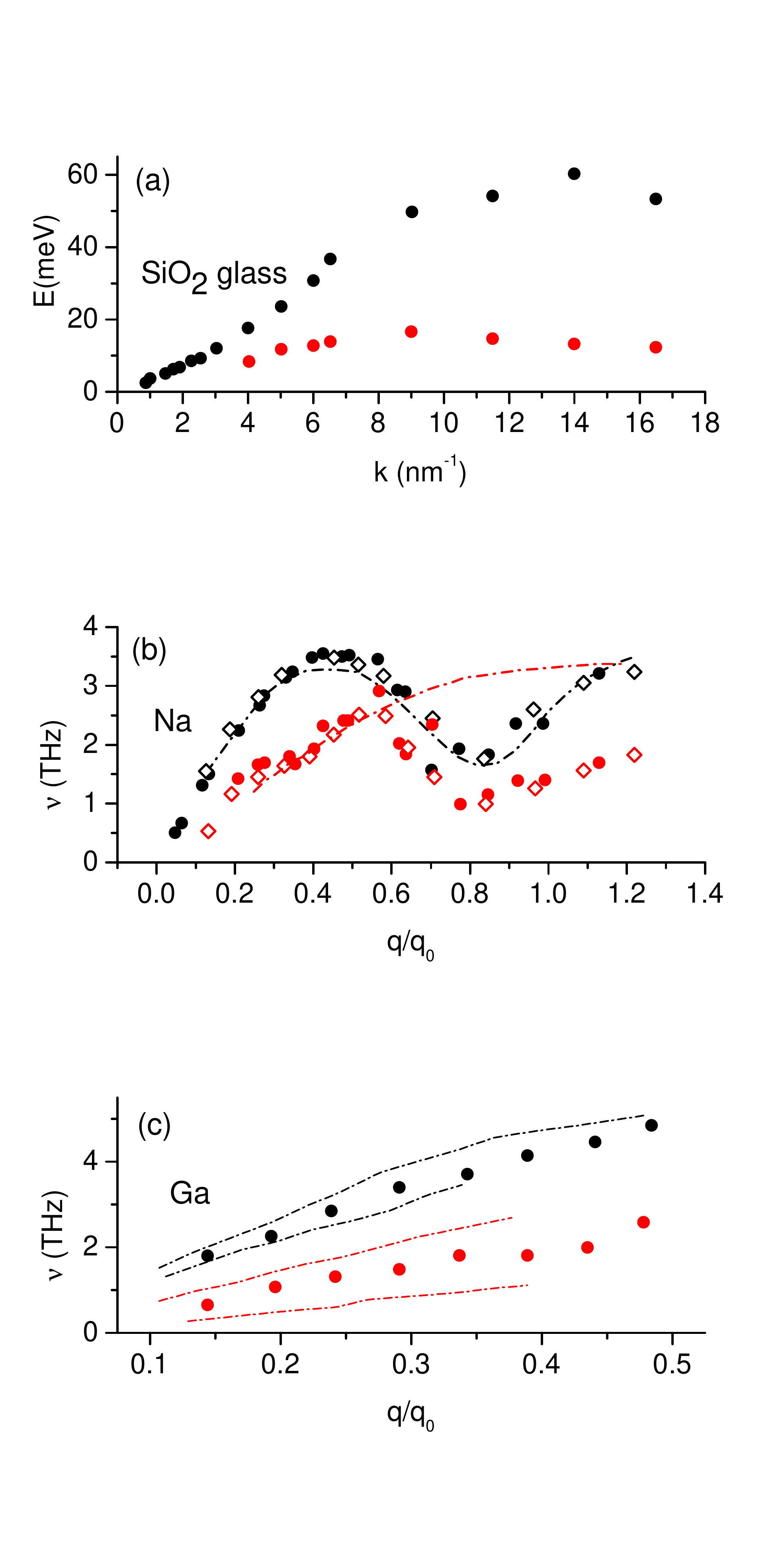}}}
\end{center}
\caption{Colour online. Experimental dispersion curves. (a): longitudinal (filled black bullets) and transverse (filled red bullets) dispersion curves in SiO$_2$ glass \protect\cite{ruzi}. (b): longitudinal (filled black bullets) and transverse (filled red bullets) excitations in liquid Na. Open diamonds correspond to longitudinal (black) and transverse (red) excitations in polycrystalline Na, and dashed-dotted lines to longitudinal (black) and transverse (red) branches along [111] direction in Na single crystal \protect\cite{mon-na}. (c): longitudinal (black bullets) and transverse (red bullets) excitation in liquid Ga. The bullets are bracketed by the highest and lowest frequency branches measured in bulk crystalline $\beta$-Ga along high symmetry directions, with black and red dashed-dotted lines corresponding to longitudinal and transverse excitations, respectively \protect\cite{mon-ga}. Dispersion curves in Na and Ga are reported in reduced zone units.}
\label{disp1}
\end{figure}

\begin{figure}
\begin{center}
{\scalebox{0.45}{\includegraphics{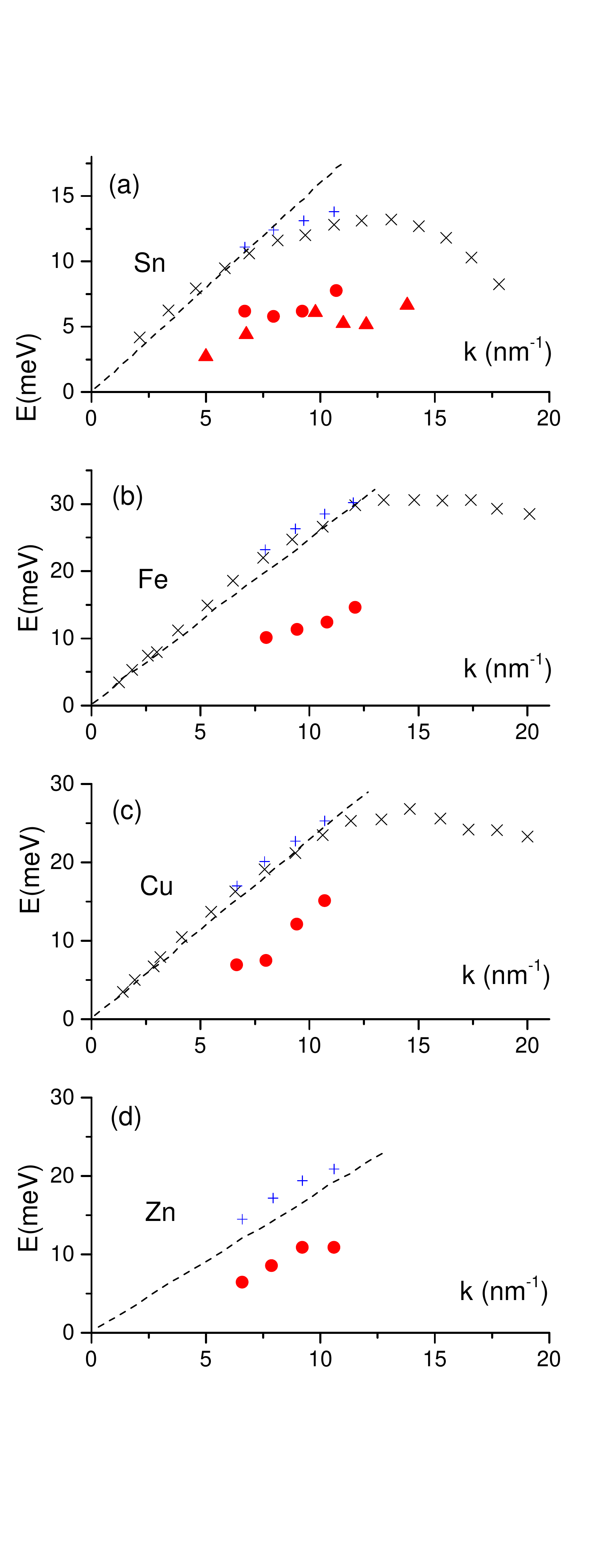}}}
\end{center}
\caption{Colour online. Longitudinal (black and blue crosses) and transverse (filled red bullets) dispersion curves in (a) liquid Sn \cite{sn}, (b) liquid Fe, (c) liquid Cu and (d) liquid Zn \cite{hoso3}. Red filled triangles in (a) are the results from ab initio simulations \cite{sn}. Blue and black crosses correspond to recent and earlier experiments, respectively. Dashed lines indicate the slope corresponding to the hydrodynamic sound in the limit of low $k$.}
\label{disp2}
\end{figure}

In Figure \ref{disp1}, we show measured dispersion curves measured in liquid Na \cite{mon-na} and liquid Ga \cite{mon-ga}, together with SiO$_2$ glass \cite{ruzi,baldi} for comparison. In Figure \ref{disp2}, we show the dispersion curves recently measured in liquid Sn \cite{sn}, Fe, Cu and Zn using the experimental setup to study liquids with high melting points \cite{hoso3}.

In Figure \ref{disp1}, we observe a striking similarity between liquids and their polycrystalline and crystalline counterparts in terms of longitudinal and transverse dispersion curves. We further note the similarity of dispersion curves in liquids and solid glasses. Overall, Figures \ref{disp1} and \ref{disp2} present an important experimental evidence regarding collective excitations in liquids. We observe that despite topological and dynamical disorder, solid-like quasi-linear dispersion curves exist in liquids in a wide range of $k$ and up to the largest $k$ corresponding to interatomic separations, as is the case in solids. Notably, this includes both high-frequency longitudinal and transverse modes.

We comment on damping of collective modes in liquids. A conservative system, crystalline or amorphous, has its eigenmodes which are non-decaying. Indeed, Eq. (\ref{eigen}) does not require system's crystallinity. For a disordered structure, Eq. (\ref{eigen}) gives eigenstates and eigenfrequencies corresponding to collective non-decaying excitations. For long wavelengths and small energies, these states are similar to harmonic plane waves and their damping in disordered systems is small. For short wavelengths, the eigenstates of the disordered system are different from the plane waves, and so damping of short-wavelength plane waves becomes appreciable. Yet the experimental dispersion curves obtained by harmonic probes such as X-rays or neutrons show that high-frequency plane waves are propagating in liquids, as witnessed by the data in Figures \ref{disp1} and \ref{disp2}. From the physical point of view, this follows from the fact that despite long-range disorder, a well-defined short-range order exists in liquids, glasses and other disordered systems, as is seen from the peaks of pair distribution functions in the short as well as medium range. Therefore, high-frequency harmonic plane waves, even though damped, are able to propagate at least the distance comparable to the typical length of the short-range order. We will find below that this length, the interatomic separation, which is also the fundamental length of the system, plays a profound role in governing the thermodynamic properties of liquids.

We have noted the similarity of vibrational properties between disordered liquids and their crystalline counterparts. Interestingly, similarity (and the lack thereof) between disordered glasses and their parent crystals have also been widely discussed. The widely discussed ``Boson'' peak in the low-frequency range has been long thought to be present in glasses only but not in crystals and to originate from disorder. However, later work \cite{chumakov,chumakov1} has demonstrated that similar vibrational features are present in crystals as well, provided glasses and crystals have similar density.

\section{Fast sound}

It is now good time to revisit the origin of fast sound mentioned earlier using detailed experimental data discussed in the previous section.

Starting from larger $k$-values, the measured speed of sound often exceeds the hydrodynamic value. This is seen in Figure \ref{disp2} where the hydrodynamic speed of sound is shown as a dashed line. The increase of the measured speed of sound over its hydrodynamic value is often called as ``fast sound'' or ``positive sound dispersion'' (PSD).

We recall Frenkel prediction discussed earlier: at high frequency where liquid's shear modulus becomes non-zero, the propagation velocity crosses over from its hydrodynamic value $v=\sqrt{\frac{B}{\rho}}$ to the solid-like elastic value $v=\sqrt{\frac{B+\frac{4}{3}G}{\rho}}$ \cite{lanstat,dyre}, where $B$ and $G$ are bulk and shear moduli, respectively.

The physical origin of the fast sound has remained controversial, including understanding relative contributions of the above mechanism and other effects such as disorder. Experimentally, the crossover of the longitudinal sound velocity from its hydrodynamic to solid-like elastic value has been been well-studied in viscous liquids where the system starts sustaining rigidity at MHz frequencies (see, e.g., Ref. \cite{jeong}, where fast sound is seen at fairly large wavelengths at which the liquid can be considered as a homogeneous medium). It is generally agreed that in this range of frequencies, fast sound originates from this mechanism \cite{jeong}.

At smaller wavelengths approaching the length of medium and short-range order, the wave feels structural inhomogeneities, and disorder of liquids and glasses starts to affect the dispersion relationship. PSD, with the relative magnitude of few per cent, was observed in a model harmonic glass and attributed to the ``instantaneous relaxation'' due to fast decay and dissipation of short-wavelength phonons in a disordered system \cite{fast1}. Later work demonstrated that starting from mesoscopic wavelengths, the effective speed of the longitudinal sound can also decrease \cite{fast2,dec1,dec2}. Different mechanisms and contributions to PSD were subsequently discussed \cite{ruocco,fast3}. The instantaneous relaxation is likely to be significant close to the zone boundary \cite{ruzi} (or the first Brillouin pseudo-zone, related to the short-range order in disordered systems \cite{ruocco}), although large PSD in silica glass may be related to the effect of mixing with the low-lying optic modes. In water, fast sound was discussed on the basis of coupling between the longitudinal and transverse excitations, and it was found that the onset of transverse excitations coincides with the inverse of liquid relaxation time \cite{water-fast,water-tran}, as predicted by (\ref{tau}).

Recent detailed experimental data discussed in the previous section enable us to directly address the origin of the fast sound and its magnitude. Combining $v_h=\sqrt{\frac{B}{\rho}}$, $v_t=\sqrt{\frac{G}{\rho}}$ and $v_l=\sqrt{\frac{B+\frac{4}{3}G}{\rho}}$ (see, e.g., Ref. \cite{dyre}), where $v_h$ is the velocity of the low-frequency hydrodynamic sound, $v_t$ is the transverse sound velocity and $v_l$ is the longitudinal velocity from the measured dispersion curves, we write

\begin{equation}
v_l^2=v_h^2+\frac{4}{3}v_t^2
\label{speeds}
\end{equation}

We note that the expression $v_l^2=\frac{B}{\rho}+\frac{4}{3}v_t^2$ is the identity for isotropic solids, and also applies to liquids in which the longitudinal speed of sound changes from the hydrodynamic to solid-like elastic value due to the onset of shear rigidity.

Using the data from Refs. \cite{sn} and \cite{hoso3}, we have taken $v_l$ and $v_t$ from the dispersion curves for Fe, Cu, Zn and Sn shown in Figure \ref{disp2} at $k$ points where the observed PSD is maximal and where $\omega(k)$ ($E(k)$) is in the quasi-linear regime before starting to curve at large $k$. For Fe, Cu, Zn, we use the new data shown in blue in Figure \ref{disp2} and consider the following $k$ points: $k=7.7$ nm$^{-1}$ (first point on the transverse branch in Figure \ref{disp2}), $k=7.8$ nm$^{-1}$ (second point on the transverse branch) and $k=8$ nm$^{-1}$ (second point on the transverse branch), respectively. For Sn, large PSD is seen at about $k=3.3$ nm$^{-1}$ corresponding to the second point on the longitudinal branch in Figure \ref{disp2}a. To find $v_t$ at this $k$, we extrapolated the higher-lying transverse points to lower $k$ while keeping them parallel to the simulation points, yielding $v_t=1220\pm 150$ m/s.

Using experimental $v_h$ and $v_t$, we have calculated $v_l$ using Eq. (\ref{speeds}). We show calculated and experimental $v_l$ in Table \ref{table} below.

\begin{table}[ht]
\begin{tabular}{l l l l l l}
\hline
       &    & $v_h$  & $v_t$  & $v_l$ (experimental) & $v_l$ (calculated) \\
       &    & [m/s]  & [m/s]  &[m/s]          &[m/s] \\

\hline
&Fe   & 3800  & 1870$\pm 50$  & 4370$\pm 30$ & 4370$\pm 50$  \\
&Cu   & 3460  & 1510$\pm 50$  & 3890$\pm 30$  & 3875$\pm 50$  \\
&Zn   & 2780  & 1620$\pm 50$  & 3330$\pm 30$  & 3350$\pm 50$  \\
&Sn   & 2440  & 1220$\pm 150$ & 2890$\pm 30$  & 2820$\pm 150$ \\
\hline
\end{tabular}
\caption{Comparison of experimental $v_l$ and $v_l$ calculated on the basis of $v_h$ and $v_t$ using Eq. (\ref{speeds}) as discussed in the text. The data for $v_l$, $v_t$ and $v_h$ is from Refs. \cite{sn} and \cite{hoso3}.
}
\label{table}
\end{table}

We observe in Table 1 that the calculated and experimental $v_l$ agree with each other very well. We therefore find that the mechanism of fast sound based on the onset of shear rigidity quantitatively accounts for the experimental data of real liquids in the wide range of $k$ spanning more than half of the first Brillouin pseudo-zone.

It is interesting to discuss pressure and temperature conditions at which the fast sound operates in this picture. The above mechanism implies that the fast sound disappears when the system loses shear resistance and transverse modes at all available frequencies. As discussed later, this takes place above the Frenkel line which demarcates liquid-like and gas-like properties at high temperature including in the supercritical region.

As already mentioned, other effects contributing to PSD can be operative, including the effects due to disorder at large $k$.

\section{Generalized hydrodynamics}

In the earlier section, we have discussed modifying (generalizing) hydrodynamic equations by including solid-like elastic effects as one way to describe both elastic and hydrodynamic response of the liquid. ``Generalized hydrodynamics'' as a distinct term refers to a number of proposals seeking to achieve essentially the same result by using a number of different phenomenological approaches \cite{boonyip,march1,baluca}. One starts with hydrodynamic equations initially applicable to low $\omega$ and $k$, and introduces a way to extend them to include the range of large $\omega$ and $k$.

From the point of view of thermodynamics, accounting for modes with high $\omega$ is important because these modes make the largest contribution to the system energy. The contribution of hydrodynamic modes is negligible by comparison.

Generalized hydrodynamics is a large field (see, e. g., \cite{boonyip,hansen2,march1,baluca}) which we can only discuss briefly emphasizing key starting equations and schemes of their modification to include higher-energy effects, with the aim to offer readers a feel for methods used and physics discussed.

The hydrodynamic description starts with viewing the liquid as a continuous homogeneous medium and constraining it with continuity equation and conservation laws such as energy and momentum conservation. Accounting for thermal conductivity and viscous dissipation using the Navier-Stokes equation, the system of equations can be linearized and solved. This gives several dissipative modes, from which the evaluation of the density-density correlation function gives the structure factor $S(k,\omega)$ in the Landau-Placzek form which includes several Lorentzians:

\begin{equation}
\begin{aligned}
&S(k,\omega)\propto\frac{\gamma-1}{\gamma}\frac{2\chi k^2}{\omega^2+\left(\chi k^2\right)^2}+\\
&\frac{1}{\gamma}\left(\frac{\Gamma k^2}{(\omega+ck)^2+\left(\Gamma k^2\right)^2}+\frac{\Gamma k^2}{(\omega-ck)^2+\left(\Gamma k^2\right)^2}\right)
\end{aligned}
\label{boon1}
\end{equation}

\noindent where $\chi$ is thermal diffusivity, $\gamma=\frac{C_p}{C_v}$ and dissipation $\Gamma$ depends on $\chi$, $\gamma$, viscosity and density.

The first term corresponds to the central Rayleigh peak and thermal diffusivity mode. The second two terms correspond to the Brillouin-Mandelstam peaks, and describe acoustic modes with the adiabatic speed of sound $c$. The ratio between the intensity of the Rayleigh peak, $I_{\rm R}$, and the Brillouin-Mandelstam peak, $I_{\rm BM}$, is the Landau-Placzek ratio: $\frac{I_{\rm R}}{I_{\rm BM}}=\gamma-1$. Applied originally to light scattering experiments, Eq. (\ref{boon1}) is also viewed as a convenient fit to high-energy experiments probing non-hydrodynamic processes where the fit that may include several Lorentzians or their modifications.

Generalizing hydrodynamic equations and extending them to large $k$ and $\omega$ is often done in terms of correlation functions. Solving the hydrodynamic Navier-Stokes equation for the transverse current correlation function $J_t(k,t)$, $\frac{\partial}{\partial t} J_t(k,t)=-\nu k^2J_t(k,t)$, where $\nu$ is kinematic viscosity, gives for the Fourier transform $J_t(k,\omega)$ a Lorentzian form similar to (\ref{boon1}):

\begin{equation}
J_t(k,\omega)=2v_0^2\frac{\nu k^2}{\omega^2+\left(\nu k^2\right)^2}
\end{equation}

\noindent where $\nu$ is kinematic viscosity and $v_0^2=J_t(k,t=0)$.

The generalization is done in terms of the memory function $K_t(k,t)$ defined in the equation for $J_t(k,\omega)$ as

\begin{equation}
\frac{\partial}{\partial t}J_t(k,\omega)=-k^2\int\limits_0^tK_t(k,t-t^\prime)J_t(k,t^\prime)dt^\prime
\label{mem}
\end{equation}

\noindent where $K_t(k,t-t^\prime)$ is the shear viscosity function or the memory function for $J_t(k,\omega)$ which describes its time dependence (``memory'').

Introducing $\tilde{J_t}(k,s)$ as the Laplace transform $J_t(k,\omega)=2\mathrm{Re}[\tilde{J_t}(k,s)]_{s=i\omega}$ and taking the Laplace transform of (\ref{mem}) gives

\begin{equation}
\tilde{J_t}(k,s)=v_0^2\frac{1}{s+k^2\tilde{K}_t(k,s)}
\label{tilde}
\end{equation}

The generalization introduces the dependence $k$ and $\omega$ by writing $\tilde{K}_t(k,s)$ as the sum of real and imaginary parts $[\tilde{K}_t(k,s)]_{s=i\omega}=K_t^\prime(k,\omega)+iK_t^{\prime\prime}(k,\omega)$. Then,

\begin{equation}
J_t(k,\omega)=2v_0^2\frac{k^2K_t^\prime(k,\omega)}{\left(\omega+k^2K_t^{\prime\prime}(k,\omega)\right)^2+\left(k^2K_t^{\prime}(k,\omega)\right)^2}
\label{genera}
\end{equation}
\noindent giving the generalized hydrodynamic description of the transverse current correlation function with a resonance spectrum.

Further analysis depends on the form of $K_t(k,t)$, which is often postulated as

\begin{equation}
K_t(k,t)=K_t(k,0)\exp\left(-\frac{t}{\tau(k)}\right)
\label{assu}
\end{equation}

Eq. (\ref{assu}) decays with time relaxation time $\tau$, and we recognize that this is essentially the same behavior described by earlier Eqs. (\ref{decay1}) or (\ref{ser}), except the postulated form also assumes $k$-dependence of $\tau$. In generalized hydrodynamics, Eq. (\ref{assu}) is used not only for $K$ but also for several types of correlation and memory functions. These often include modifications such as including more exponentials with different decay times in order to improve the fit to experimental or simulation data.

Mode-coupling schemes consider correlation functions for density and current density, factorise higher-order correlation functions by expressing them as the product of two time correlation functions with coupling coefficients in the form of static correlation functions, and give a better agreement for the relaxation function as compared to the single exponential decay model.

Neglecting $k$-dependence of $\tau$ for the moment, taking the Laplace transforms of (\ref{assu}) to find $K_t^\prime(k,\omega)$ and $K_t^{\prime\prime}(k,\omega)$ and using them in (\ref{genera}) gives $J_t(k,\omega)$ as \cite{boonyip}

\begin{equation}
J_t(k,\omega)\propto\frac{1}{\left(\omega^2-\left(k^2K_t(k,0)-\frac{1}{2\tau^2}\right)\right)^2+f(\tau,K_t(k,0))}
\label{genera1}
\end{equation}
\noindent where $f$ is the non-essential function of $\tau$ and $K_t(k,0)$.

The resonance frequency in (\ref{genera1}) corresponds to the propagation of shear modes provided $k^2K_t(k,0)>\frac{1}{2\tau^2}$. This condition defines the high-frequency regime of wave propagation in the solid-like elastic medium. Importantly, this condition is essentially the same as the one we derived from the generalized hydrodynamic equation (\ref{gener3}), as follows from the discussion between Eqs. (\ref{gener3}) and (\ref{hydro-gen}).

Similar expressions can be derived for the longitudinal current correlation function which also includes a static time-independent term which does not decay. This term corresponds to non-zero bulk modulus which gives propagating longitudinal waves in the hydrodynamic regime, as discussed in the previous section.

An alternative approach to generalize hydrodynamics is to make a phenomenological assumption that a dynamical variable in the liquid is described by the generalized Langevin equation:

\begin{equation}
\frac{{\partial}a(t)}{\partial t}+i\Omega a(t)+\int\limits_0^t a(t^\prime)K(t-t^\prime)dt^\prime=f(t)
\end{equation}

\noindent where the first two terms reflect the possibility of propagating modes, the third term plays the role of friction with the memory function $K$ and $f$ is the random force.

This approach proceeds by treating $a(t)$ not as a single variable but as a collection of variables of choice so that $a(t)$ becomes a vector including, in its simplest forms, conserved density, current density and energy variables. These variables are further generalized to include their dependence on wavenumber $k$. This gives a set of coupled equations solved in the matrix form. The set of dynamical variables can be extended to include the stress tensor and heat currents. In this case, the generalized viscosity is found to have the same exponential decay as in (\ref{assu}) once the stress tensor is explicitly introduced as a dynamical variable, the assumption is made regarding stress correlation function and a number of approximations are made. Then, similar viscoelastic effects are found as in the previous approach \cite{boonyip}.

Propagation of shear and longitudinal modes is also discussed in the mode-coupling theories mentioned above. The theory seeks to take a more general approach in the following sense. Considering that correlation functions are due to density and current density correlators, the theory represents $\tilde{K}_t(k,s)$  in (\ref{tilde}) by the second-order memory functions $M_t(k,t)$ and $M_l(k,t)$ for transverse and longitudinal currents, so that the transverse function $\tilde{J}_t(k,s)$ and longitudinal function $\tilde{J}_l(k,s)$ acquire the forms of damped oscillators. $\tilde{J}_l(k,s)$ differs from $\tilde{J}_t(k,s)$ by the presence of non-zero static term, giving a finite static restoring force for the longitudinal mode. As in the previous considerations, this gives propagating longitudinal modes in the hydrodynamic regime. Rather than postulating the relaxation functions $M_t(k,t)$ and $M_l(k,t)$ as in (\ref{assu}), the mode-coupling theory considers higher-order correlation functions and approximates them by the products of two-time correlation functions. Memory functions can then be calculated using the results from molecular dynamics simulations such as static correlation functions and other parameters required as the input. For simple systems, the onset of shear wave propagation can be related to certain shoulder-like features in the calculated memory function.

The amount of current research in generalized hydrodynamics has markedly decreased as compared to several decades ago \cite{boonyip}. Interestingly, the steer towards going beyond the hydrodynamic description and generalized hydrodynamics came from the experimental, and not theoretical, community after the solid-like properties of liquids were discovered and problems related to the hydrodynamic description of those properties became apparent \cite{pilgrim2,mon-na}. Some of the more recent examples include exploring how hydrodynamic description gives rise to a single underlying relaxation process and accounting for the viscoelastic effects using several first frequency moments (see \cite{mok1,mok2,mok3} and references therein). Other approaches assume ad hoc that more dynamical variables and their second and third derivatives are involved in extrapolating the hydrodynamic regime to high $k$ and $\omega$ \cite{bryk} and, following earlier proposals \cite{schepper}, use the generalized collective modes schemes where the sum of exponentials such as (\ref{assu}) is assumed to describe the decay of correlations. General disadvantages of this and similar schemes are related to the phenomenological and empirical nature of the method \cite{pilgrim2,mon-na}.

Continuing interest in generalized hydrodynamics is stimulated by fitting the experimental spectra where, for example, the second-order memory function is assumed to take the exponential form (\ref{assu}) \cite{ruocco} or as a sum of two or more exponentials \cite{verkerk}.

\section{Comment on the hydrodynamic approach to liquids}

Challenges involved in generalized hydrodynamics were appreciated by practitioners at the early stages of development \cite{boonyip}, including often phenomenological and empirical ways involved in extrapolating hydrodynamic description into the solid-like elastic regime. We do not review these here, although we note the following. Generalized hydrodynamics introduces $k$ and $\omega$-dependencies in the liquid properties such as diffusion, viscosity, thermal conductivity, heat capacity and so on, with the aim to calculate and discuss these functions in the non-hydrodynamic regime. It is not entirely clear what is the physical meaning of concepts such as diffusion or viscosity at large $\omega$ where the system's response is elastic rather than viscous. Understanding physical effects at these frequencies is important because short-wavelength modes govern most important system properties such as energy.

We question a more fundamental premise of the hydrodynamic description of liquids: ``The advantage of approaching the large ($k$,$\omega$) region by generalizing the hydrodynamic description is that one maintains contact with the long-wavelength, low-frequency region at all stages of the development. This gives insight to the structure of the resulting equation'' \cite{boonyip}. Although being able to track the evolution of equations may be insightful in some cases, it may not be advantageous in general. There is no fundamental reason to designate the hydrodynamic approach as the universally correct starting point. The traditional reason for the hydrodynamic approach to liquids is that they are flowing systems and therefore obey hydrodynamic equations. As we have discussed above, this applies to times $t>\tau$ ($\omega<\omega_{\rm F}$) only whereas for $t<\tau$ ($\omega>\omega_{\rm F}$) the system is solid-like and can be described by solid-like equations. Furthermore, we have seen that the same properties of collective modes are obtained by either starting with the hydrodynamic equations and incorporating solid-like elastic effects or starting with the elasticity equations and incorporating the hydrodynamic fluidity.

Instead, we propose that for the purposes of fundamental microscopic description, liquids should be considered for what they are: systems with molecular dynamics of both types, solid-like oscillatory motion and diffusive jumps, with relative weights of these motions changing with temperature. As discussed below, these relative weights govern most important system properties. In this approach, the hydrodynamic regime ($\omega\tau<1$) and solid-like elastic regime ($\omega\tau>1$) can, and in many cases should, be considered separately and without necessarily seeking to extrapolate one regime onto the other. In addition to avoiding problems of ad-hoc extrapolation assumptions often present in generalized hydrodynamics, this approach has the added benefit of rigorously delineating different regimes of liquid dynamics where important properties are qualitatively different. This will become particularly apparent when we discuss the change of dynamics in the supercritical region at the Frenkel line, the effect that the hydrodynamic description misses.

The hydrodynamic and solid-like elastic descriptions of liquids apply in their respective domains. It turns out that it is the solid-like description that is relevant for constructing the thermodynamic theory of liquids discussed in the next section. This is because high-frequency modes make the largest contribution to the system energy due to quadratic density of states $\propto\omega^2$, and propagate in the solid-like elastic regime $\omega\tau>1$. Importantly, this does not require extrapolations involved in the generalized hydrodynamics approach.

\section{Phonon theory of liquid thermodynamics}

\subsection{Harmonic theory}

We have seen above that collective modes in liquids include one longitudinal mode and two transverse modes propagating at frequency $\omega>\omega_{\rm F}=\frac{1}{\tau}$ in the solid-like elastic regime. The energy of these modes is the liquid vibrational energy. In addition to oscillating, particles in the liquids undergo diffusive jumps between quasi-equilibrium positions as discussed above. We write the total liquid energy as

\begin{equation}
E=K+P_l+P_t(\omega>\omega_{\rm F})+P_d
\label{en0}
\end{equation}

In Eq. (\ref{en0}), $K$  is the sum of all kinetic terms including vibrational and diffusional components. In the classical case, $K=\frac{3}{2}Nk_{\rm B}T$, and does not depend on how the kinetic energy partitions into oscillating and diffusive components. $P_l$ and $P_t(\omega>\omega_{\rm F})$ are potential energies of the longitudinal mode and transverse phonons with frequency $\omega>\omega_{\rm F}$, respectively. For now, we tentatively include in Eq. (\ref{en0}) the term $P_d$, related to the energy of interaction of diffusing particles with other parts of the system. $P_d$ is understood to be part of system's potential energy which is not already contained in the potential energy of the phonon terms, $P_l$ and $P_t(\omega>\omega_{\rm F})$. $P_d$ is small compared to other terms in (\ref{en0}) as discussed below.

The smallness of $P_d$ can be discussed by approaching the liquid from either gas or solid state. Lets consider a dilute interacting gas where system's potential energy is entirely given by the potential energy of the longitudinal mode, $P_l$, with the available wavelengths that depend on pressure and temperature. The remaining energy in the system is the kinetic energy corresponding to the free particle motion, giving $P_d=0$. Density increase (and temperature decrease) result in decreasing wavelength of the longitudinal mode until it reaches values comparable to solid-like interatomic separation $a$ (see the earlier section ``Experimental evidence for high-frequency collective modes in liquids). In this dense gas regime, the system's potential energy is still given by $P_l$, which is the energy of longitudinal mode but now with the full solid-like spectrum of wavelengths ranging from the system size to $a$. Further density increase or temperature decrease result in the appearance of the solid-like oscillatory component of motion. This process is most conveniently discussed above the critical point where no liquid-gas phase transition intervenes and where the crossover from purely diffusive motion to combined diffusive and solid-like oscillatory motion takes place at the Frenkel line discussed in later sections. The emergence of solid-like oscillatory component of particle motion is related to the emergence of transverse modes with frequency $\omega>\omega_{\rm F}$ in Eq. (\ref{en0}). The potential energy of transverse modes now contributes to the system's potential energy, and the remaining energy corresponds to the free particle motion ($P_d=0$ in Eq. (\ref{en0})) as in the dense gas.

We can also approach the liquid from the solid state. In the solid, the potential energy is the sum of potential components of longitudinal and transverse modes. The emergence of diffusive motion in the liquid results in the disappearance of transverse modes with frequency $\omega<\omega_{\rm F}$ according to (\ref{tau}) and modifies the potential energy of transverse modes to $P_t(\omega>\omega_{\rm F})$ in Eq. (\ref{en0}). This implies smallness of low-frequency potential energy of transverse modes: $P_t(\omega<\omega_{\rm F})\ll P_t(\omega>\omega_{\rm F})$, where $P_t(\omega<\omega_{\rm F})$ is the potential energy of low-frequency transverse modes. Instead of low-frequency transverse vibrations with potential energy $P_t(\omega<\omega_{\rm F})$ in a solid, atoms in a liquid ``slip'' and undergo diffusive motions with frequency $\omega_{\rm F}$ and associated potential energy $P_d$, hence $P_d\approx P_t(\omega<\omega_{\rm F})$. Combining this with $P_t(\omega<\omega_{\rm F})\ll P_t(\omega>\omega_{\rm F})$, $P_d\ll P_t(\omega>\omega_{\rm F})$ follows. Re-phrasing this, were $P_d$ large and comparable to $P_t(\omega>\omega_{\rm F})$, strong restoring forces at low frequency would result, and lead to the existence of low-frequency vibrations instead of diffusion. We also note that because $P_l\approx P_t$, $P_d\ll P_t(\omega>\omega_{\rm F})$ gives $P_d\ll P_l$, further implying that $P_d$ can be omitted in Eq. (\ref{en0}).

We note that in the regime $\tau\gg\tau_{\rm D}$, the justification for the smallness of $P_d$ in the two previous paragraphs becomes unnecessary. Indeed, using a rigorous statistical-mechanical argument it is easy to show that the total energy of diffusing atoms (the sum of their kinetic and potential energy) can be ignored to a very good approximation if $\tau\gg\tau_{\rm D}$. This is explained in the ``Viscous liquids'' section below in detail (see Eqs. \ref{1},\ref{2},\ref{ave1} and discussion around them), where we also remark that $\tau\gg\tau_{\rm D}$ corresponds to almost entire range of $\tau$ in which liquids exist as such.

Neglecting small $P_d$ in Eq. (\ref{en0}) is the only approximation in the theory; subsequent transformations serve to make the calculations convenient only. Eq. (\ref{en0}) becomes

\begin{equation}
E=K+P_l+P_t(\omega>\omega_{\rm F})
\label{en01}
\end{equation}

Eq. (\ref{en01}) can be re-written using the virial theorem $P_l=\frac{E_l}{2}$ and $P_t(\omega>\omega_{\rm F})=\frac{E_t(\omega>\omega_{\rm F})}{2}$ (here, $P$ and $E$ refer to their average values) and by additionally noting that the total kinetic energy $K$ is equal to the value of the kinetic energy of a solid and can therefore be written, using the virial theorem, as the sum of kinetic terms related to longitudinal and transverse waves: $K=\frac{E_l}{2}+\frac{E_t}{2}$, giving

\begin{equation}
E=E_l+\frac{E_t(\omega>\omega_{\rm F})}{2}+\frac{E_t}{2}
\label{en001}
\end{equation}

Noting that $E_t$ can be represented as $E_t=E_t(\omega<\omega_{\rm F})+E_t(\omega>\omega_{\rm F})$, liquid energy reads

\begin{equation}
E=E_l+E_t(\omega>\omega_{\rm F})+\frac{E_t(\omega<\omega_{\rm F})}{2}
\label{en1}
\end{equation}

The first two terms in (\ref{en1}) give the energy of propagating phonon states in the liquid. The second term is the energy of two transverse modes which decreases with temperature. This decrease includes both kinetic and potential parts, however the total kinetic energy of the system stays the same as in Eq. (\ref{en0}). The last term ensures that the decrease of the energy of transverse waves does not change the total kinetic energy, rather than points to the existence of low-frequency transverse waves (these are non-propagating in liquids).

Either (\ref{en001}) or (\ref{en1}) can now be used to calculate the liquid energy. Each term in Eqs. (\ref{en001}) or (\ref{en1}) can be calculated as the phonon energy, $E_{ph}$:

\begin{equation}
E_{ph}=\int E(\omega,T)g(\omega)d\omega
\label{phonen}
\end{equation}
\noindent where $g(\omega)$ is the phonon density of states.

Lets consider Eq. (\ref{en1}) and let $Z_2$ be the partition function associated with the first two terms in Eq. (\ref{en1}). Then, $Z_2$ is:

\begin{equation}
\begin{aligned}
&Z_2=(2\pi\hbar)^{-N^\prime}\int\exp\left(-\frac{1}{2T}\sum\limits_{i=1}^N(p_i^2+\omega_{li}^2q_i^2)\right) dpdq\\\times
&\int\exp\left(-\frac{1}{2T}\sum\limits_{\omega_{ti}>\omega_{\rm F}}^{2N}(p_i^2+\omega_{ti}^2q_i^2)\right)dpdq
\label{part1}
\end{aligned}
\end{equation}

\noindent where $\omega_{\rm F}=\frac{1}{\tau}$, $\omega_{li}$ and $\omega_{ti}$ are frequencies of longitudinal and transverse waves, $N$ is the number of atoms and $N^\prime$ is the number of phonon states that include longitudinal waves and transverse waves with frequency $\omega>\omega_{\rm F}$. Here and below, $k_{\rm B}=1$.

We recall our earlier discussion that the longitudinal mode propagates in two different regimes: hydrodynamic regime $\omega\tau<1$ or solid-like elastic regime $\omega\tau>1$. This gives different dissipation laws in the two regimes, but this circumstance is unimportant for calculating the energy. Indeed, (\ref{phonen}) makes no reference to dissipation, and includes the mode energy and the density of states only. These are the same in the two regimes, and hence for the purposes of calculating the energy, the longitudinal mode can be considered as one single mode with Debye density states. This statement is not entirely correct because the mode is not well described in the regime $\omega\tau\approx 1$, however this circumstance is not essential because, as we will see later, almost entire energy is due to the modes with high frequency propagating in the solid-like elastic regime anyway.

Integrating (\ref{part1}), we find

\begin{equation}
Z_2=T^N\left(\prod\limits_{i=1}^N\hbar\omega_{li}\right)^{-1}T^{N_1}\left(\prod\limits_{\omega_{ti}>\omega_0}^{2N}\hbar\omega_{si}\right)^{-1}
\label{part2}
\end{equation}

\noindent where $N_1$ is the number of transverse modes with $\omega>\omega_{\rm F}$.

In the harmonic approximation, frequencies $\omega_{li}$ and $\omega_{ti}$ are considered to be temperature-independent, in contrast to anharmonic case discussed in the next section. Then, Eq. (\ref{part2}) gives the energy $E=T^2\frac{d}{dT}\ln Z=NT+N_1T$.

$N_1$ can be calculated using the quadratic density of states in the Debye model, as is done in solids \cite{landau}. Here and below, the developed theory is at the same level of approximation as Debye theory of solids. The density of states of transverse modes is $g_t(\omega)=\frac{6N}{\omega_{mt}^3}\omega^2$, where $\omega_{mt}$ is Debye frequency of transverse modes and we have taken into account that the number of transverse modes in the solid-like density of states is $2N$. $\omega_{mt}$ can be somewhat different from the longitudinal Debye frequency; for simplicity we assume $\omega_{mt}\approx\omega_{\rm D}$. Then, $N_1=\int\limits_{\omega_{\rm F}}^{\omega_{\rm D}}g_t(\omega)d\omega=2N\left(1-\left(\frac{\omega_{\rm F}}{\omega_{\rm D}}\right)^3\right)$.

To calculate the last term in Eq. (\ref{en1}), we note that similarly to $E_t(\omega>\omega_{\rm F})=N_1 T$, $E_t(\omega<\omega_{\rm F})$ can be calculated to be $E_t(\omega<\omega_{\rm F})=N_2 T$, where $N_2$ is the number of shear modes with $\omega<\omega_{\rm F}$. Because $N_2=2N-N_1$, $N_2=2N\left(\frac{\omega_{\rm F}}{\omega_{\rm D}}\right)^3$. The total liquid energy is $E=(N+N_1+\frac{N_2}{2})T$ according to Eq. (\ref{en1}), giving finally \cite{prb1}:

\begin{equation}
E=NT\left(3-\left(\frac{\omega_{\rm F}}{\omega_{\rm D}}\right)^3\right)
\label{harmo}
\end{equation}

At low temperature where $\tau\gg\tau_{\rm D}$, or $\omega_{\rm F}\ll\omega_{\rm D}$, Eq. (\ref{harmo}) gives $c_v=\frac{1}{N}\frac{dE}{dT}=3$, the harmonic solid result. At high temperature when $\tau\rightarrow\tau_{\rm D}$ and $\omega_{\rm F}\rightarrow\omega_{\rm D}$, Eq. (\ref{harmo}) gives $c_v=2$, consistent with the experimental result in Figure \ref{mercury}. As the number of transverse modes with frequency above $\omega_{\rm F}$ decreases with temperature, $c_v$ decreases from about 3 to 2. A quantitative agreement in the entire temperature range can be studied by using Eq. (\ref{maxrel}) or $\omega_{\rm F}=\frac{G_\infty}{\eta}$, where $\eta$ is taken from the independent experiment. This way, $E$ in Eq. (\ref{harmo}) and $c_v$ have no free fitting parameters. The agreement of Eq. (\ref{harmo}) with the experimental $c_v$ of liquid Hg is good at this level of approximation already \cite{prb1}.

In this picture, the decrease of $c_v$ with temperature is due to the evolution of collective modes in the liquid, namely the reduction of the number of transverse modes above the frequency $\omega_{\rm F}=\frac{1}{\tau}$. We will discuss experimental data of $c_v$ for several types of liquids in more detail below, including metallic, noble and molecular liquids, and will find that their $c_v$ similarly decreases with temperature as Eq. (\ref{harmo}) predicts. The same trend, the decrease of $c_v$ with temperature, has been experimentally found in complex liquids, including such systems as toluene, propane, ether, chloroform, benzene, methyl cyclohexane and cyclopentane, hexane, heptane, octane and so on \cite{dexter}.

We have focused on calculating liquid energy and resulting heat capacity that have contributions from collective modes and diffusing atoms. We have not discussed liquid entropy which includes the configurational entropy measuring the total phase space available to the system, the phase space sampled by diffusive particle jumps. Unlike entropy, the energy is not related to exploring the phase space, and corresponds to the instantaneous state of the system (in the microcanonical ensemble, or averaged over fluctuations in the canonical ensemble). We will return to this point below when we discuss thermodynamic properties of viscous liquids.

We make two remarks related to using the Debye model. First, the Debye model is particularly relevant for disordered isotropic systems such as glasses \cite{landau}, which are known to be nearly identical to liquids from the structural point of view \cite{dyre}. Furthermore, we have seen earlier that the dispersion curves in liquids are very similar to those in solids (including crystals, poly-crystals and glasses). Therefore, the Debye model can be used in liquids to the same extent as in solids. One important consequence of this is that high-frequency modes in liquids make the largest contribution to the energy, as they do in solids including disordered solids. This is re-iterated elsewhere in this paper.

Second, recall our earlier observation that $\omega$ gradually increases from 0 to $\omega=ck$ around $\omega_{\rm F}$ with a square-root dependence (see Eq. (\ref{hydro-gen}) and discussion below). Writing $\int\limits_{\omega_{\rm F}}^{\omega_{\rm D}}g_t(\omega)d\omega$ in the previous paragraph assumes a sharp lower frequency cutoff at $\omega_{\rm F}$, and is an approximation in this sense. The approximation is justified because it is the highest frequency modes above $\omega_{\rm F}$ that make the most contribution to the liquid energy, and because Debye density of states we employ is already an approximation to the frequency spectrum, the approximation that may be larger than the one involved in substituting the square-root crossover with a sharper cutoff.

\subsection{Comment on the phonon theory of liquid thermodynamics}

We pause for the moment to make several comments about Eq. (\ref{harmo}) and its relationship to our starting Eq. (\ref{01}) in the Introduction. $g(r)$ and $U(r)$ featuring in Eq. (\ref{01}) are not generally available apart from simple systems such as Lennard-Jones liquids. For simple liquids, $g(r)$ and $U(r)$ can be determined from experiments or simulations and subsequently used in Eq. (\ref{01}). Unfortunately, neither $g(r)$ nor $U(r)$ are available for liquids with any larger degree of complexity of structure or interactions. For example, many-body correlations \cite{born,hender} and network effects can be strong in familiar liquid systems such as olive oil, SiO$_2$, Se, glycerol, or even water \cite{emilio}, resulting in complicated structural correlation functions that cannot be reduced to the simple two- or even three-body correlations that are often used. As discussed in Ref. {\cite{march}, approximations become difficult to control when the order of correlation functions already exceeds three-body correlations. Similarly, it is challenging to extract multiple correlation functions from the experiment. The same problems exist for interatomic interactions, which can be equally multibody and complex, and consequently not amenable to determination in experiments or simulations. On the other hand, $\omega_{\rm F}$ ($\tau$) is available much more widely as discussed above, enabling us to calculate and understand liquid $c_v$ readily.

Next, expressing the liquid energy in terms of $\omega_{\rm F}$ in Eq. (\ref{harmo}) represents a more general description of liquids as compared to Eq. (\ref{01}). In Eq. (\ref{01}), the energy strongly depends on interactions. It was for this reason that Landau and Lifshitz state that the liquid energy is strongly system-dependent and therefore cannot be calculated in general form \cite{landau}. Let us now consider liquids with very different structural correlations and interatomic interactions such as, for example, H$_2$O, Hg, AsS, olive oil, and glycerol. As long as $\omega_{\rm F}$  of the above liquids is the same at a certain temperature, Eq. (\ref{harmo}) predicts that their energy is the same (in molecular liquids, we are referring to the inter-molecular energy as discussed below in more detail). In this sense, expressing the liquid energy as a function of $\omega_{\rm F}$ only is a more general description because $\omega_{\rm F}$ is a uniformly common property for all liquids.

Finally, Eq. (\ref{harmo}), as well as its modifications below, are simple. This makes it fairly easy to understand and interpret experimental data as discussed in the later section.

An objection could be raised that, although our approach explains the experimental $c_v$ of liquids as discussed below, the approach is based on $\omega_{\rm F}$, the emergent property rather than on the ostensibly lower-level data such as $g(r)$ and $U(r)$ in Eq. (\ref{01}). This brings us to an important question of what we aim to achieve by a physical theory. According to one view, ``The point of any physical theory is to make statements about the outcomes of future experiments on the basis of results from the previous experiment'' \cite{pierls}. This emphasizes relationships between experimental properties. In this sense, Eq. (\ref{harmo}) provides a relationship between liquid thermodynamic properties such as energy and $c_v$ on one side and its dynamical and oscillatory properties such as $\omega_{\rm F}$ on the other.

%Interestingly, Eq. (\ref{harmo}) allows to explore an important relationship between $\omega_{\rm F}$ on one hand and $g(r)$ and $U(r)$ on the other, i.e. the relationship between dynamical and structural properties of liquids. Indeed, equating Eqs. (\ref{harmo}) and (\ref{01}) gives the relationship between the dynamical property $\omega_{\rm F}$ and structural property $g(r)$. Such as relationship has not been hitherto explored, and can be an interesting subject for future work. Fundamentally, $\omega_{\rm F}$ and $\tau$ depend on liquid structure and interactions, but calculating this dependence from first principles is not feasible due to the exponential complexity of the first-principles description of liquids discussed earlier.

\subsection{Including anharmonicity and thermal expansion}

In calculating the energy $E=T^2\frac{d}{dT}\ln Z$, we have assumed that the phonon frequencies are temperature-independent. Generally, the phonon frequencies reduce with temperature. This takes place at both constant pressure and constant volume. At constant volume, reduction of frequencies is related to inherent anharmonicity and increased vibration amplitudes. If frequencies are temperature-dependent, applying $E=T^2\frac{d}{dT}\ln Z$ to Eq. (\ref{part2}) gives

\begin{equation}
E_2=NT-T^2\sum\limits_{i=1}^N\frac{1}{\omega_{li}}\frac{d\omega_{li}}{dT}+N_1T-T^2\sum\limits_{i=1}^{N_1}\frac{1}{\omega_{ti}}\frac{d\omega_{ti}}{dT}
\label{anharm}
\end{equation}

\noindent where the derivatives are at constant volume. Eq. (\ref{anharm}) gives the first two terms in Eq. (\ref{en1}).

Using Gr\"{u}neisen approximation, it is possible to derive a useful approximate relation: $\frac{1}{\omega}\left(\frac{{\rm d}\omega_i}{{\rm d}T}\right)_v=-\frac{\alpha}{2}$, where $\alpha$ is the coefficient of thermal expansion \cite{left,gla-tr}. Using this in (\ref{anharm}) gives

\begin{equation}
E_2=(N+N_1)T\left(1+\frac{\alpha T}{2}\right)
\label{anharm1}
\end{equation}

The last term in Eq. (\ref{en1}), $\frac{E_t(\omega<1/\tau)}{2}$, can be calculated in the same way, giving $\frac{1}{2}N_2T\left(1+\frac{\alpha T}{2}\right)$, where $N_2$ is the number of shear modes with $\omega<\omega_{\rm F}$ calculated in the previous section. Adding this term to Eq. (\ref{anharm1}) and using $N_1$ and $N_2$ from the previous section gives the anharmonic liquid energy:

\begin{equation}
E=NT\left(1+\frac{\alpha T}{2}\right)\left(3-\left(\frac{\omega_{\rm F}}{\omega_{\rm D}}\right)^3\right)
\label{anharm2}
\end{equation}

\noindent which reduces to (\ref{harmo}) when $\alpha=0$.

Eq. (\ref{anharm2}) has been found to quantitatively describe $c_v$ of 5 commonly studied liquid metals in a wider temperature range where $c_v$ decreases from about 3 around the melting point to 2 at high temperature \cite{prb2}. The presence of the anharmonic term in Eq. (\ref{anharm2}), $\left(1+\frac{\alpha T}{2}\right)$, explains why experimental $c_v$ of liquids may exceed the Dulong-Petit value $c_v=3$ close to the melting point \cite{grimvall,wallace}.

At low temperature when $\tau\gg\tau_{\rm D}$, Eq. (\ref{anharm2}) gives

\begin{equation}
E=3NT\left(1+\frac{\alpha T}{2}\right)
\label{enesol}
\end{equation}

\noindent and $c_v$ is

\begin{equation}
c_v=3\left(1+\alpha T\right)
\label{sol}
\end{equation}

Eq. (\ref{sol}) is equally applicable to solids and viscous liquids where $\tau\gg\tau_{\rm D}$, and has been found consistent with several simulated crystalline and amorphous systems \cite{left}.

We note that Eqs. (\ref{enesol}) and (\ref{sol}) don't need to be derived from Eq. (\ref{anharm2}), and also follow from considering the solid as a starting point where all three modes are present.

\subsection{Including quantum effects}

If the temperature range includes low temperature where $\frac{\hbar\omega_{\rm D}}{T}\ll 1$ does not hold, effects related to quantum excitations become important. In this case, each term in Eq. (\ref{en1}) can be calculated using the phonon free energy \cite{landau} as

\begin{equation}
F_{ph}=E_0+T\sum\limits_i\ln\left(1-\exp\left(-\frac{\hbar\omega_i}{T}\right)\right)
\end{equation}

\noindent where $E_0$ is the energy of zero-point vibrations. In calculating the energy, $E_{ph}=F_{ph}-T\frac{d F_{ph}}{d T}$, we assume $\frac{d\omega_i}{d T}\ne 0$ as in the previous section, giving for the phonon energy

\begin{equation}
E_{ph}=E_0+\hbar\sum\limits_i\frac{\omega_i-T\frac{d\omega_i}{d T}}{\exp\left(\frac{{\hbar\omega_i}}{T}\right)-1}
\label{en2}
\end{equation}

Using $\frac{{\rm d}\omega_i}{{\rm d}T}=-\frac{\alpha\omega_i}{2}$ as before gives

\begin{equation}
E_{ph}=E_0+\left(1+\frac{\alpha T}{2}\right)\sum\limits_i\frac{\hbar\omega_i}{\exp\left(\frac{{\hbar\omega_i}}{T}\right)-1}
\label{en3}
\end{equation}

In this form, Eq. (\ref{en3}) can be used to calculate each of the three terms in (\ref{en1}). The energy of one longitudinal mode, the first term in Eq. (\ref{en1}), can be calculated by substituting the sum in Eq. (\ref{en3}), $\sum$, with Debye vibrational density of states for longitudinal phonons, $g(\omega)=\frac{3N}{\omega_{\rm D}^3}\omega^2$, where $\omega_{\rm D}$ is Debye frequency. Integrating from 0 to $\omega_{\rm D}$ gives $\sum=NTD\left(\frac{\hbar\omega_{\rm D}}{T}\right)$, where $D(x)=\frac{3}{x^3}\int\limits_0^x\frac{z^3{\rm d}z}{\exp(z)-1}$ is Debye function \cite{landau}. The energy of two transverse modes with frequency $\omega>\omega_{\rm F}$, the second term in Eq. (\ref{en1}), can be similarly calculated by substituting $\sum$ with density of states $g(\omega)=\frac{6N}{\omega_{\rm D}^3}\omega^2$, where the normalization accounts for the number of transverse modes of $2N$. Integrating from $\omega_{\rm F}$ to $\omega_{\rm D}$ gives $\sum=2NTD\left(\frac{\hbar\omega_{\rm D}}{T}\right)-2NT\left(\frac{\omega_{\rm F}}{\omega_{\rm D}}\right)^3D\left(\frac{\hbar\omega_{\rm F}}{T}\right)$. Finally, $E_t(\omega<\omega_{\rm F})$ in the last term in Eq. (\ref{en1}) is obtained by integrating $\sum$ from 0 to $\omega_{\rm F}$ with the same density of states, giving $\sum=2NT\left(\frac{\omega_{\rm F}}{\omega_{\rm D}}\right)^3D\left(\frac{\hbar\omega_{\rm F}}{T}\right)$. Putting all terms in Eq. (\ref{en3}) and then Eq. (\ref{en1}) gives finally the liquid energy

\begin{equation}
\begin{aligned}
&E=E_0+NT\left(1+\frac{\alpha T}{2}\right)\left(3D\left(\frac{\hbar\omega_{\rm D}}{T}\right)-\left(\frac{\omega_{\rm F}}{\omega_{\rm D}}\right)^3D\left(\frac{\hbar\omega_{\rm F}}{T}\right)\right)
\label{enf}
\end{aligned}
\end{equation}
\noindent

In general, $E_0$ is temperature-dependent because it depends on $\omega_{\rm F}$ and therefore $T$. However, this becomes important at temperatures of several K only, whereas below we deal with significantly higher temperatures where $E_0$ and its derivative in (\ref{enf}) are small compared to the second temperature-dependent term. In the subsequent comparison of (\ref{enf}) with experimental $c_v$, we therefore do not include $E_0$.

In the high-temperature classical limit where $\frac{\hbar\omega_{\rm D}}{T}\ll 1$ and, therefore, $\frac{\hbar\omega_{\rm F}}{T}\ll 1$ ($\omega_{\rm F}<\omega_{\rm D}$), Debye functions become 1, and (\ref{enf}) reduces to the energy of the classical liquid, Eq. (\ref{anharm2}).

For some of the liquids discussed in the next section, the high-temperature classical approximation $\frac{\hbar\omega_{\rm D}}{T}\ll 1$ does not hold in the temperature range considered \cite{scirep}. In this case, quantum effects at those temperatures become significant, and Eq. (\ref{enf}) should be used to calculate liquid $c_v$.

\subsection{Comparison with experimental data}

The most straightforward comparison of the above theory to experiments is to calculate the energy using Eqs. (\ref{harmo}), (\ref{anharm2}) or (\ref{enf}) and experimental $\omega_{\rm F}$. This is often done by fitting $\omega_{\rm F}$ to function such as the VFT law, using it to calculate the energy and differentiating it to find $c_v$ and compare it to the experimental data. Eqs. (\ref{harmo}), (\ref{anharm2}) or (\ref{enf}) involve no free fitting parameters, and contain parameters related to system properties only. If $\omega_{\rm F}$ is calculated from experimental viscosity as $\omega_{\rm F}=\frac{G_\infty}{\eta}$, Eq. (\ref{harmo}) contains $G_\infty$ and $\tau_{\rm D}$ which enter as the product $G_\infty\tau_{\rm D}$. Eq. (\ref{anharm2}) contains parameters $G_\infty\tau_{\rm D}$ and $\alpha$. In Eq. (\ref{enf}), $G_\infty$ and $\tau_{\rm D}$ feature separately.

In the last few years, we have compared theoretical and experimental $c_v$ of over 20 different systems, including metallic, noble, molecular and network liquids \cite{prb1,prb2,scirep}. We aimed to check our theoretical predictions in the widest temperature range possible, and therefore used the data at pressures exceeding the critical pressures from the National Institute of Standards and Technology (NIST) database \cite{nist}. As a result, many studied liquids are supercritical. In Figure \ref{cv3}, we show the comparison of theoretical and experimental data for several representative liquids. We have included three liquids in each class: metallic, noble and molecular liquids.

\begin{figure}
\begin{center}
{\scalebox{0.4}{\includegraphics{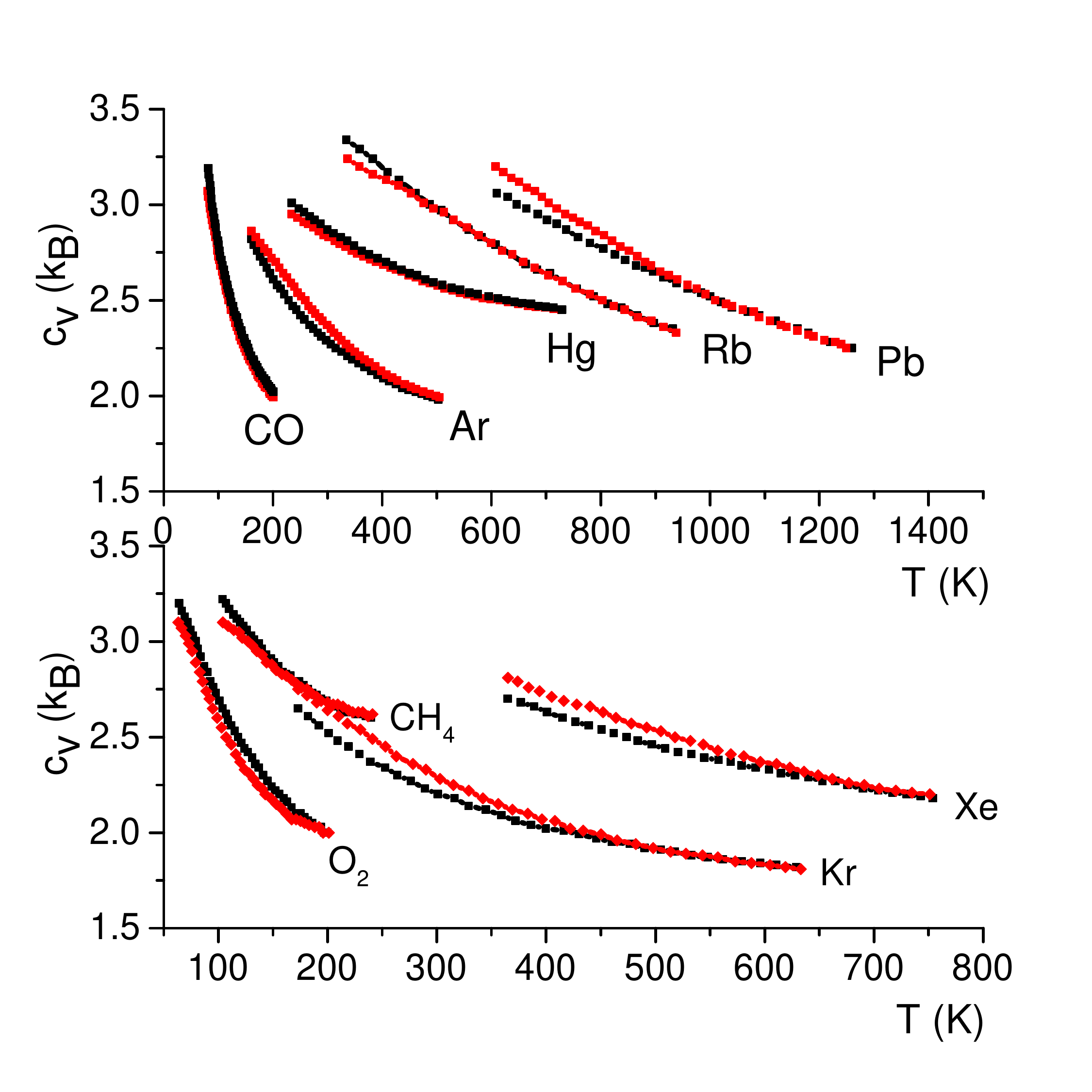}}}
\end{center}
\caption{Colour online. Experimental $c_v$ (black color) in metallic, noble and molecular liquids ($k_{\rm B}=1$). Experimental $c_v$ are measured on isobars. Theoretical $c_v$ (red color) was calculated using Eq. (\ref{enf}). The data are from Ref. \cite{scirep}. The data for molecular and noble liquids are taken at high pressure to increase the temperature range where these systems exist in the liquid form \cite{nist}. We show the data in two graphs to avoid overlapping.}
\label{cv3}
\end{figure}

We observe good agreement between experiments and theoretical predictions in a wide temperature range of about 50-–1300 K in Figure \ref{cv3}. The agreement supports the interpretation of the universal decrease of $c_v$ with temperature: the decrease is due to the reduction of the number of transverse modes propagating above frequency $\frac{1}{\tau}$.

We note that Debye model is not a good approximation in molecular and hydrogen-bonded systems where the frequency of intra-molecular vibrations considerably exceeds the rest of frequencies in the system (e.g. 3572 K in CO and 2260 K in O$_2$). However, the intra-molecular modes are not excited in the temperature range of experimental $c_v$ (see Figure {\ref{cv3}). Therefore, the contribution of intra-molecular motion to $c_v$ is purely rotational, $c_{\rm rot}$. The rotational motion is excited in the considered temperature range, and is classical, giving $c_{\rm rot}=R$ for linear molecules such as CO and O$_2$ and $c_{\rm rot}=\frac{3R}{2}$ for molecules with three rotation axes such as CH$_4$. Consequently, $c_v$ for liquid CO shown in Figure ({\ref{cv3}) corresponds to the heat capacity per molecule, with $c_{\rm rot}$ subtracted from the experimental data. In this case, $N$ in Eqs. (\ref{harmo}), (\ref{anharm2}) or (\ref{enf}) refers to the number of molecules.

\subsection{Phonon excitations at low temperature}

The number of excited phonon states increases with temperature. At low temperature, this results in the well-known increase of $c_v$: $c_v\propto T^3$. This increase can compete with the decrease of $c_v$ to the progressive loss of transverse modes discussed above. In practice, all liquids solidify at low temperature and room pressure except helium. In liquid helium under pressure, $c_v$ can first increase with temperature due to the phonon excitation effects. This is followed by the decrease of $c_v$ at higher temperature, similar to the behavior of classical liquids in Figure \ref{cv3}. As a result, $c_v$ can have a maximum \cite{hel-pres}.

An interesting assertion can be made about the operation of transverse modes in a hypothetical liquid in the limit of zero temperature: transverse modes do not contribute to liquid's energy and specific heat in this limit \cite{prb1}. Indeed, let us consider a liquid with a certain $\omega_{\rm F}$ and calculate the quantum energy of two transverse modes with frequency above $\omega_{\rm F}$ as $E^t_T(\omega>\omega_{\rm F})=\int\limits_{\omega_{\rm F}}^{\omega_{\rm D}}\frac{\hbar\omega}{\exp\frac{\hbar\omega}{T}-1}g_t(\omega)d\omega$. $E^t_T(\omega>\omega_{\rm F})$ can be written as

\begin{equation}
E^t_T(\omega>\omega_{\rm F})=\int\limits_0^{\omega_{\rm D}}\frac{\hbar\omega g_t(\omega)d\omega}{\exp\frac{\hbar\omega}{T}-1}-\int\limits_0^{\omega_{\rm F}}\frac{\hbar\omega g_t(\omega)d\omega}{\exp\frac{\hbar\omega}{T}-1}
\label{q1}
\end{equation}

Integrating (\ref{q1}) with Debye density of states $g_t(\omega)=\frac{6N}{\omega_{\rm D}^3}\omega^2$ gives:

\begin{equation}
E^t_T(\omega>\omega_{\rm F})=2NTD\left(\frac{\hbar\omega_{\rm D}}{T}\right)-2NT\left(\frac{\omega_{\rm F}}{\omega_{\rm D}}\right)^3D\left(\frac{\hbar\omega_{\rm F}}{T}\right)
\label{q2}
\end{equation}

In the low-temperature limit where $D(x)=\frac{\pi^4}{5x^3}$, the two terms cancel exactly, giving $E^t_T(\omega>\omega_{\rm F})=0$. The same result follows without relying on the Debye model and from observing that in the low-temperature limit, the upper integration limits in both terms in (\ref{q1}) can be extended to infinity due to fast convergence of integrals. Then, $E^t_T(\omega>\omega_{\rm F})$ in (\ref{q1}) is the difference between two identical terms and is zero \cite{prb1}.

Physically, the reason for $E^t_T(\omega>\omega_{\rm F})=0$ is that only high-frequency transverse modes exist in a liquid according to (\ref{tau}), but these are not excited at low temperature.

We will re-visit this result in the later section discussing solid-like approaches to quantum liquids such as liquid helium.

\section{Heat capacity of supercritical fluids}

In the above discussion, $c_v$ decreases from about 3 at low temperature to 2 at high, corresponding to the complete loss of solid-like transverse modes. It is interesting to ask how $c_v$ changes on further temperature increase. On general grounds, one expects to find the gas-like value $c_v=\frac{3}{2}$ at high temperature where the kinetic energy dominates.

If the system is below the critical point (see Figure 1), further temperature increase involves boiling and the first-order transition, with $c_v$ discontinuously decreasing to $\frac{3}{2}$ in the gas phase. The intervening phase transition excludes the state of the liquid where $c_v$ can gradually change from 2 to $\frac{3}{2}$ and where interesting physics operates. However, this becomes possible above the critical point. This brings us to the interesting discussion of the supercritical state of matter.

\subsection{Frenkel line}

Supercritical fluids started to be widely deployed in many important industrial processes \cite{sup1,sup2} once their high dissolving and extracting properties were appreciated. These properties are unique to supercritical fluids and primarily result from the combination of high density and high particle mobility. Theoretically, little was known about the supercritical state, apart from the general assertion that supercritical fluids can be thought of as high-density gases or high-temperature fluids whose properties change smoothly with temperature or pressure and without qualitative changes of properties. This assertion followed from the known absence of a phase transition above the critical point.

We have recently proposed that this picture should be modified, and that a new line, the Frenkel line (FL), exists above the critical point and separates two states with distinct properties (see Figure \ref{frenline}) \cite{pre,phystoday,prl,ufn}. The main idea of the FL lies in considering how particle dynamics changes in response to pressure and temperature. Recall that particle dynamics in the liquid can be separated into solid-like oscillatory and gas-like diffusive components. This separation applies equally to supercritical fluids as it does to subcritical liquids: increasing temperature reduces $\tau$, and each particle spends less time oscillating and more time jumping; increasing pressure reverses this and results in the increase of time spent oscillating relative to jumping. Increasing temperature at constant pressure (or decreasing pressure at constant temperature) eventually results in the disappearance of the solid-like oscillatory motion of particles; all that remains is the diffusive gas-like motion. This disappearance represents the qualitative change in particle dynamics and gives the point on the FL in Figure \ref{frenline}. Notably, the FL exists at arbitrarily high pressure and temperature, as does the melting line.

\begin{figure}
\begin{center}
{\scalebox{0.45}{\includegraphics{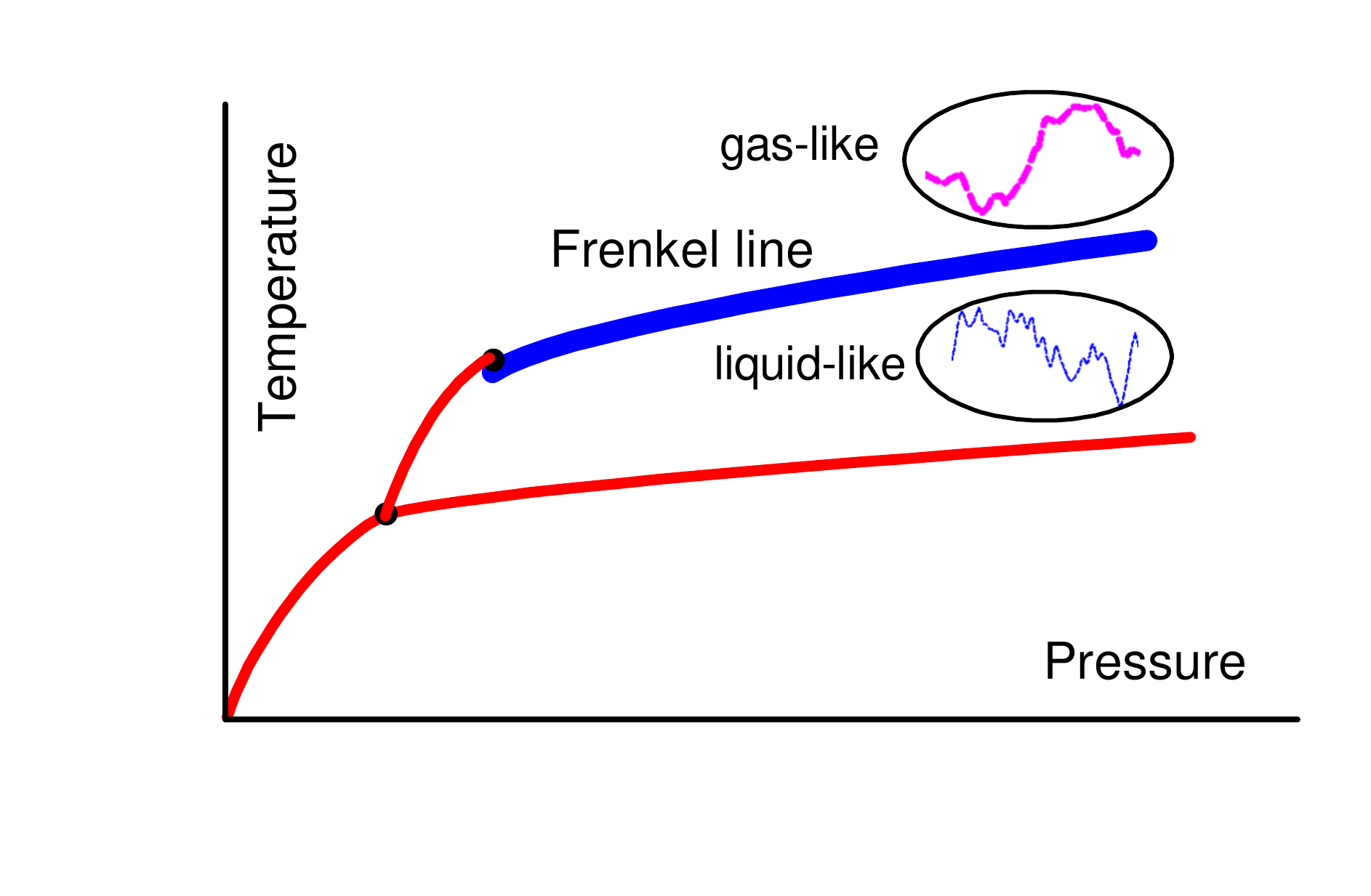}}}
\end{center}
\caption{Colour online. The Frenkel line in the supercritical region. Particle dynamics includes both oscillatory and diffusive components below the line, and is purely diffusive above the line. Below the line, the system is able to support rigidity and transverse modes at high frequency. Above the line, particle motion is purely diffusive, and the ability to support rigidity and transverse modes is lost at all available frequencies. Crossing the Frenkel line from below corresponds to the transition between the ``rigid'' liquid to the ``non-rigid'' gas-like fluid.}
\label{frenline}
\end{figure}

Qualitatively, the FL corresponds to $\tau\rightarrow\tau_{\rm D}$ (here, $\tau_{\rm D}$ refers to the minimal period of transverse modes), implying that particle motion loses its oscillatory component. Quantitatively, the FL can be rigorously defined by pressure and temperature at which the minimum of the velocity autocorrelation function (VAF) disappears \cite{prl}. Above the line defined in such a way, velocities of a large number of particles stop changing their sign and particles lose the oscillatory component of motion. Above the line, VAF is monotonically decaying as in a gas \cite{prl}.

Another criterion for the FL which is important for our discussion of thermodynamic properties and which coincides with the VAF criterion is $c_v=2$ \cite{prl}. Indeed, $\tau=\tau_{\rm D}$ corresponds to the complete loss of two transverse modes at all available frequencies (see Eq. (\ref{tau})). The ability to support transverse waves is associated with solid-like rigidity. Therefore, $\tau=\tau_{\rm D}$ corresponds to the crossover from the ``rigid'' liquid to the ``non-rigid'' gas-like fluid where no transverse modes exist \cite{pre,phystoday,prl,ufn,condmat}, corresponding to the qualitative change of the excitation spectrum.

According to Eq. (\ref{harmo}), $\omega_{\rm F}=\omega_{\rm D}$ or $\tau=\tau_{\rm D}$ gives $c_v=2$. This corresponds to the qualitative change of the excitation spectrum in the liquid, the loss of transverse modes. Therefore, we expect to find an interesting behavior of $c_v$ around $c_v=2$ and its crossover to a new regime. This is indeed the case as discussed in the next section.

Due to the qualitative change of particle dynamics, the FL separates the states with different macroscopic properties, consistent with experimental data \cite{nist}. This includes diffusion constant, viscosity, thermal conductivity, speed of sound and other properties \cite{pre,prl,ufn}. For example, the fast sound discussed earlier disappears above the FL due to the loss of shear resistance at all available frequencies. Depending on the temperature and pressure path on the phase diagram, the crossover of a particular property may not take place on the FL directly but close to it.

We note a different proposal to define a line above the critical point, the Widom line. At the critical point, thermodynamic functions have divergent maxima. Above the critical point, these maxima broaden and persist in the limited range of pressure and temperature. This enables one to define lines of maxima of different properties such as heat capacity, thermal expansion, compressibility and so on. Close to the critical point, system properties can be expressed in terms of the correlation length, the maxima of which is the Widom line \cite{stanley-widom}.

The physical significance of the Widom line was originally attributed to the effect of persisting critical fluctuations on system's dynamical properties \cite{stanley-widom}. Following the detection of PSD above the critical point \cite{benci,disu1}, the Widom line was proposed to separate two supercritical states where PSD does and does not operate \cite{disu2} (see \cite{pre} for the discussion of extrapolating the line to high pressure and temperature where no maxima exist). The states with and without PSD were called ``liquid-like'' and ``gas-like'' because they resemble the presence and absence of PSD in subcritical liquids and gases. The discussion of the effect of the Widom line on thermodynamic, dynamical and transport properties followed (see, e.g., \cite{imre,gallo}).

Persisting critical anomalies and fluctuations related to the Widom line certainly affect system properties close to the critical point. At the same time, the physical origin of the Widom line and the FL is different, as evident from the above discussion. A detailed discussion of this point is outside the scope of this review. Here, we include two brief remarks: (a) the FL is not physically related to the critical point and critical fluctuations and exists in systems where the boiling line and the critical point are absent such as the soft-sphere system \cite{prl}; and (b) the persisting maxima of thermodynamic functions and the Widom line decay around (1.5-2)$T_c$ and strongly depend on the property (e.g. heat capacity, compressibility and so on) and on the path on the phase diagram (i.e. the location of the Widom line depends on whether the property is calculated along isobars, isotherms and so on) \cite{braw1,braw2,braw3,braw4}. This is in contrast to the FL which exists at arbitrarily high temperature and pressure and is property- and path-independent.

\subsection{Heat capacity above the Frenkel line}

A confirmation of the above theoretical proposal that the specific heat undergoes a crossover around $c_v=2$ comes from molecular dynamics simulations in the supercritical state \cite{pre,natcom}. $c_v$ of the model Lennard-Jones liquid is shown in Figure \ref{cv-fr}. We first observe a fairly sharp decrease of $c_v$ from about 3 to 2, similar to the previously discussed behavior in Figure \ref{cv3}. This is followed by the flattening and slower decrease at higher temperature. The crossover takes place at around $c_v=2$ as predicted.

\begin{figure}
\begin{center}
{\scalebox{0.35}{\includegraphics{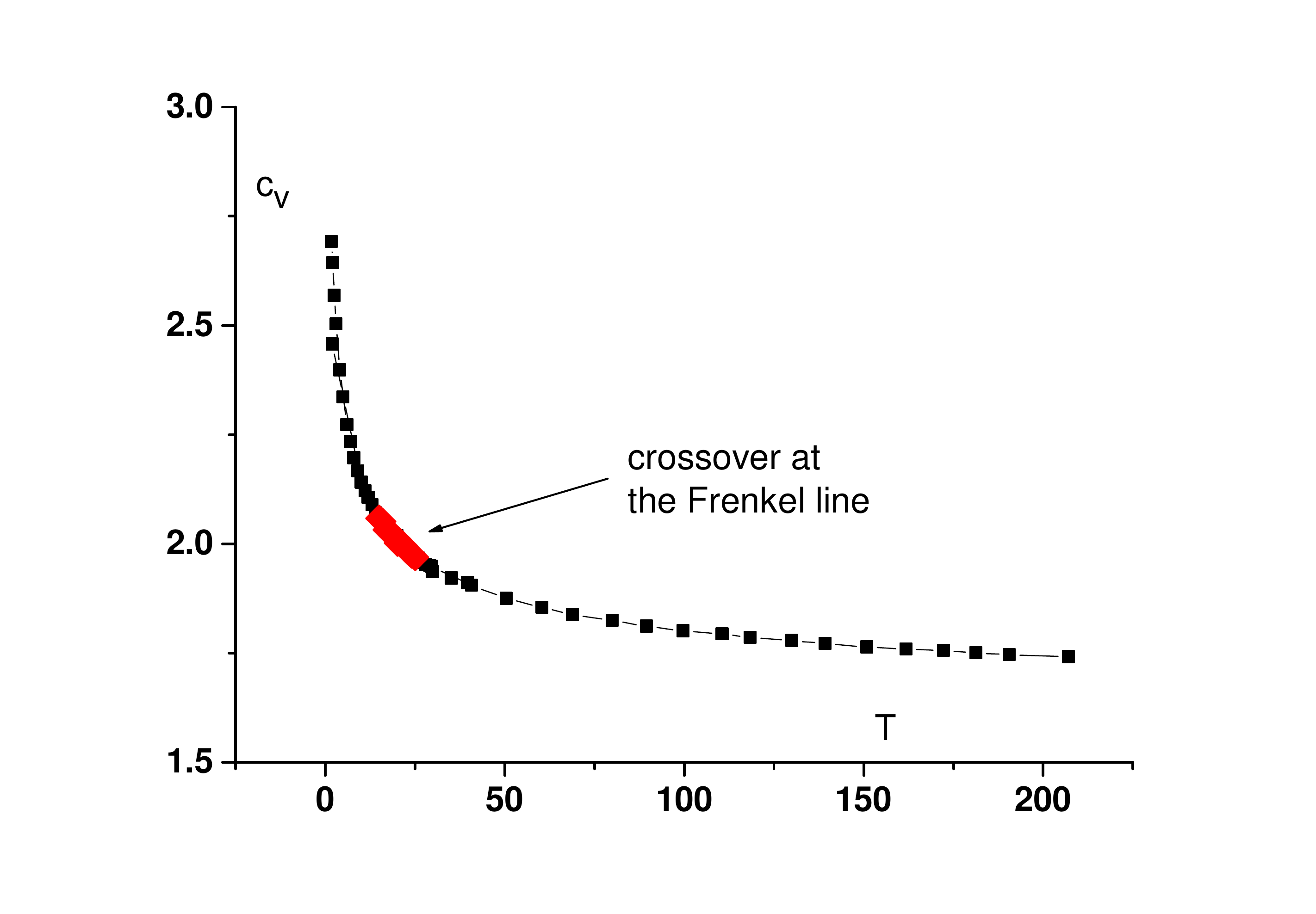}}}
\end{center}
\caption{Colour online. $c_v$ ($k_{\rm B}=1$) as a function of temperature from the molecular dynamics simulation of the Lennard-Jones (LJ) liquid using the data from Ref. \cite{pre}. Temperature is in LJ units. Density is $\rho=1$ in LJ units. The region of dynamical crossover at $c_v=2$ is highlighted in red and by the arrow.}
\label{cv-fr}
\end{figure}

Understanding the slower decrease of $c_v$ above the FL involves the discussion of how the remaining longitudinal mode evolves with temperature (recall that two transverse modes disappear at the FL). When the FL is crossed from below, particles lose the oscillatory motion around their quasi-equilibrium positions, and start undergoing purely diffusive jumps with distances comparable with the interatomic distance $a$. Further increase of particle energy at higher temperature increases the mean free path of particles $L$, the average distance which the particles travel before colliding. $L$ sets the minimal wavelength of the remaining longitudinal mode, $\lambda_L$: indeed, oscillation wavelength can only be larger than $L$. Therefore, the propagating longitudinal mode above the FL has the wavelengths satisfying

\begin{equation}
\lambda>L
\label{longl}
\end{equation}

We observe that the oscillations of the longitudinal mode in (\ref{longl}) disappear with temperature starting with the highest frequency (smallest wavelength) above the FL, in interesting contrast to the evolution of transverse modes in (\ref{tau}) where transverse modes disappear starting with the smallest frequency. The difference of temperature evolution of collective modes below and above the FL is responsible for the crossover of $c_v$ at the FL discussed below.

The energy of the above longitudinal mode, $E_l$, can be calculated using Eq. (\ref{phonen}) as

\begin{equation}
E_l=\int\limits_0^{\omega_L}E(\omega,T)g(\omega)d\omega
\end{equation}

\noindent where $\omega_L=\frac{2\pi}{\lambda_L}c=\frac{2\pi}{L}c$ is the minimal frequency.

Taking $g(\omega)=\frac{3N}{\omega_{\rm D}^3}\omega^2$ as before and $E(\omega,T)=T$ in the classical case gives $E_l=NT\left(\frac{\omega_L}{\omega_{\rm D}}\right)^3$, or $NT\left(\frac{a}{L}\right)^3$. The total energy of the system is the sum of the kinetic energy, $\frac{3}{2}NT$, and potential energy. Using the equipartition theorem, the potential energy can be written as $\frac{E_l}{2}$. This gives the total energy of the non-rigid gas-like fluid above the FL as

\begin{equation}
E=\frac{3}{2}NT+\frac{1}{2}NT\left(\frac{a}{L}\right)^3
\label{supe1}
\end{equation}

Just above the FL, $L\approx a$. According to Eq. (\ref{supe1}), this gives $c_v=2$, the result that also follows from the Equation (\ref{harmo}) describing the rigid liquid. When $L\gg a$ at high temperature, Eq. (\ref{supe1}) gives $c_v=\frac{3}{2}$ as expected.

The crossover of $c_v$ seen in Figure {\ref{cv-fr}} is therefore attributed to two different mechanisms governing the decrease of $c_v$. Below the FL, $c_v$ decreases from the solid value of 3 to 2 due to the progressive disappearance of two transverse modes with frequency $\omega>\omega_{\rm F}$. Above the FL, $c_v$ decreases from $2$ to the ideal-gas value of $\frac{3}{2}$ due to the disappearance of the remaining longitudinal mode starting with the shortest wavelength governed by $L$. Remaining long-wavelength longitudinal oscillations, sound, make only small contribution to the system energy and heat capacity.

The softening of the phonon frequencies with temperature can be accounted for in the same way as in the case of subcritical fluids above (see Eqs. (\ref{anharm}, \ref{anharm1})), giving \cite{natcom}:

\begin{equation}
E=\frac{3}{2}NT+\frac{1}{2}NT\left(1+\frac{\alpha T}{2}\right)\left(\frac{a}{L}\right)^3
\label{supe2}
\end{equation}

The actual decrease of $c_v$ between $c_v=2$ and $c_v=\frac{3}{2}$ can be calculated if temperature dependence of $L$ is known. This dependence can be taken from the independent measurement of the gas-like viscosity of the supercritical fluid:

\begin{equation}
\eta=\frac{1}{3}\rho\bar{u}L
\label{suvisc}
\end{equation}

\noindent where $\bar{u}$ is the average velocity defined by temperature.

Taking $\eta$ from the experiment, calculating $L$ using (\ref{suvisc}) and using it in Eq. (\ref{supe1}) or Eq. (\ref{supe2}) enables us to calculate $E$ and $c_v$. This gives good agreement with the experimental $c_v$ for several supercritical systems, including noble and molecular fluids \cite{natcom}. In these systems, $c_v$ slowly decreases with temperature as is seen in Figure \ref{cv-fr} in the high-temperature range.

We note that in our discussion of liquid thermodynamics throughout this paper, we assumed that the mode energy is $T$ (in the classical case). This applies to harmonic waves. In weakly-anharmonic cases, the anharmonicity can be accounted for in the Gr\"{u}neisen approximation (see Eq. \ref{anharm2} and related discussion). If the anharmonicity is strong, the mode energy can substantially differ from $T$. This can include the case of very high temperature or inherently anharmonic systems such as the hard-spheres system as an extreme example where heat capacity is equal to the ideal-gas value.

\section{Heat capacity of liquids and system's fundamental length}

The behavior of liquid $c_v$ in its entire range from the solid value, $c_v=3$, to the ideal-gas value, $c_v=\frac{3}{2}$, can be unified and generalized in terms of wavelengths.

Lets consider $c_v$ in the rigid liquid state, Eq. (\ref{harmo}) and in the non-rigid gas-like fluid, (\ref{supe1}). We do not consider anharmonic effects related to phonon softening: these give small corrections ($\alpha T\ll 1$) to the energy in Eqs. (\ref{anharm2}), (\ref{supe2}). An interesting insight comes from combining Eqs. (\ref{harmo}) and (\ref{supe1}) and interpreting both of them in terms of wavelengths  \cite{jphyschem} (see Figure \ref{length}).

\begin{figure}
\begin{center}
{\scalebox{0.35}{\includegraphics{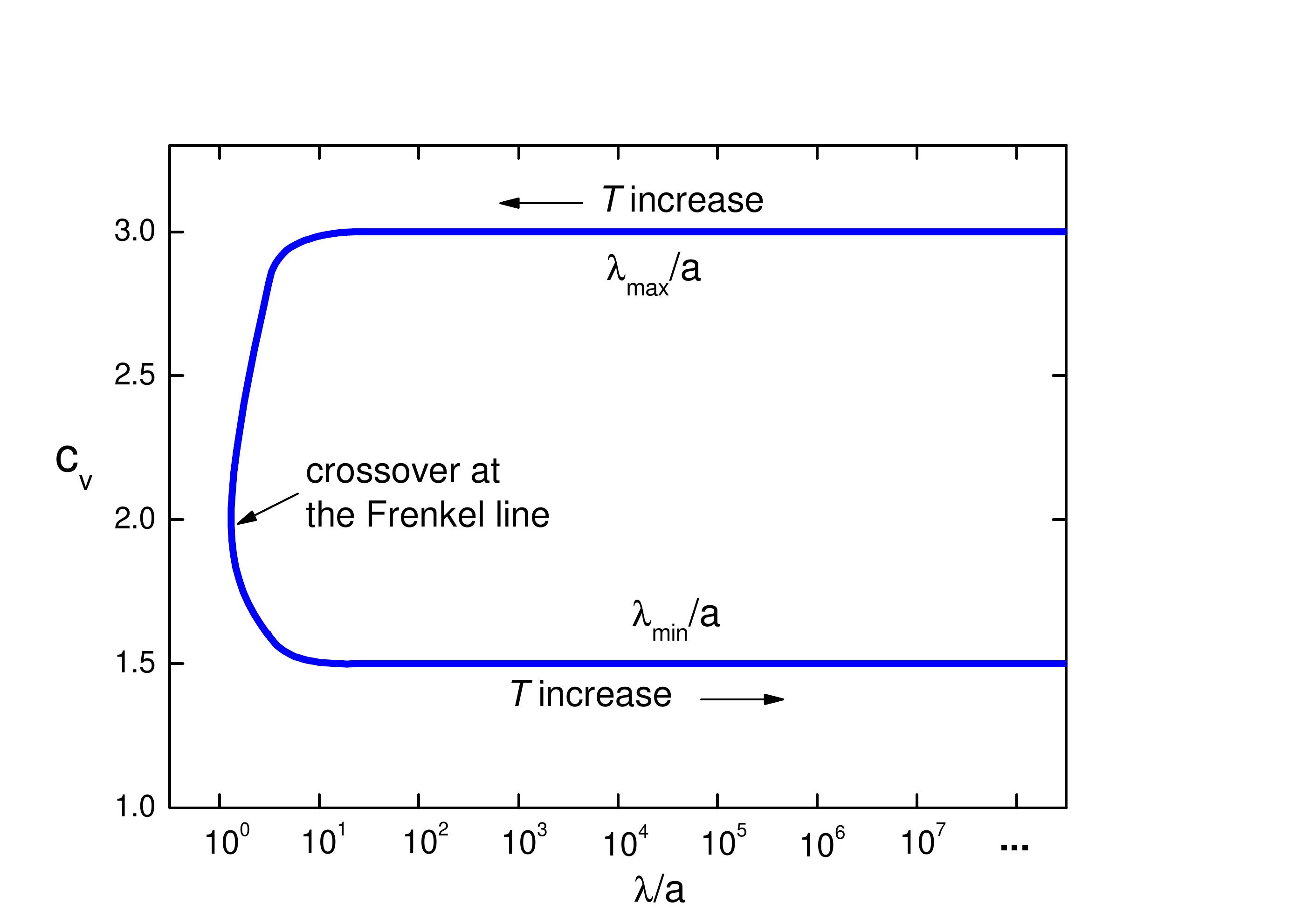}}}
\end{center}
\caption{$c_v$ as a function of the characteristic wavelengths $\lambda_{\rm max}$ (maximal transverse wavelength in the system) and $\lambda_{\rm min}$ (minimal longitudinal wavelength in the system) illustrating that most important changes of thermodynamics of the disordered system take place when both wavelengths become comparable to the fundamental length $a$.}
\label{length}
\end{figure}

The minimal frequency of transverse modes that a liquid supports, $\omega_{\rm F}$, corresponds to the maximal transverse wavelength, $\lambda_{\rm max}$, $\lambda_{\rm max}=a\frac{\omega_{\rm D}}{\omega_{\rm F}}=a\frac{\tau}{\tau_{\rm D}}$, where $a$ is the interatomic separation, $a\approx 1-2$ \AA. According to Eq. (\ref{harmo}), $c_v$ remains close to its solid-state value of 3 in almost entire range of available wavelengths of transverse modes until $\omega_{\rm F}$ starts to approach $\omega_{\rm D}$, including in the viscous regime discussed below, or when $\lambda_{\rm max}$ starts to approach $a$. When $\lambda_{\rm max}=a$, $c_v$ becomes $c_v=2$ according to Eq. (\ref{harmo}) and undergoes a crossover to another regime given by Eq. (\ref{supe1}). In this regime, the minimal wavelength of the longitudinal mode supported by the system is $\lambda_{\rm min}=L$. According to Eq. (\ref{supe1}), $c_v$ remains close to the ideal gas value of $\frac{3}{2}$ in almost entire range of the wavelengths of the longitudinal mode until $\lambda_{\rm min}$ approaches $a$. When $\lambda_{\rm max}=a$, $c_v$ becomes $c_v=2$, and matches its low-temperature value at the crossover as schematically shown in Figure \ref{length}.

Consistent with the above discussion, Figure \ref{length} shows that $c_v$ remains constant at either $3$ or $\frac{3}{2}$ over many orders of magnitude of $\frac{\lambda}{a}$, including the regime of viscous liquids and glasses discussed below, except when $\frac{\lambda}{a}$ becomes close to 1 by order of magnitude.

Figure \ref{length} emphasizes a transparent physical point: modes with the smallest wavelengths comparable to interatomic separations $a$ contribute most to the energy and $c_v$ in the disordered systems (as they do in crystals) because they are most numerous. Consequently, conditions $\lambda_{\rm max}\approx a$ for two transverse modes and $\lambda_{\rm min}\approx a$ for one longitudinal mode, corresponding to the disappearance of modes with wavelengths comparable to $a$, give the largest changes of $c_v$ as is seen in Figure \ref{length}.

The last result is tantamount to the following general assertion: the most important changes in thermodynamics of the disordered system are governed by its {\it fundamental length} $a$ only. Because this length is not affected by disorder, this assertion holds equally in ordered and disordered systems.

Interestingly, the above assertion does not follow from the hydrodynamic approach to liquids. The hydrodynamic approach works well at large wavelengths, but may not correctly describe effects at length scales comparable to $a$. Yet, as we have seen, this scale which plays an important role in governing system's thermodynamic properties.

\section{Evolution of collective modes in liquids: summary}

We can now summarize the above discussion of how collective modes change in liquids with temperature. This is illustrated in Figure \ref{ricky}.

\begin{figure*}
\begin{center}
{\scalebox{0.6}{\includegraphics{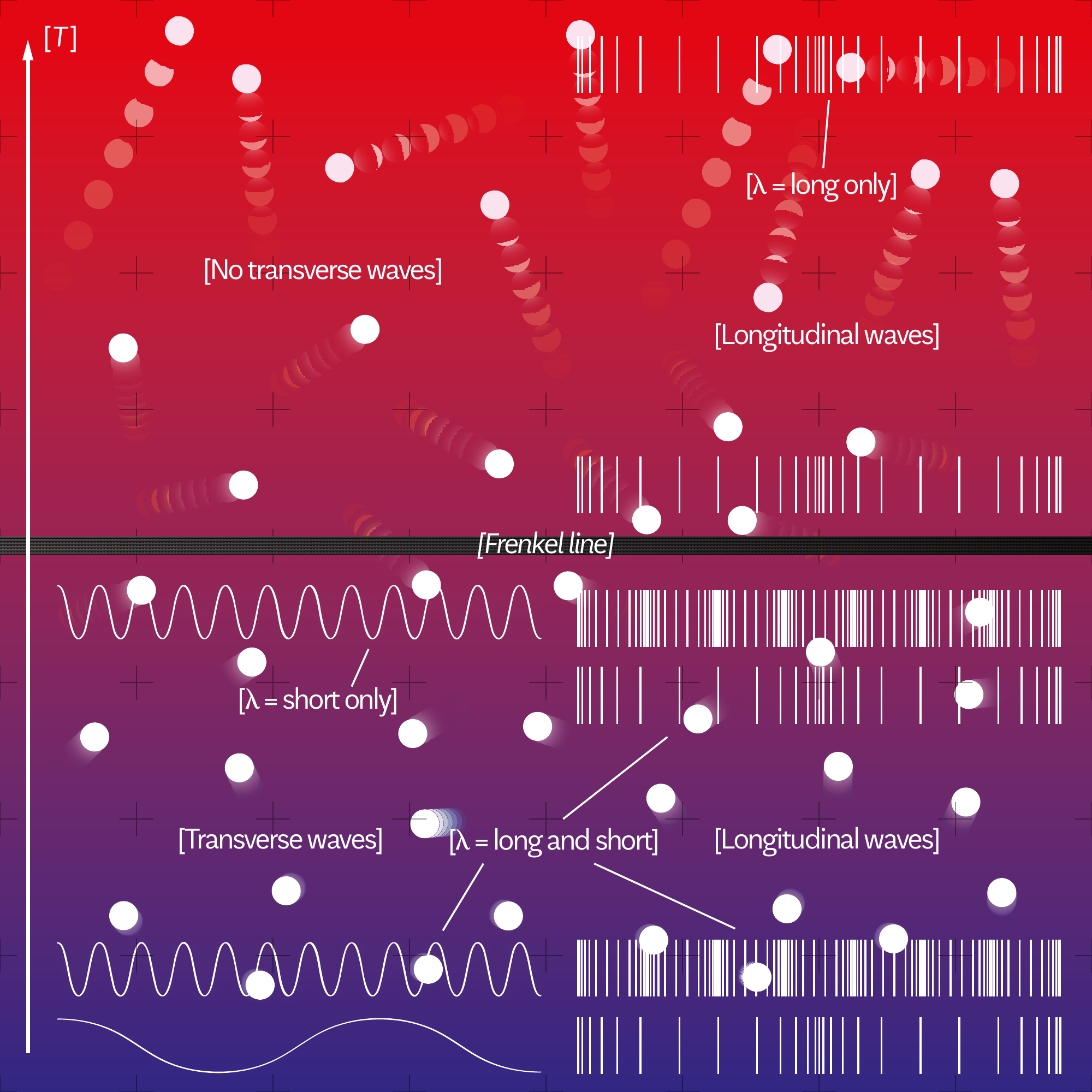}}}
\end{center}
\caption{Colour online. Evolution of transverse and longitudinal waves in disordered matter, from viscous liquids and glasses at low temperature to gases at high. The variation of colour from deep blue at the bottom to light red at the top corresponds to temperature increase. The Figure shows that the transverse waves (left) start disappearing with temperature starting with long wavelength modes and completely disappear at the Frenkel line. The longitudinal waves (right) do not change up to the Frenkel line but start disappearing above the line starting with the shortest wavelength, with only long-wavelength longitudinal modes propagating at high temperature.}
\label{ricky}
\end{figure*}

Figure \ref{ricky} illustrates that at low temperature, liquids have the same set of collective modes as in solids: one longitudinal mode and two transverse modes. In the viscous regime at low enough temperature where $\tau\gg\tau_{\rm D}$ or $\omega_{\rm F}\ll\omega_{\rm D}$, the liquid energy is almost entirely given by the vibrational energy due to these modes, as discussed in the next section. On temperature increase, the number of transverse modes propagating above the frequency $\omega_{\rm F}$ decreases. At the FL where particles lose the oscillatory component of motion and start moving diffusively as in a gas, the two transverse modes disappear. This picture is consistent with the results of molecular dynamics simulations where transverse modes are directly calculated from the transverse current correlation functions \cite{condmat}.

Above the FL, the collective mode is the remaining longitudinal mode with the wavelength larger than $L$, and its energy progressively decreases with temperature until it becomes close to the ideal gas.

The evolution of collective modes and related changes of liquid energy and heat capacity are intimately related to the change of microscopic dynamics of particles and the relative weights of diffusive and oscillatory components. We will return to this point below when we discuss the mixed state of liquid dynamics as contrasted to pure dynamical states of solids and gases.

\section{Viscous liquids}

In this section, we discuss how energy and heat capacity of viscous liquids can be understood on the basis of collective modes. ``Viscous'' or ``highly-viscous'' liquids are loosely defined as liquids where

\begin{equation}
\tau\gg\tau_{\rm D}
\label{3}
\end{equation}

More generally, viscous liquids are discussed as systems that avoid crystallization and enter the glass transformation range. When $\tau$ exceeds the experimental time scale of $10^2-10^3$ s and particle jumps stop operating during the observation time (in the field of glass transition, particle jumps are often referred to as ``alpha-relaxation''), the system forms glass. This defines glass transition temperature as $\tau(T_g)=10^2-10^3$ s \cite{dyre}.

Properties of viscous liquids have been widely discussed due to the interest in the problem of liquid-glass transition, the problem which consists of several unusual effects and includes persisting controversies (see, e.g., \cite{dyre,angell,ngai,phillips,gl4,gl5,gl6,gl7,gl8}). Understanding viscous liquids above $T_g$ is thought to facilitate explaining effects involved in the actual liquid-glass transition at $T_g$ and possibly effects below $T_g$ \cite{dyre,angell,ngai,gl4,gl6,gl7,gl8}.

We find that in some respects, the glass transition problem is more controversial that it needs to be. This is partly because the controversies emerged before good-quality experimental data became available. For example, the crossover from the VFT to the Arrhenius (or nearly Arrhenius) behavior at low temperature \cite{cro1,cro2,cro3,cro4,cro5} removes the basis for discussing possible divergence and associated ideal glass transition at the VFT temperature $T_0$, as discussed above.

\subsection{Energy and heat capacity}

Perhaps unexpectedly, understanding basic thermodynamic properties of viscous liquids such as energy and heat capacity is easier than of low-viscous liquids. It does not involve expanding the energy into the oscillatory and diffusive parts as in Eq. (\ref{en0}) or integrating over the operating phonon states as in Eqs. (\ref{phonen}-\ref{part2}). The main results can be obtained on the basis of one parameter only, $\frac{\tau_{\rm D}}{\tau}$ (or $\frac{\omega_{\rm F}}{\omega_{\rm D}}$) using a simple yet rigorous statistical-mechanical argument \cite{scirep2}.

The jump probability for a particle is the ratio between the time spent diffusing and oscillating. The jump event lasts on the order of Debye vibration period $\tau_{\rm D}\approx 0.1$ ps. Recall that $\tau$ is the time between two consecutive particle jumps, and therefore is the time that the particle spends oscillating. Therefore, the jump probability is $\frac{\tau_{\rm D}}{\tau}$. In statistical equilibrium, this probability is equal to the ratio of diffusing atoms, $N_{\rm dif}$, and the total number of atoms, $N$. Then, at any given moment of time:

\begin{equation}
\frac{N_{\rm dif}}{N}=\frac{\tau_{\rm D}}{\tau}
\label{1}
\end{equation}

If $E_{\rm dif}$ is the energy associated with diffusing particles, $E_{\rm dif}\propto N_{\rm dif}$. Together with $E_{\rm tot}\propto N$, Eq. ({\ref{1}) gives

\begin{equation}
\frac{E_{\rm dif}}{E_{\rm tot}}=\frac{\tau_{\rm D}}{\tau}
\label{2}
\end{equation}

Eq. (\ref{2}) implies that under condition (\ref{3}), the contribution of $E_{\rm dif}$ to the total energy at any moment of time is negligible.

We note that Eq. (\ref{2}) corresponds to the instantaneous value of $E_{\rm dif}$ which, from the physical point of view, is given by the smallest time scale of the system, $\tau_{\rm D}$. During time $\tau_{\rm D}$, the system is not in equilibrium. The equilibrium state is reached when the observation time exceeds system relaxation time, $\tau$. After time $\tau$, all particles in the system undergo jumps. Therefore, we need to calculate $E_{\rm dif}$ that is averaged over time $\tau$.

Let us divide time $\tau$ into $m$ time periods of duration $\tau_{\rm D}$ each, so that $m=\frac{\tau}{\tau_{\rm D}}$. Then, $E_{\rm dif}$, averaged over time $\tau$, $E^{\rm av}_{\rm dif}$, is

\begin{equation}
E^{\rm av}_{\rm dif}=\frac{E^1_{\rm dif}+E^2_{\rm dif}+...+E^m_{\rm dif}}{m}
\label{ave}
\end{equation}

\noindent where $E^i_{\rm dif}$ are instantaneous values of $E_{\rm dif}$ featured in Eq. (\ref{2}). $\frac{E^{\rm av}_{\rm dif}}{E_{\rm tot}}$ is

\begin{equation}
\frac{E^{\rm av}_{\rm dif}}{E_{\rm tot}}=\frac{E^1_{\rm dif}+E^2_{\rm dif}+...+E^m_{\rm dif}}{{E_{\rm tot}}\cdot m}
\label{ave01}
\end{equation}

Each of the terms $\frac{E^i_{\rm dif}}{E_{\rm tot}}$ in Eq. (\ref{ave01}) is equal to $\frac{\tau_{\rm D}}{\tau}$, according to Eq. (\ref{2}). There are $m$ terms in the sum in Eq. (\ref{ave01}). Therefore,

\begin{equation}
\frac{E^{\rm av}_{\rm dif}}{E_{\rm tot}}=\frac{\tau_{\rm D}}{\tau}
\label{ave1}
\end{equation}

We therefore find that under the condition (\ref{3}), the ratio of the average energy of diffusion motion to the total energy is negligibly small, as in the instantaneous case. Consequently, the energy of the liquid under the condition (\ref{3}) is, to a very good approximation, given by the remaining vibrational part. Similarly, the liquid constant-volume specific heat, $c_{v,\rm {l}}=\frac{1}{N}\frac{{\rm d}E_{\rm l}}{{\rm d}T}$ is entirely vibrational in the regime (\ref{3}):

\begin{equation}
\begin{aligned}
&E_{\rm l}=E_{\rm l}^{\rm vib}\\
&c_{v,\rm {l}}=c_{v,\rm {l}}^{\rm vib}
\end{aligned}
\label{energy}
\end{equation}

The vibrational energy and specific heat of liquids in the regime (\ref{3}) is readily found. When regime (\ref{3}) is operative, $E_{\rm l}^{\rm vib}$ to a very good approximation is $E_{\rm l}^{\rm vib}=3NT$ (here and below, $k_{\rm B}=1$). Indeed, a solid supports one longitudinal mode and two transverse waves in the range $0<\omega<\frac{1}{\tau_{\rm D}}$. The ability of liquids to support shear modes with frequency $\omega>\frac{1}{\tau}$, combined with $\tau\gg\tau_{\rm D}$ in Eq. (\ref{3}), implies that a viscous liquid supports most of the shear modes present in a solid. Furthermore and importantly, it is only the high-frequency shear modes that make a significant contribution to the liquid vibrational energy, because the vibrational density of states is approximately proportional to $\omega^2$. Hence in the regime (\ref{3}), $E_{\rm l}^{\rm vib}=3NT$ to a very good approximation, as in a solid.

We now now consider Eqs. (\ref{energy}) in harmonic and anharmonic cases. In the harmonic case, Eqs. (\ref{energy}) give the energy and specific heat of a liquid as $3NT$ and $3$, respectively, i.e. the same as in a harmonic solid:

\begin{equation}
\begin{aligned}
&E_{\rm l}^{\rm h}=E_{\rm s}^{\rm h}=3NT\\
&c_{v,\rm {l}}^{\rm h}=c_{v,\rm {s}}^{\rm h}=3
\label{henergy}
\end{aligned}
\end{equation}

\noindent where ${\rm s}$ corresponds to the solid and ${\rm h}$ to the harmonic case.

In the anharmonic case, Eqs. (\ref{energy}) are modified by the intrinsic anharmonicity related to softening of phonon frequencies, and become Eqs. (\ref{enesol}) and (\ref{sol}) as discussed above.

Three pieces of evidence support the above picture. First, experimental specific heat of liquid metals at low temperature is close to 3, consistent with the above predictions \cite{grimvall,wallace}. As experimental techniques advanced and gave access to high pressure and temperature, specific heats of many noble, molecular and network liquids were measured in a wide range of parameters including in the supercritical region \cite{nist}. Similarly to liquid metals, the experimental $c_v$ of these liquids was found to be close to 3 at low temperature where Eq. (1) applies (see Ref. \cite{scirep} for a compilation of the NIST and other data of $c_v$ for over 20 liquids of different types).

Second, condition $\tau\gg\tau_{\rm D}$ becomes particularly pronounced in viscous liquids approaching liquid-glass transition where
$\frac{\tau_{\rm D}}{\tau}$ becomes as small as $\frac{\tau_{\rm D}}{\tau}\approx 10^{-15}$. Experiments have shown that in the highly viscous regime just above $T_g$, $C_p$ measured at high frequency and representing the vibrational part of heat capacity coincides with the total low-frequency heat capacity usually measured \cite{johari1,johari2}, consistent with Eq. (\ref{energy}). In the glass transformation range close to $T_g$, the two heat capacities start to differ due to non-equilibrium effects and freezing of configurational entropy, and coincide again below $T_g$ in the solid glass.

Third, representing $c_v$ by its vibrational term in the highly viscous regime above $T_g$ gives the experimentally observed change of heat capacity in viscous liquids above $T_g$ as compared to glasses below $T_g$. This is discussed in the next section.

We recall that the only condition used to make the above assertions is Eq. (\ref{3}). For practical purposes, this condition is satisfied for $\tau\gtrsim 10\tau_{\rm D}$. Perhaps not widely recognized, the condition $\tau\approx 10\tau_{\rm D}$ holds even for low-viscous liquids such as liquid metals (Hg, Na, Rb and so on) and noble liquids such as Ar near their melting points, let alone for more viscous liquids such as room-temperature olive or motor oil, honey and so on.

Notably, the condition $\tau\gtrsim 10\tau_{\rm D}$ corresponds to almost the entire range of $\tau$ at which liquids exist. This fact was not fully appreciated in earlier theoretical work on liquids. Indeed, on lowering the temperature, $\tau$ increases from its smallest limiting value of $\tau=\tau_{\rm D}\approx 0.1$ ps to $\tau\approx 10^{3}$ s where, by definition, the liquid forms glass at the glass transition temperature $T_g$. Here, $\tau$ changes by 16 orders of magnitude. Consequently, the condition $\frac{\tau_{\rm D}}{\tau}\ll 1$, Eq. (\ref{3}), or $\tau\gtrsim 10\tau_{\rm D}$, applies in the range $10^3-10^{-12}$ s, spanning 15 orders of magnitude of $\tau$. This constitutes almost entire range of $\tau$ where liquids exist as such.

\subsection{Entropy}

Although Eq. (\ref{ave1}), combined with Eq. (\ref{3}), implies that the energy and $c_v$ of a liquid are entirely vibrational as in a solid, this does {\it not} apply to entropy: the diffusional component to entropy is substantial, and can not be neglected \cite{scirep2}.

Indeed, if $Z_{\rm vib}$ and $Z_{\rm dif}$ are the contributions to the partition sum from vibrations and diffusion, respectively, the total partition sum of the liquid is $Z=Z_{\rm vib}\cdot Z_{\rm dif}$. Then, the liquid energy is $E=T^2\frac{\rm d}{{\rm d}T}\left(\ln(Z_{\rm vib}\cdot Z_{\rm dif})\right)=T^2\frac{\rm d}{{\rm d}T}\ln Z_{\rm vib}+T^2\frac{\rm d}{{\rm d}T}\ln Z_{\rm dif}=E_{\rm vib}+E_{\rm dif}$ (here and below, the derivatives are taken at constant volume). Next, $\frac{E^{\rm av}_{\rm dif}}{E_{\rm tot}}\ll 1$ from Eq. (\ref{ave1}) also implies $\frac{E_{\rm dif}}{E_{\rm vib}}\ll 1$, where, for brevity, we dropped the subscript referring to the average. Therefore, the smallness of diffusional energy, $\frac{E_{\rm dif}}{E_{\rm vib}}\ll 1$, gives

\begin{equation}
\frac{\frac{\rm d}{{\rm d}T}\ln Z_{\rm dif}}{\frac{\rm d}{{\rm d}T}\ln Z_{\rm vib}}\ll 1
\label{ent1}
\end{equation}

The liquid entropy, $S=\frac{\rm d}{{\rm d}T}\left(T\ln(Z_{\rm vib}\cdot Z_{\rm dif})\right)$, is:

\begin{equation}
S=T\frac{\rm d}{{\rm d}T}\ln Z_{\rm vib}+\ln Z_{\rm vib}+T\frac{\rm d}{{\rm d}T}\ln Z_{\rm dif}+\ln Z_{\rm dif}
\label{ent2}
\end{equation}

The condition (\ref{ent1}) implies that the third term in Eq. ({\ref{ent2}) is much smaller than the first one, and can be neglected, giving

\begin{equation}
S=T\frac{\rm d}{{\rm d}T}\ln Z_{\rm vib}+\ln Z_{\rm vib}+\ln Z_{\rm dif}
\label{ent3}
\end{equation}

Eq. (\ref{ent3}) implies that the smallness of $E_{\rm dif}$, expressed by Eq. (\ref{ent1}), does {\it not} lead to the disappearance of all entropy terms that depend on diffusion because the term $\ln Z_{\rm dif}$ remains. This term is responsible for the excess entropy of liquid over the solid. On the other hand, the smallness of $E_{\rm dif}$ {\it does} lead to the disappearance of terms depending on $Z_{\rm dif}$ in the specific heat. Indeed, $c_{v,\rm {l}}=T\frac{{\rm d}S}{{\rm d}T}$ (here, $S$ refers to entropy per atom or molecule), and from Eq. (\ref{ent3}), we find:

\begin{equation}
c_{v,\rm {l}}=T\frac{\rm d}{{\rm d}T}\left(T\frac{\rm d}{{\rm d}T}\ln Z_{\rm vib}\right)+T\frac{\rm d}{{\rm d}T}\ln Z_{\rm vib}+T\frac{\rm d}{{\rm d}T}\ln Z_{\rm dif}
\label{ent4}
\end{equation}

Using Eq. (\ref{ent1}) once again, we observe that the third term in Eq. (\ref{ent4}) is small compared to the second term, and can be neglected, giving

\begin{equation}
c_{v,{\rm l}}=T\frac{\rm d}{{\rm d}T}\left(T\frac{\rm d}{{\rm d}T}\ln Z_{\rm vib}\right)+T\frac{\rm d}{{\rm d}T}\ln Z_{\rm vib}
\label{ent5}
\end{equation}
\noindent

As a result, $c_v$ does not depend on $Z_{\rm dif}$, and is given by the vibrational term that depends on $Z_{\rm vib}$ only. As expected, Eq. (\ref{ent5}) is consistent with $c_v$ in Eq. (\ref{henergy}).

Physically, the inequality of liquid and solid entropies, $S_{\rm l}\ne S_{\rm s}$, is related to the fact that the entropy measures the total phase space available to the system, which is larger in the liquid due to the diffusional component present in Eq. (\ref{ent3}). However, the diffusional component, $\ln Z_{\rm dif}$, although large, is slowly varying with temperature according to Eq. (\ref{ent1}), resulting in a small contribution to $c_v$ (see Eqs. (\ref{ent4}) and (\ref{ent5})). On the other hand, the energy corresponds to the instantaneous state of the system (or averaged over $\tau$), and is not related to exploring the phase space. Consequently, $E_{\rm l}=E_{\rm vib}$, yielding Eq. (\ref{ent1}) and the smallness of diffusional contribution to $c_v$ despite $S_{\rm l}\ne S_{\rm s}$.

We note that the common thermodynamic description of entropy does not involve time: it is assumed that the observation time is long enough for the total phase space to be explored. In a viscous liquid with large $\tau$, this exploration is due to particle jumps, and is complete at long times $t\gg\tau$ only, at which point the system becomes ergodic.

If extrapolated below $T_g$, configurational entropy of viscous liquids reaches zero at a finite temperature, constituting an apparent problem known as the widely discussed Kauzmann paradox. More recently, issues involved in separating configurational and vibrational entropy and interpreting experimental data became apparent, affecting the way the Kauzmann paradox is viewed and extent of the problem (see, e.g., Refs. \cite{dyre}, \cite{mckenna} and references therein).

\section{Liquid-glass transition}

In the previous section, we have ascertained that in the viscous regime $\tau\gg\tau_{\rm D}$, liquid energy and heat capacity are essentially given by the vibrational terms. What happens to heat capacity when temperature drops below the glass transition temperature $T_g$ and we are dealing with the solid glass, the non-equilibrium system where $\tau$ exceeds observation time?

Figure \ref{cpfig} gives a typical example of the change of the constant-pressure specific heat, $c_p$, in the glass transformation range around $T_g$. If $c_p^{\rm l}$ and $c_p^{\rm g}$ correspond to the specific heat above and below $T_g$ on both sides of the glass transformation range, $\frac{c_p^{\rm l}}{c_p^{\rm g}}=1.1-1.8$ for various liquids \cite{gl7,wang1}. The change of $c_p$ at $T_g$ is considered as the ``thermodynamic'' signature of the liquid-glass transition, and serves to define $T_g$ in the calorimetry experiments. $T_g$ measured as the temperature of the change of $c_p$ coincides with the temperature at which $\tau$ reaches $10^2-10^3$ s and exceeds the observation time.

\begin{figure}
\begin{center}
{\scalebox{0.4}{\includegraphics{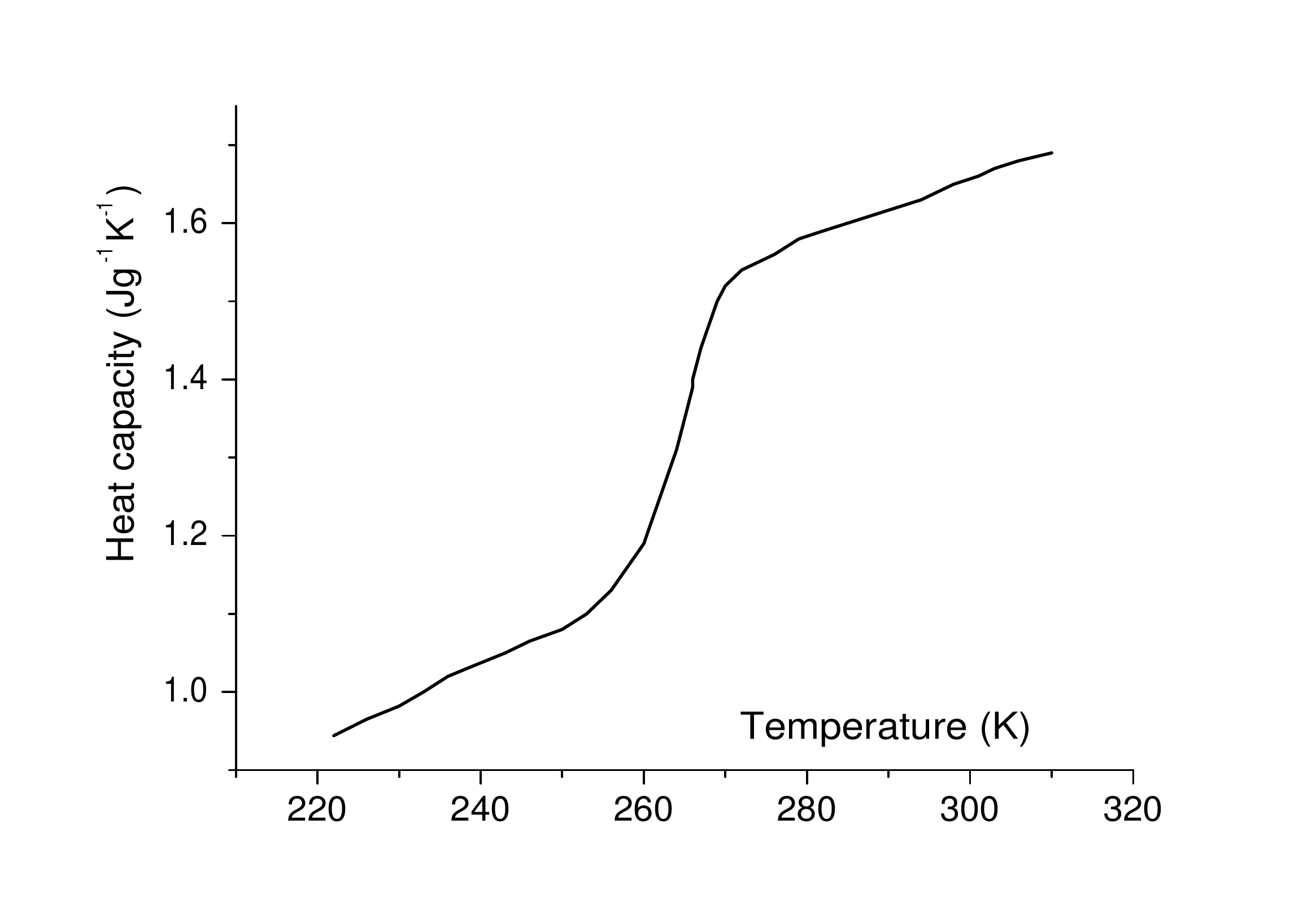}}}
\end{center}
\caption{Heat capacity of Poly(a-methyl styrene) measured in calorimetry experiments \cite{mckenna}. Glass transformation range operates in the interval of about $260-270$ K.}
\label{cpfig}
\end{figure}

Most researchers do not consider the change of $c_p$ as a signature of the phase transition. This is supported by the numerous data testifying that the structure of the viscous liquid above $T_g$ and the structure of glass are nearly identical. What causes the change of $c_p$ at $T_g$?

Recall that liquid response includes viscous response related to diffusive jumps and solid-like response. When the viscous response stops at $T_g$ during the experimental time scale (from the definition of $T_g$) and only the elastic response remains, system's bulk modulus $B$ and thermal expansion coefficient $\alpha$ change. This results in different $c_p$ above and below $T_g$ \cite{gla-tr}.

Lets consider that pressure $P$ is applied to a liquid. According to the Maxwell-Frenkel viscoelastic picture, the change of liquid volume, $v$, is $v=v_{\rm el}+v_{\rm r}$, where $v_{\rm el}$ and $v_{\rm r}$ are associated with solid-like elastic deformation and viscous relaxation process. Lets now define $T_g$ as the temperature at which $\tau$ exceeds the observation time $t$. This implies that particle jumps are not operative at $T_g$ during the time of observation. Therefore, $v$ at $T_g$ is given by purely elastic response as in elastic solid. Then, we write

\begin{equation}
\begin{aligned}
&P=B_{\rm l}\frac{v_{\rm el}+v_{\rm r}}{V_{\rm l}^0} \\
&P=B_{\rm g}\frac{v_{\rm g}}{V_{\rm g}^0}
\end{aligned}
\label{moduli}
\end{equation}

\noindent where $V_{\rm l}^0$ and $V_{\rm g}^0$ are initial volumes of the liquid and the glass, $v_{\rm g}$ is the elastic deformation of the glass and $B_{\rm l}$ and $B_{\rm g}$ are bulk moduli of the liquid and glass, respectively.

Let $\Delta T$ be a small temperature interval that separates the liquid from the glass such that $\tau$ in the liquid, $\tau_{\rm l}$, is
$\tau_{\rm l}=\tau(T_{\rm g}+\Delta T)$ and $\frac{\Delta T}{T_g}\ll 1$. Then, $V_{\rm l}^0\approx V_{\rm g}^0$. Similarly, the difference between the elastic response of the liquid and the glass can be ignored for small $\Delta T$, giving $v_{\rm el}\approx v_{\rm g}$. Combining the two expressions in (\ref{moduli}), we find:

\begin{equation}
B_{\rm l}=\frac{B_{\rm g}}{\epsilon_1+1}
\label{22}
\end{equation}

\noindent where $\epsilon_1=\frac{v_{\rm r}}{v_{\rm el}}$ is the ratio of relaxational and elastic response to pressure.

The coefficients of thermal expansion of the liquid and the glass, $\alpha_{\rm l}$ and $\alpha_{\rm g}$, can be related in a similar way. Lets consider liquid relaxation in response to the increase of temperature by $\Delta T$. We write

\begin{equation}
\begin{aligned}
&\alpha_{\rm l}=\frac{1}{V_0^{\rm l}}\frac{v_{\rm el}+v_{\rm r}}{\Delta T}\\
&\alpha_{\rm g}=\frac{1}{V_0^{\rm g}}\frac{v_{\rm g}}{\Delta T}
\end{aligned}
\label{alphas}
\end{equation}

\noindent where $v_{\rm el}$ and $v_{\rm r}$ are volume changes due to solid-like elastic and relaxational response as in Eq. (\ref{moduli}) but now in response to temperature variation and $v_{\rm g}$ is elastic response of the glass. Combining the two expressions for $\alpha_{\rm l}$ and $\alpha_{\rm g}$ and assuming $V_{\rm l}^0=V_{\rm g}^0$ and $v_{\rm el}=v_{\rm g}$ as before, we find

\begin{equation}
\alpha_{\rm l}=(\epsilon_2+1)\alpha_{\rm g}
\label{33}
\end{equation}

\noindent where $\epsilon_2=\frac{v_{\rm r}}{v_{\rm el}}$ is the ratio of relaxational and elastic response to temperature.

Eqs. (\ref{22},\ref{33}) describe the relationships between $B$ and $\alpha$ in the liquid and the glass due to the presence of particle jumps in the liquid above $T_g$ and their absence in the glass at $T_g$, insofar as $T_g$ is the temperature at which $t<\tau$. Consistent with experimental observations, these equations predict that liquids above $T_g$ have larger $\alpha$ and smaller $B$ as compared to below $T_g$.

We are now ready to calculate $c_p$ below and above $T_g$. In the previous section, we have seen that in the highly viscous regime, $c_v$ is given by the vibrational component of motion only (see Eq. (\ref{ent5})). Hence, we use $c_v=3(1+\alpha T)$ from Eq. (\ref{sol}) which accounts for phonon softening due to inherent anharmonicity. Writing constant-pressure specific heat $c_p$ as $c_p=c_v+nT\alpha^2B$, where $n$ is the number density, we find $c_p$ above and below $T_g$ as

\begin{equation}
\begin{aligned}
&c_p^{\rm l}=3\left(1+\alpha_{\rm l}T\right)+nT\alpha_{\rm l}^2B_{\rm l},~~T>T_g\\
&c_p^{\rm g}=3\left(1+\alpha_{\rm g}T\right)+nT\alpha_{\rm g}^2B_{\rm g},~~T<T_g
\end{aligned}
\label{cp}
\end{equation}

Eq. (\ref{cp}) predicts that temperature dependence of $c_p$ should follow that of $\alpha$, in agreement with simultaneous measurements of $c_p$ and $\alpha$ showing that both quantities closely follow each other across $T_g$ \cite{take}.

From Eq. (\ref{cp}), $\frac{c_p^{\rm l}}{c_p^{\rm g}}$ can be calculated using experimental $B$ and $\alpha$ above and below $T_g$. This gives good agreement with the experimentally observed $\frac{c_p^{\rm l}}{c_p^{\rm g}}$ \cite{gla-tr}.

We have related the change of $c_p$ at $T_g$ to the change of system's thermal and elastic properties when the liquid falls out of equilibrium at $T_g$. It is important to note that $T_g$ is not a fixed temperature. $T_g$ decreases with the observation time, or increases with the quench rate $q$ (see, e.g., \cite{logar1,logar2}). This is a generic effect involved in many glass transition phenomena where a typical relaxation time exceeds the experimental time scale.

In Figure \ref{tg}, we show an example of how $T_g$, defined as the temperature of the jump of heat capacity, increases with the quench rate $q$ in an appreciably large temperature range.

\begin{figure}
\begin{center}
{\scalebox{0.45}{\includegraphics{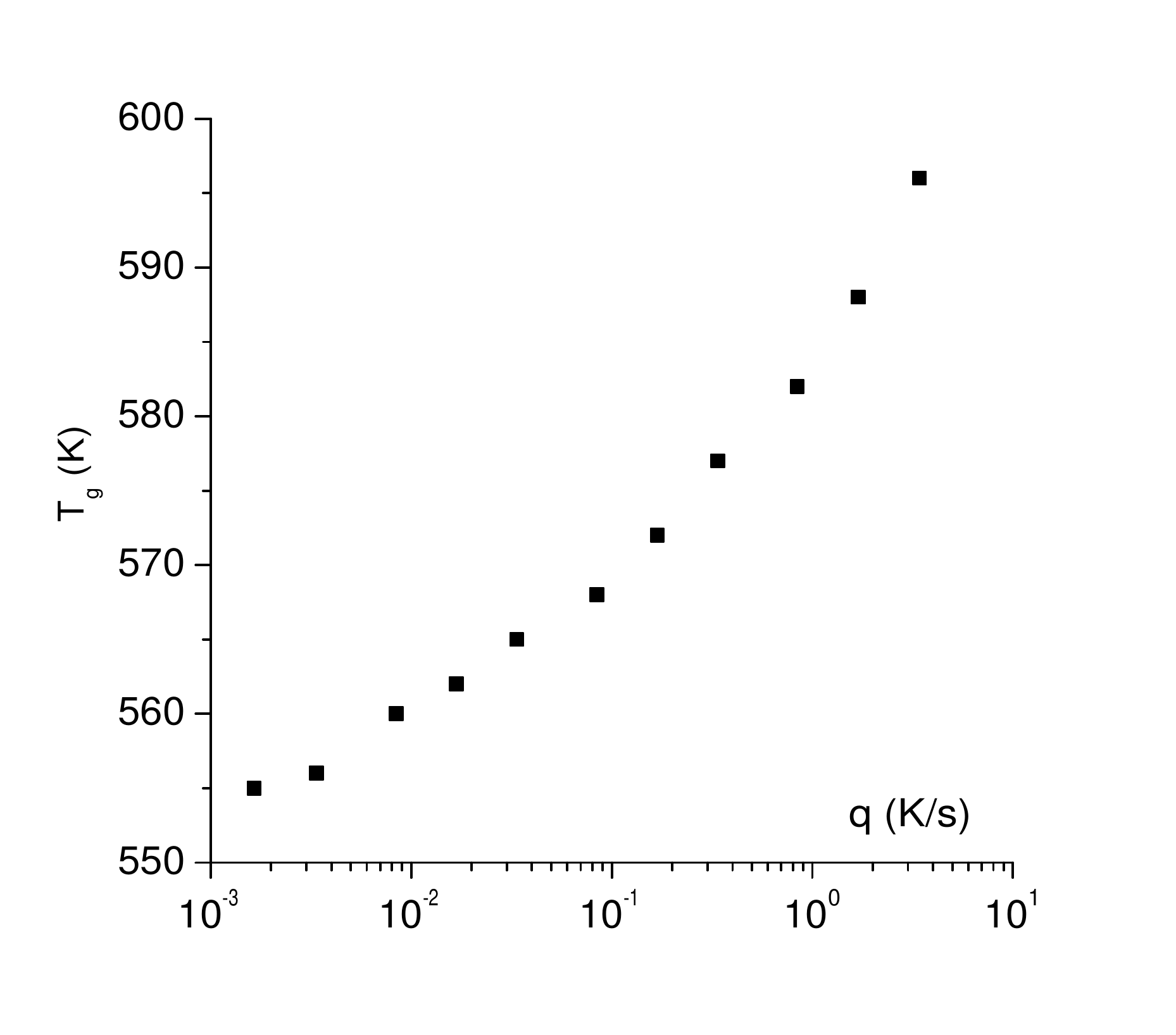}}}
\end{center}
\caption{Increase of $T_g$ in Pd$_{40}$Ni$_{40}$P$_{19}$Si$_{1}$ glass with the quench rate $q$ \cite{logar1}.}
\label{tg}
\end{figure}

This effect can be explained as follows. Recall that the jump of $c_p$ at $T_g$ takes place when the observation time $t$ crosses liquid relaxation time $\tau$. This implies that because $q=\frac{\Delta T}{t}$, $\tau$ at which the jump of heat capacity takes place is $\tau(T_g)=\frac{\Delta T}{q}$, where $\Delta T$ is the temperature interval of glass transformation range. Combining this with $\tau(T_g)=\tau_{\rm D}\exp(U/T_g)$ (here $U$ is approximately constant because $\tau$ is nearly Arrhenius around $T_g$ as discussed in the previous section ``Continuity of solid and liquid states...'') gives

\begin{equation}
T_g=\frac{U}{\ln\frac{\Delta T}{\tau_0}-\ln q}
\label{71}
\end{equation}

According to Eq. (\ref{71}), $T_g$ increases with the logarithm of the quench rate $q$. In particular, this increase is predicted to be faster than linear with $\ln q$. This is consistent with experiments \cite{logar1,logar2} and the data in Figure \ref{tg}. We note that Eq. (\ref{71}) predicts no divergence of $T_g$ because the maximal physically possible quench rate is set by the minimal internal time, Debye vibration period $\tau_{\rm D}$, so that $\frac{\Delta T}{\tau_{\rm D}}$ in Eq. (\ref{71}) is always larger than $q$.

Can the ``glass transition line'' be identified on the phase diagram separating the combined oscillatory and diffusive particle motion above the line from the purely oscillatory motion observed below $T_g$ during the experimental time scale? This would serve as the opposite to the Frenkel line which separates the combined oscillatory and diffusive particle motion below the line from the purely diffusive particle motion above the line at high temperature. As we have seen above, $T_g$ depends on the observation time (or frequency), and so no well-defined glass transition line exists on the phase diagram because the low-temperature state is a non-equilibrium liquid. The Frenkel line, on the other hand, separates two equilibrium states of matter.

To summarize this section, we have seen that several important experimental results of the glass transition, including the heat capacity jump and dependence of $T_g$ on the quench rate can be understood in the picture viewing the glass as the viscous liquid that falls out of equilibrium at $T_g$.

There are several other interesting non-equilibrium effects involved in liquid physics. These serve as good examples and case studies that inspire thinking about more general issues, including the foundations of statistical mechanics and their modification and extension to non-equilibrium conditions (see, e.g., \cite{evans}).

\section{Phase transitions in liquids}

We have ascertained several solid-like properties of liquids in the above discussion. The important basic property that immediately follows from this picture is that during time shorter than $\tau$, the local structure of the liquid does not change. This implies the presence of well-defined short- and medium-range order and transverse-like excitations as in solids and gives the possibility for the structure to undergo a phase transition. Phase transitions in liquids have been indeed found, although they were discovered fairly recently and their exploration started much later than of phase transitions in solids.

First-order transitions in liquids demarcate different local structures and thermodynamic properties. The transitions are common in multi-component systems and liquid crystals where composition and molecular orientation serve as order parameters. On the other hand, the possibility of the first-order transition in simple isotropic liquids was not known until about 2-3 decades ago, except for earlier theoretical works (see, e.g, Ref. \cite{tr1}).

In recent years, phase transformations have been found \cite{braball,tr2,tr3,tr4,tr5,tr6,tr7,tr8,tr9,tr10,tr11,tr12,tr13,tr15,tr16,tr17,tr18,tr181} in several different types of liquids, including in elementary liquids (e.g., P \cite{tr2,tr5,tr6,tr18}, Se \cite{tr2,tr4,tr7,tr8}, S \cite{tr2,tr4}, Bi \cite{tr2,tr4}, Te \cite{tr2,tr4}), oxide liquids (e.g., H$_2$O \cite{tr2,tr181}, Y$_2$O$_3$-Al$_2$O$_3$ \cite{tr5}, GeO$_2$ \cite{tr9}, B$_2$O$_3$ \cite{tr15}, P$_2$O$_5$ \cite{tr17}), halogenides (e.g., AlCl$_3$ \cite{tr10}, ZnCl$_2$ \cite{tr10}, AgI \cite{tr11}), and chalcogenides (e.g., AsS \cite{braball}, As$_2$S$_3$ \cite{tr16}, GeSe$_2$ \cite{tr12}). Pressure-induced transformations are accompanied by structural changes in both short-range and intermediate-range order as well as changes of all physical properties. Moreover, multiple pressure-induced phase transitions may take place in one system: for example, AsS undergoes the transformation between the molecular and covalent liquid, followed by the transformation to the metallic phase \cite{braball}. The transformations take place in the narrow pressure range and with large changes of structure and major properties such as viscosity.

Transformations in simple liquids can be both sharp and smeared. The analysis suggests that sharp transitions take place in liquids whose parent crystals undergo phase transitions with large changes of the short-range order structure and bonding type \cite{tr19}.

One of the first examples of sharp liquid-liquid transitions is the semiconductor-metal transformation in liquid Se \cite{tr7}. The transition is accompanied by the change of the short-range order structure, volume and enthalpy jumps as well as by very large jump of conductivity. Near the melting curve, the transition occurs at 700 C and 4 GPa. At very high temperatures this transition becomes smooth and finally almost disappears.

Another clear example of the sharp liquid-liquid transition in a simple isotropic system is the transition in liquid phosphorus \cite{tr6,tr18}. In Figure \ref{phos} we show sharp changes of the structure factor taking place in a narrow range of pressure and temperature. An abrupt and reversible structural transformation takes place between the low-pressure molecular liquid and the high-pressure polymeric liquid. This is shown in the phase diagram in Figure \ref{phos}b. As for Se, the line of liquid-liquid transformation is terminated at very high temperature only and above 2200\celsius.

\begin{figure}
\begin{center}
{\scalebox{0.4}{\includegraphics{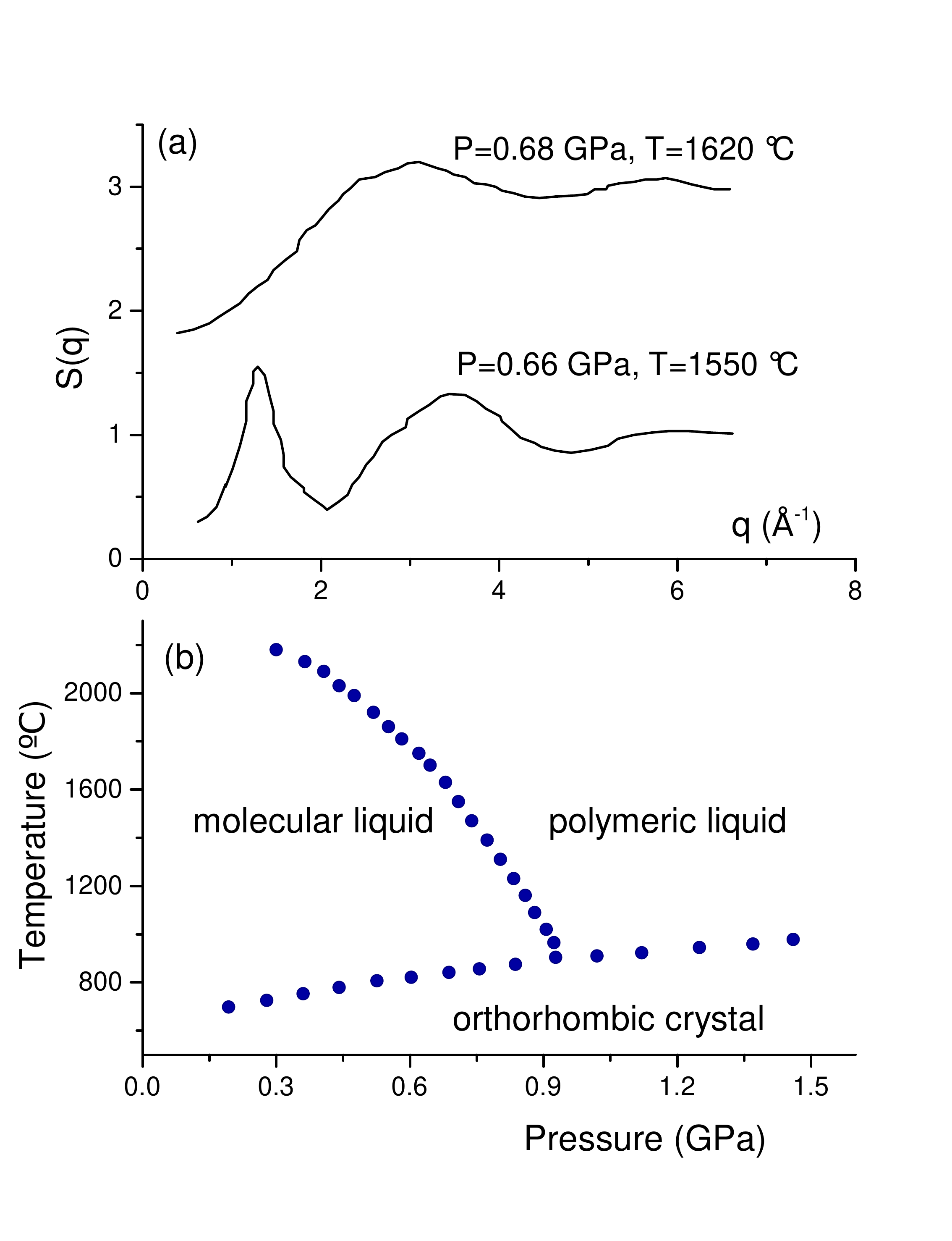}}}
\end{center}
\caption{(a) Sharp changes in the structure factor in liquid phosphorus in a narrow range of pressure and temperature. (b) Phase diagram showing the transition line between molecular and polymeric phosphorus. The data are from Ref. \cite{tr18}.}
\label{phos}
\end{figure}

The key to understanding these transitions lies in liquid dynamical properties. Indeed, if, as was often the case in the field, we consider a liquid as a structureless dense gas, no sharp phase transformations are possible. On the other hand, the oscillatory-diffusive picture of liquid dynamics based on $\tau$ offers a different insight. In the regime $\tau\gg\tau_{\rm D}$, particles perform many oscillations around fixed positions before jumping to the nearby quasi-equilibrium sites. Therefore, a well-defined short- and medium-range order exists during time $\tau$. In this case, liquids not far above the melting point can support pressure-induced sharp or smeared structural changes, similarly to their solid analogues.

Interestingly, the critical point of the liquid-gas transition in phosphorus is around 695$\celsius$ and 8.2 MPa. This means that the transition between the molecular and polymeric fluids takes place in the supercritical state. This may have come as a surprising finding in view of the perceived similarity of the supercritical state in terms of physical properties. Yet, as discussed above, well-defined short- and medium-range order exist in liquids above the critical point as long as the system is in the ``rigid''-liquid state below the FL where $\tau\gg\tau_{\rm D}$. We can therefore predict that liquid-liquid transitions in the supercritical state operate in the rigid-liquid below the FL but not in the non-rigid gas-like fluid state above the line.

Not surprisingly and similar to solids, liquid-liquid phase transitions are accompanied by the change of spectrum of collective modes. Recently, the evidence for this has started to come from experiments and modeling \cite{tr20}.

Understanding liquid structure and its response to pressure and temperature will continue to benefit from the development of experimental techniques and in-situ experiments in particular (see, e.g., Refs. \cite{salmon1,salmon2} for review).

\section{Quantum liquids: solid-like and gas-like approaches}

A quantum liquid is a liquid at temperature low enough where the effects related to particle statistics, Bose-Einstein or Fermi-Dirac, become operable. Quantum liquids is a large area of research (for review, see, e.g. \cite{landau,hydro,griffin,pines,leggett,annett}), largely stimulated by superfluidity in liquid helium. Here, we point to gas-like and solid-like approaches to quantum liquids and to similarities and differences of these approaches to those used in classical liquids discussed earlier.

%Thermodynamics of quantum liquids is often said to be well-understood \cite{griffin,pines,leggett,annett}), the assertion that may appear surprising given that thermodynamics of classical liquids is not.

The solid-like approach to the thermodynamics of quantum liquids is due to Landau. Emphasizing strong interactions in the liquid and rejecting earlier proposals which did not, Landau asserted that the energy of a low-temperature quantum liquid, such as liquid helium at room pressure, is the energy of the longitudinal phonon mode \cite{landau}.

In this consideration, the quantum nature of the liquid simplifies the understanding of its thermodynamics: Landau argued that any weakly perturbed state of the quantum system is a set of elementary excitations, or quasi-particles. In the low-temperature quantum liquid, the quasi-particles are phonons and are the lowest energy states in the system. This gives the solid-like heat capacity of a quantum liquid equal to that in the quantum solid but with one longitudinal mode only \cite{landau}:

\begin{equation}
c_v=\frac{2\pi^2n}{15\left(\hbar u\right)^3}T^3
\label{chelium}
\end{equation}

\noindent where $n$ is the number density and $u$ is the speed of sound.

Eq. (\ref{chelium}) is in agreement with the experimental heat capacity of liquid helium at room pressure.

Interestingly, Landau assumed that only one longitudinal mode contributes to the energy of a low-temperature quantum liquid and did not consider high-frequency transverse modes predicted earlier by Frenkel. This has been consistent with the absence of direct experimental evidence of transverse modes in liquid helium at room pressure. However, it is interesting to ask whether one should generally consider transverse modes in a hypothetical low-temperature liquid. As we have seen earlier (see section ``Phonon excitations at low temperature'' above), transverse modes do not contribute to the liquid energy in the limit of zero temperature. Hence Landau's assumption turned out to be correct.

In addition to explaining the experimental heat capacity, the solid-like phonon picture of liquid helium explained superfluidity. Superfluidity emerges due to the impossibility to excite phonons in the liquid moving slower than the critical velocity. In the original Landau theory, the critical velocity is the speed of sound. Considerably lower critical velocity found experimentally was later attributed to other effects such as energy-absorbing vortices.

The above low-temperature picture is discussed in the linear dispersion regime, $\epsilon=cp$. At higher temperatures, higher phonon branches become excited, including the roton part of the spectrum. Interestingly, the roton part, originally thought to be specific to helium, later discussed in the context of the Bose-Einstein condensate (BEC) \cite{griffin} and thought to be unusual in more recent discussions \cite{annett}, is seen in many classical high-temperature liquids (see, e.g., \cite{ruocco}, \cite{mon-na} and Figure \ref{disp1}).

In addition to the solid-like approach to liquid helium mentioned above, the hydrodynamic approach has been widely used to discuss hydrodynamic effects (naturally) such as density waves (first sound) and temperature or entropy waves (second sound) and their velocities \cite{hydro,pines}. Interestingly and similar to the classical liquids, two regimes of wave propagation and two sounds are distinguished depending on $\omega$. Waves with $\omega\tau<1$ are in the hydrodynamic regime, and are referred to as the first sound. Regime $\omega\tau_q>1$ corresponds to the ``quasi-particle'' sound, where $\tau_q$ is the lifetime of the quasi-particle excitation \cite{pines}, and is analogous to the solid-like elastic modes in classical liquids discussed above.

Interesting problems related to gas-like versus solid-approach emerge when the question of BEC in liquid helium is considered. As in the previous discussion, we can identify two approaches: gas-like and solid-like. The gas-like approach is due to Bogoliubov, and starts with the Hamiltonian describing weakly perturbed states of the Bose gas:

\begin{equation}
H=\sum_{\bf p}\frac{p^2}{2m}a_{\rm p}^+a_{\rm p}+\frac{1}{2}\sum U_{{\rm p}_1{\rm p}_2}^{{\rm p}^{\prime}_1{\rm p}^{\prime}_2}a^+_{{\bf p}^\prime_1}a^+_{{\bf p}^\prime_2}a_{{\bf p}_2}a_{{\bf p}_1}
\label{bogol}
\end{equation}

\noindent where the first and second terms represent kinetic and potential energy, $a^+_{\bf p}$, $a_{\bf p}$ are creation and annihilation operators and $U_{{\rm p}_1{\rm p}_2}^{{\rm p}^{\prime}_1{\rm p}^{\prime}_2}$ is the matrix element of the pair interaction potential $U(r)$.

Without the second term, the ground state of the system is the BEC gas state. For weak interactions, the energy levels of the system can be calculated in the perturbation theory. As a result, the diagonalised Hamiltonian reads \cite{landau}:

\begin{equation}
\begin{aligned}
&H=E_0+\sum\epsilon(p)b_{\rm p}^+b_{\rm p}\\
&\epsilon(p)=\sqrt{u^2p^2+\left(\frac{p^2}{2m}\right)^2}\\
&u=\sqrt{\frac{4\pi\hbar^2na}{m^2}}
\end{aligned}
\label{bogol1}
\end{equation}

\noindent where $n$ is concentration, $a=\frac{m}{4\pi\hbar^2}U_0$ and $U_0$ is the volume integral of the pair interaction potential.

According to Eq. (\ref{bogol1}), the presence of interactions modifies the energy spectrum of the Bose gas and results in the emergence of the low-energy collective mode with the propagation speed $u$. At small momenta, $\epsilon=up$.

This result is analogous to the gas-like approach to classical liquids where the weak interactions result in the low-frequency sound. What happens when interactions are strong as in liquid helium and when the perturbation theory does not apply? Here, we face the same problem of strong interactions as in the classical case.

Landau rejected the possibility of BEC in a strongly-interacting system: in his view, the low-energy states of the strongly-interacting system are collective modes rather than single-particle states as in gases, the picture similar to quantum solids where phonons are the lowest energy states and where BEC is irrelevant. In later developments, BEC was generalized for the case of strongly-interacting system on the basis of macroscopic occupation of some one-particle state. It was estimated that in low-temperature liquid helium, about 10\% of atoms are in the BEC state while the rest is in the normal state (in this picture, the interactions ``deplete'' BEC) \cite{pines}. The BEC component is then related to the superfluid component, and its weight changes with temperature.

It is probably fair to say that compared to well-studied effects of BEC in gases, operation of BEC in liquids is not understood in a consistent and detailed picture. Pines and Nozieres remark \cite{pines} that a quantitative microscopic theory of liquid helium is yet to emerge. Leggett comments on the challenge of obtaining direct experimental evidence of BEC in liquid helium as compared to gases \cite{leggett}.

We now recall our starting picture of liquids where particle motion includes two components: oscillatory and diffusive. Can quantum liquids be understood on the basis of these two types of motion only, similarly to classical liquids? An interesting insight has come from path-integral simulations \cite{ceperley}: the emergence of macroscopic exchanges of diffusing atoms contributes to the $\lambda$-peak in the heat capacity, confirming the earlier Feynman result \cite{feynman}, and is related to momentum condensation and superfluidity.

This picture enables one to adopt the solid-like approach to quantum liquids (instead of the commonly discussed gas-like approach): we approach the system from the solid state where strong interactions and resulting collective modes are considered as a starting point, and introduce diffusive particle jumps as in the classical case. From the thermodynamic point of view, these jumps only modify the phonon spectrum in classical liquids. In quantum liquids, they additionally contribute to the exchange energy because particle jumps enable the effect of quantum exchange \cite{leggett1}.

Can the exchange energy be related to exchange frequency $\omega_{\rm F}$, as is the case for liquid energy in Eqs. (\ref{harmo}), (\ref{anharm2}) or (\ref{enf})? This would amount to a Frenkel reduction discussed earlier but applied to the exchange energy. We think interesting insights may follow. We feel that generally developing closer ties between the areas and tools of classical and quantum liquids should result in new understanding.

\section{Mixed and pure dynamical states: liquids, solids, gases}

The emphasis of our review has been on understanding experimental and modeling data and on providing relationships between different physical properties. In addition to this rather practical approach, we can revisit a more general question alluded to in the Introduction: how are we to view and classify liquids in terms of their proximity to gases or solids? Throughout the history of liquid research, different ways of addressing this question were discussed \cite{frenkel,landau,ziman,boonyip,march,march1,baluca,faber,hansen2,eyring,landmech}.

On the basis of discussion in this paper, our answer is that liquids do not need such a classification, or any other compartmentalizing for that matter. With their interesting and unique properties, liquids belong to a state of their own. Throughout this review, we have seen that most important properties of liquids and supercritical fluids can be consistently understood in the picture where we are compelled to view them as distinct systems in the notably {\it mixed} dynamical state. Particles undergo both oscillatory motions and diffusive jumps, and the relative weight of the two components of motion changes with temperature. As discussed in the section ``Viscous liquids'' above, this relative weight is quantified by the ratio $\frac{\tau_{\rm D}}{\tau}$, which we now define as parameter $R$:

\begin{equation}
R=\frac{\tau_{\rm D}}{\tau}=\frac{\omega_{\rm F}}{\omega_{\rm D}}
\label{r}
\end{equation}
\noindent

We have seen that $R$ enters the energy of both low-viscous liquids (see Eqs. (\ref{harmo}),(\ref{anharm2}),(\ref{enf})) and highly-viscous liquids including in the glass transformation range (see Eqs. (\ref{1}),(\ref{2}),(\ref{ave1})) and is implicitly present throughout our discussion.

In liquids, $R$ varies between 0 and 1, and defines the liquid's proximity to the solid or gas state. This enables us to delineate solids and gases as two limiting cases in terms of dynamics and thermodynamics.

In solids, particle motion is purely oscillatory, corresponding to $R=0$. Indeed, $\tau\rightarrow\infty$ in ideal crystals or becomes astronomically large in familiar glasses such as SiO$_2$ at room temperature \cite{gla-tr}.

In gases, particle motion is purely diffusive. This corresponds to $R=1$, as is the case in the supercritical state above the FL where the oscillatory component of particle motion is lost and where $\tau\approx\tau_{\rm D}$.

We note that $R=1$ at the FL above the critical point or in subcritical liquids constitutes a microscopic and physically transparent criterion of the difference between liquids and gases \cite{phystoday}. Indeed, existing common criteria include distinctions such as that gas fills available volume but liquid does not, or that gas does not possess a cohesive state but liquid does. These criteria are either not microscopic, are tied to a particular pressure range or can not be implemented in practice \cite{phystoday}. On the other hand, asserting that the gas state is characterized by purely diffusive dynamics of particles whereas the liquid state includes both diffusive and oscillatory components of particle motion gives a microscopic and physically transparent criterion.

%In terms of energy, $R$ is related to the ratio of kinetic and potential energy. Indeed, small $R$ in viscous liquids and glasses signifies the smallness of kinetic energy relative to potential energy as in solids. At the FL separating liquid-like and gas-like dynamics where $R=1$, particle kinetic and potential energies are approximately equal to each other \cite{pre}.

We therefore find that $R=0$ and $R=1$ give solids and gases as two limiting cases of dynamical properties. In this sense, gases and solids are {\it pure} states of matter in terms of their dynamics. It is for this reason that they have been well understood theoretically. Liquids, on the other hand, are a {\it mixed} state in terms of their dynamics, the state that combines solid-like and gas-like motions. It is the mixed state which has been the ultimate problem for the theory of liquids.

On the basis of $R$-parameter, we see that liquids can only be viewed as solid-like or gas-like when $R$ is either close to 0 or 1. In all other cases, liquids are thermodynamically close to neither state. This becomes apparent from looking at the experimental thermodynamic data such as in Figure \ref{cv3}. This highlights our earlier point about the distinct mixed dynamical state of liquids and associated rich physics.

Once the last assertion is appreciated, theorists become better informed about what approach to liquids is more appropriate. The best starting point for liquid theory is to make no assumptions regarding the proximity of liquids to gases or solids and seek no extrapolations of the hydrodynamic regime to the solid-like regime and vice versa. Instead, the best starting point is to consider the microscopic picture of liquid dynamics and its mixed character from the outset, and recognize that the relative weights of diffusive and oscillatory components change with temperature. Depending on the property in question, we can encounter several possibilities.

If we are concerned with long-time and low-energy observables only ($t>\tau$ or $\omega\tau<1$), the relevant equations are hydrodynamic \cite{hydro}. Well-understood, these equations describe hydrodynamic properties independently and separately from the solid-like regime. The solid-like approach to liquids does not apply to hydrodynamic effects.

If we are interested in thermodynamic properties such as energy and heat capacity, it is the solid-like properties of liquids that matter most because high-frequency modes contribute to the liquid energy almost entirely and propagate in the solid-like regime $\omega\tau>1$. In this case, we can focus on the solid-like regime of liquid dynamics from the outset and treat it separately and independently from the hydrodynamic regime. In this approach, we do not need to extrapolate the hydrodynamic description into the solid-like regime as is done in generalized hydrodynamics and where extrapolation schemes may be an issue.

Each regime, hydrodynamic or solid-like, can be analyzed separately. There are also mixed cases where, for example, we observe solid-like high-frequency modes ($\omega\tau>1)$ at long times ($t>\tau$) because we are interested in their propagation length. Here, we can start with either hydrodynamic or elasticity equations and modify them appropriately. This gives the same results as we have seen above.

\section{Conclusions and outlook}

Our important conclusion regarding the theoretical view of liquids has already been made in the previous section. In this review, we discussed how this view evolved and how different ideas proposed at very different times were developing. With the recent evidence for high-frequency solid-like modes in liquids, it has now become possible to use the solid-like approach to liquids and discuss their most important thermodynamic properties such as energy and heat capacity. We have reviewed how this can be done for liquids in different regimes: low-viscous subcritical liquids, high-temperature supercritical gas-like fluids, viscous liquids in the glass transformation range and systems at the liquid-glass transition. In each case, we have noted limitations and caveats of this approach throughout this review.

As alluded to in the Introduction, liquids have been viewed as inherently complicated systems lacking useful theoretical concepts such as a small parameter. New understanding of liquids, including the increasing amount of high-energy experimental data and its quantitative agreement with predicted thermodynamic properties, change this traditional perspective. We are beginning to understand liquid thermodynamics on the basis of high-frequency collective modes. Contrary to the pessimistic and vague picture often drawn about them, liquids are emerging as exciting and unique systems amenable to theoretical understanding in a consistent picture.

Several points can be mentioned that may advance liquid research. The evidence for high-frequency collective modes and transverse modes in particular has started to emerge fairly recently. It will be interesting to widen the number of systems with mapped solid-like dispersion curves and go beyond simple liquids. It will also be interesting to extend the experiments to high temperature and pressure including the supercritical state and to follow the evolution of collective modes as predicted theoretically. This can be directly compared to the concomitant variation of thermodynamic properties such as heat capacity.

We have not reviewed molecular dynamics (MD) simulations although in places we discussed modeling aimed at backing up experiments and theory. MD simulations of liquids are as old as the method itself: indeed, the need for MD simulations was originally rationalized by the difficulty to construct liquid theory \cite{allen}, with simulations playing the role of testing the theory. With the first simulation of liquids performed in 1957, the generated data exceeds what is feasible to review. For more recent examples, an interested reader can consult liquid textbooks and review papers cited throughout this review. A common issue faced by computer simulations is the same as in experiments: understanding and interpreting the data. With reliable interatomic potentials existing today, it is not hard to calculate $c_v$ shown in Figure \ref{cv3}, but understanding the results requires a physical model. As far as liquid heat capacity is concerned, it is fair to say that MD simulations have not resulted in understanding liquid $c_v$ such as shown in Figure \ref{cv3}. Once liquid thermodynamics is better understood, MD simulations will provide interesting microscopic insights and potentially uncover novel effects. These can include the operation of collective modes in the solid-like elastic regime and their evolution at conditions not currently sampled by experiments including in the supercritical state.

We feel that bringing concepts and tools from classical and quantum liquids closer may result in new understanding, particularly in the area of thermodynamics and operation of BEC in real strongly-interacting liquids. Exploring the mixed state of liquid dynamics and the separation of solid-like oscillatory and gas-like diffusive particle dynamics in quantum liquids may bring unexpected new insights.

We are grateful to S. Hosokawa and A. Mokshin for discussions and providing data and to EPSRC for support.

\end{document}